\documentclass[preprint,aps,tightenlines,showpacs]{revtex4}
\usepackage[dvips]{graphicx}
\usepackage{dcolumn}
\begin{document}
\title{Quark-Hadron Duality and Parity Violating Asymmetry \\
 of Electroweak Reactions in the $\Delta$ Region }
\author{
 K. Matsui$^a$, T. Sato$^a$, and
T.-S. H. Lee$^b$
}
\affiliation{
$^a$
Department of Physics, Osaka University, Toyonaka, Osaka 560-0043,
 Japan\\
$^b$
Physics Division, Argonne National Laboratory, Argonne,
Illinois 60439}

\begin{abstract}
A dynamical model\cite{sl1,sl2,sl3}
of electroweak pion production
reactions in the $\Delta$(1232) region
has been extended to include the neutral current contributions for examining
the local Quark-Hadron Duality in neutrino-induced reactions and for
investigating how the axial $N$-$\Delta$ form factor
can be  determined by the parity violating
asymmetry of $N(\vec{e},e^\prime)$ reactions.
We first show that the recent data of
 $(e,e^\prime)$ structure functions $F_1$ and
$F_2$, which exhibit the Quark-Hadron Duality, are in good 
agreement with our predictions. For possible future experimental tests,
we then predict
that the structure functions $F_1$, $F_2$, and $F_3$ for $(\nu,e)$ and
$(\nu,\nu^\prime)$ processes
also show the similar Quark-Hadron Duality.
The spin dependent structure functions $g_1$ and $g_2$
of $(e,e^\prime)$ have also been calculated from our model.
It is found that the local Quark-Hadron Duality is not seen in
the calculated  $g_1$ and $g_2$, while our results for $g_1$ and
some polarization observables associated with the exclusive 
$p(\vec{e},e^\prime \pi)$
and $\vec{p}(\vec{e},e^\prime\pi)$ reactions   
are in reasonably good agreement with the recent data.
In the study of parity violating 
asymmetry  $A$ of $N(\vec{e},e^\prime)$ reactions, the relative 
importance between the nonresonant mechanisms and the $\Delta$ excitation is
investigated by taking into account the unitarity condition.
Predictions are made for using the data of $A$ to
test the axial $N$-$\Delta$ form factors determined previously in
the studies of $N(\nu_\mu,\mu^-\pi)$ reactions.
The predicted asymmetry $A$ are also compared with
the Parton Model predictions for future experimental investigations
of Quark-Hadron Duality.

\end{abstract}
                                                                                
\maketitle

\section{Introduction}

In recent years, a dynamical model had been developed\cite{sl1,sl2}
to investigate electromagnetic pion production reactions in the region 
near the $\Delta$(1232) resonance. The model was subsequently extended\cite{sl3}
to also investigate neutrino-induced reactions $(\nu_\mu, \mu^-\pi)$.
Fairly consistent descriptions
of all of the available data in the $\Delta$ region have been obtained. 
In this work, we further extend this
model (called the SL model ) 
to address two questions of current interest : 
(1) Will  Hadron-Quark
Duality first observed by Bloom and Gilman\cite{bg} in inclusive $p(e,e')$
be also seen in
the neutrino-induced $N(\nu,e)$ and $N(\nu,\nu^\prime)$ reactions ?
(2) How can  parity violating asymmetry
of inclusive $N(\vec{e},e)$ reaction be used to improve
 our knowledge about the axial $N$-$\Delta$ form factor ?
Experimental data for testing our predictions can be obtained
at new neutrino facilities\cite{fermi} and Jefferson
Laboratory (JLab)\cite{well,bosted}.
 
Our first step is to construct neutral currents within
the SL model.  This then allows us to extend our previous
calculations\cite{sl3} of $N(\nu_\mu, \mu^-\pi)$ reactions to
also predict $N(\nu,\nu^\prime \pi)$ reactions for examining some
questions concerning the Quark-Hadron Duality.
Simply speaking, an inclusive lepton
scattering observable exhibits Quark-Hadron Duality when 
an average of this quantity over an appropriately chosen
scaling variable in the resonance region
is close to that in the Deeply Inelastic Scattering (DIS) region.
More details on this subject can be found 
in a recent review by Melnitchouk, Ent and Keppel\cite{meln}.
The Quark-Hadron Duality was first observed\cite{bg}
 for the $p(e,e^\prime)$ structure function
$F_2=\omega W_2$.
Recent experimental data\cite{nicu,liang} for both
$F_2$ and $F_1=m_N W_1$  have further
confirmed more quantitatively this interesting observation.
Theoretical attempts in
understanding the Quark-Hadron Duality within QCD
were  already made\cite{nach,gp,rgp} in 1970's. More recent
works are reported in, for example,  
Refs.\cite{xi1,xi2,mueller,carlson,edel,meln-1}. 
Within the Standard model\cite{holstein}, it is natural to ask whether
the Quark-Hadron Duality should also exist in the neutrino-induced
$N(\nu, e)$
and $N(\nu,\nu^\prime)$ reactions (from nowon, $\nu$ and $e$ denote
the neutrinos and charged leptons, respectively,
of any generation within the Standard Model).
In the absence of necessary neutrino data,
this can be explored theoretically in the $\Delta$ region using
the extended SL model presented in this paper.
Obviously, we can only explore the 'local' Quark-Hadron Duality in the
$\Delta$ region.

We will also report on our investigations of
the spin dependent structure functions $g_1$ and $g_2$
of $(e,e^\prime)$ processes. To see the accuracy of our model in predicting
these quantities, we first show that some recent data\cite{biselli,joo}
of the polarization
observables associated with the exclusive $p(\vec{e},e^\prime \pi)$
and $\vec{p}(\vec{e},e^\prime \pi)$ reactions agree well with 
the predictions from the considered SL model.
We then find that $g_1$ calculated from our model
agree reasonably well with the recent data\cite{fatemi}, but do not show
Quark-Hadron Duality when compared with the
DIS data\cite{abe}. 
The calculated $g_2$ is also found to be rather different from
the DIS data which can be 
described reasonably well by the Wandzura-Wilczek formula\cite{wwil}.

Our second task in this work is to apply
the extended SL model to address some
questions concerning  the $N$-$\Delta$  form factors.
It has been well recognized that these form factors are important information 
for testing current hadron models and
also lattice QCD calculations in the near future. The vector parts of
the $N$-$\Delta$  form factors in the  $Q^2\leq 4$ (GeV/c)$^2$ region
have been rather well determined 
by analyzing very extensive high precision
data of electromagnetic pion production reactions,
as reviewed in Ref.\cite{burkertlee}. On the other hand, the axial vector
$N$-$\Delta$ form factor is not well determined mainly
because of the short of high precision data of neutrino-induced
pion production reactions in the $\Delta$ region.
For example, 
the axial $N$-$\Delta$ from factor determined in Ref.\cite{sl3}
is rather different from the one determined
previously\cite{kitagaki}. This leads to
uncertainties in interpreting the axial $N$-$\Delta$ form factor
in terms of the hadron structure calculations such as those
reported in Refs.\cite{nimi-1,nimi-2,golli,bruno}.
The situation
will be improved when the data from new neutrino facilities become
available in the near future. Alternatively, progress can be made
by following Refs.\cite{cg,jones,isha,hwang,nath,musolf,hammer,nimi-3,rekalo}
to investigate the parity violating asymmetry of inclusive
$N(\vec{e},e^\prime)$ reactions. 
This polarization observable
is due to the interference between the
electromagnetic currents and neutral currents and hence can
be used to explore the axial $N$-$\Delta$ form factor.
Experimental effort in this direction is being made\cite{well}
at JLab. To facilitate this
study, we have applied
 the extended SL model
to explore the dependence of the
parity violating asymmetry of
$N(\vec{e},e^\prime)$ on the axial $N$-$\Delta$ form factor.

In section II, we will specify the considered models
of electromagnetic currents ($em$), weak charged currents ($cc$),
and weak neutral currents ($nc$). We first recall their forms in the 
Standard model\cite{holstein}
and then specify how such currents are
defined in terms of hadronic degrees of freedom of the SL model.
In section III, we give expressions of the inclusive cross sections
and indicate how the structure functions
are calculated within our hadronic model
and Parton Model.
Section IV is devoted to present results for studying the
Quark-Hadron Duality in the $\Delta$ region.
The predicted  $N(e,e^\prime)$ structure functions will also 
be compared with the recent data.
The results for the parity violating
asymmetry of the inclusive $N(\vec{e},e^\prime)$
will be given in section V. 
A summary is given in section VI.

\section{Models of Electroweak Currents}

We first recall the electroweak currents defined in the Standard 
Model\cite{holstein}.
In the considered $\Delta$ excitation region, we can 
eliminate heavy $W$ and $Z$ bosons and keep interactions involving
only up ($u$) and down ($d$) quarks.
The interaction Lagrangian for our study can then be written as
\begin{eqnarray}
L_{eff} & =&-\sqrt{4\pi\alpha}
[\bar{e}\gamma^\mu e + j^\mu_{em}]  A_\mu \nonumber \\
& & - \frac{G_F V_{ud}}{\sqrt{2}}[
  \bar{\nu} \gamma_\mu(1-\gamma_5)e j^{\mu\dagger}_{cc}
 + \bar{e} \gamma_\mu(1-\gamma_5)\nu j^\mu_{cc}]
 \nonumber \\
& & - \frac{G_F}{\sqrt{2}}[
   \bar{\nu}\gamma_\mu(1-\gamma_5)\nu
 + \bar{e}(2g_V^e\gamma_\mu - 2 g_A^e \gamma_\mu\gamma^5)e
 ]j^\mu_{nc} \,,
\end{eqnarray}
where $\alpha=1/137$, $ G_F =1.1664\times 10^{-5}$ GeV$^{-2}$,
$g_V^e = - 1/2 + 2 \sin^2\theta_W$, $g_A^e = -1/2$, 
$A_\mu$ is the photon field,
$e$ and
$\nu$ are the field operators of the charged leptons and neutrinos, 
respectively. The Weinberg angle
$\theta_W$  is 
 known empirically to be   
 $\sin^2\theta_W=0.231$ and
$V_{ud} = 0.974$ is the the Cabibbo-Kobayashi-Maskawa (CKM) coefficient.
The electromagnetic current ($j^\mu_{em}$), weak charge current
($j^\mu_{cc}$) and weak neutral current ($j^\mu_{nc}$)
carried by $u$ and $d$  quarks can be written as
\begin{eqnarray}
j^\mu_{em} &= & \frac{2}{3}\bar{u}\gamma^\mu u -
 \frac{1}{3}\bar{d}\gamma^\mu d \,, \\
j^\mu_{cc} & = &  \bar{u}\gamma^\mu(1 - \gamma_5) d \,, \\
j^\mu_{nc} & = & \frac{1}{2}[\bar{u}\gamma^\mu(1-\gamma_5)u
 - \bar{d}\gamma^\mu(1-\gamma_5)d ] - 2 \sin^2\theta_W j_{em}^\mu \,.
\end{eqnarray}

With the simplification that only $u$ and $d$ quarks are kept, it is
well known\cite{musolf} that the above currents
can be classified according 'strong' isospin.
We thus can write
\begin{eqnarray}
j^\mu_{em} &=& V^\mu_3+ V^\mu_{isoscalar} \,, \\
j^\mu_{cc} &=& (V^\mu_1+i V^\mu_2) - (A^\mu_1 + i A^\mu_2) \,, \\
j^{\mu\dagger}_{cc} &=& (V^\mu_1-i V^\mu_2) - (A^\mu_1 - iA^\mu_2) \,,\\
j^\mu_{nc} &=& (1-2 \sin^2\theta_W) j^\mu_{em} - V^\mu_{isoscalar}
-A^\mu_3 \,,
\end{eqnarray}
where
the isospin components of the
vector ($V$) and axial vector ($A$) currents are defined as
\begin{eqnarray}
V^\mu_i &=& \bar{q}\gamma^\mu \frac{\tau_i}{2} q \,, \\
A^\mu_i &=& \bar{q}\gamma^\mu\gamma_5 \frac{\tau_i}{2} q \,, \\
V^\mu_{isoscalar} &=&\bar{q}\frac{1}{6}\gamma^\mu q  \,.
\end{eqnarray}
Here we have defined an isospin doublet field operator
 $\bar{q}=q^+\gamma^0=(u^+,d^+)\gamma^0$ and 
$\tau_i$ is the usual Pauli operator.

The above expressions will allow us to calculate electroweak structure
functions of deeply inelastic
$(e, e^\prime)$, $(\nu, e)$ and $(\nu,\nu^\prime)$
processes within the Parton Model, as
explained in, for example, Ref.\cite{tw}.
 However, they can not be used directly
for investigating meson production reactions in the resonance region
where the perturbative QCD is not applicable.
At the present time, the most tractable ways for investigating 
these reactions are in terms of
hadronic degrees of freedom.
The starting point is  a hadronic effective Lagrangian
constrained by the symmetry properties of the Standard Model.
Accordingly,  the resulting electroweak  currents have the forms of
Eqs.(5)-(8), but are written in terms of hadronic field operators.
This can be achieved by
using the standard effective chiral lagrangian methods\cite{park,gl,holstein}.
In the SL model developed in 
Refs.\cite{sl1,sl2,sl3}, the constructed electroweak
 currents are expressed in terms of
the field operators of the nucleon ($N$), Delta ($\Delta$),
pion ($\vec{\pi}$), omega meson ($\omega$), and rho meson ($\vec{\rho}$).
They can be written as
\begin{eqnarray}
\vec{V}^\mu\cdot\vec{v}_\mu & = & \bar{N}
   {}   [\gamma^\mu \vec{v}_\mu - \frac{\kappa^V}{2m_N}\sigma^{\mu\nu}
\partial_\nu\vec{v}_\mu]\cdot
    \frac{\vec{\tau}}{2}N
 + \frac{g_A}{2F}\bar{N} \gamma^\mu\gamma_5
[\vec{v}_\mu\cdot \vec{\tau}]N \times \vec{\pi} \nonumber \\
& &  + [\vec{\pi} \times \partial^\mu \vec{\pi}]\cdot\vec{v}_\mu
  - \frac{g_{\omega\pi V}}{m_\pi}\epsilon_{\alpha\mu\nu\delta}
[\partial^\alpha\vec{v}^\mu]\cdot \vec{\pi}
[\partial^\nu\omega^\delta ] 
 -i \bar{\Delta}_\mu \vec{T}\cdot \vec{v}_\nu \Gamma^{\mu\nu}_VN \,, \\
\vec{A}^{\mu}\cdot \vec{v}_\mu & = & 
g_A \bar{N} \gamma^\mu \gamma_5\frac{\vec{\tau}}{2}\cdot \vec{v}_\mu N
     - F \partial^\mu \vec{\pi} \cdot \vec{v}_\mu
     - f_{\rho\pi A}(\vec{\rho}^\mu \times \vec{\pi})\cdot\vec{v}_\mu
 +  \bar{\Delta}_\nu \vec{T}\cdot \vec{v}_\mu \Gamma^{\mu\nu}_A N \,,
\end{eqnarray}
where $\vec{v}_\mu$ is an arbitrary isovector function,
$F=93$ MeV is the pion decay constant, and $g_A=1.26$ is
the axial coupling strength of the nucleon.
 The other parameters
as well as the vertex $\Gamma^{\mu\nu}_{A,V}$ and isospin 
operator $T$ of 
the $N$-$\Delta$ transition
are given in Refs.\cite{sl1,sl2,sl3}.
With the electromagnetic current $j^\mu_{em}$
given explicitly in Ref.\cite{sl1},
the relation Eq.(5) and Eq.(12) can be used to get $V^\mu_{isoscalar}$
 and hence
$j^\mu_{nc}$ of Eq.(8) is also completely determined within the SL
model. The neutral current $j^\mu_{nc}$ is needed to extend the calculations
of Refs.\cite{sl1,sl2,sl3} to also study $(\nu,\nu^\prime\pi )$ reactions and
the parity violating asymmetry
  of the $(\vec{e},e^\prime)$ reaction.
 
The electroweak currents defined by Eqs.(5)-(8) and (12)-(13)  must be
supplemented by additional effective Lagrangians describing the haronic 
interactions, such as $\pi NN$, $\pi N \Delta$, $\rho NN$, 
$\rho\pi\pi$, and $\omega NN$ couplings,  to 
calculate the meson  production amplitudes within the SL model.
The details of
such a dynamical approach have been given
in Refs.\cite{sl1,sl2,sl3}. Here it is sufficient
to just illustrate schematically the
basic meson production  mechanisms of the SL model.
In Fig.1, we show the constructed non-resonant pion production
mechanisms.  For the $em$ current contributions, the
waved-line is photon and all diagrams contribute.
For $cc$ and $nc$ current contributions, the
wave-line is the vector current $V$ or axial vector current $A$.
 Like the $em$ case, all terms in Fig.1 contribute to
$nc$ amplitude but with different isospin weighting factors defined by
Eqs.(5) and (8). For the $cc$ contributions,
 all terms contribute to the amplitudes induced by the vector current
$V$ except that 
the $\rho$ exchange in (c) should be excluded because it is an isoscalar 
interaction. The $cc$'s amplitude due to axial vector current
$A$ contains only mechanisms
(a), (b), (e) with only $\rho$ exchange, and (f).
In both $cc$ and $nc$ cases, the pion pole term due to the
second term $-F\partial^\mu\vec{\pi}\cdot \vec{v}_\mu$ of axial current
Eq.(13) has additional contributions, as explained in Ref.\cite{sl3}.

The $\Delta$ excitation mechanism is illustrated in Fig.2.  
For the later discussions in this paper, here we recall that the
matrix element of the
axial $N$-$\Delta$ coupling  $\Gamma_A^{\mu\nu}$ in Eq.(13) can be written
as
\begin{eqnarray}
<p_\Delta \mid \Gamma^{\mu\nu}_A \mid p_N >
&=& d_1(q^2)g^{\mu\nu} + \frac{d_2(q^2)}{m^2_N}P_\alpha(q^\alpha g^{\mu\nu}
-g^{\alpha\mu}q^\nu) \nonumber \\
& &-\frac{d_3(q^2)}{m^2_N}q^\mu p_N^\nu - i\frac{d_4(q^2)}{m^2_N}
    \epsilon^{\mu\nu\alpha\beta}P_\alpha q_\beta\gamma_5 \,,
\end{eqnarray}
where $p_\Delta$ and $p_N$ are the momenta of the $\Delta$ 
and $N$, respectively, $q=p_\Delta-p_N$, and $ P=(p_\Delta+p_N)$.
The strengths of the form factors  at
$Q^2=-q^2=0$ are identified with the
 quark model predictions\cite{nimi-1} and are
found to be $d_i(0)= D_i \frac{6}{\sqrt{2}5}g_A$
with $g_A=1.26$, $D_1=0.969$, $D_2=-0.180$, $D_3=0.836$, and $d_4=0$.
In Ref.\cite{sl3}, it was found that all of the available data
of $N(\nu_\mu,\mu^-\pi)$ data in the $\Delta$ region can be well
described if the $q^2$-dependence of each form factor in
Eq.(14) is taken to be ($Q^2 = -q^2$)
\begin{eqnarray}
d_{1,2}(Q^2) & = & d_{1,2}(0)G^A_{N,\Delta}(Q^2) \,, \nonumber \\
d_3(Q^2) & = & d_3(0)\frac{m_N^2}{Q^2 + m_\pi^2}G^A_{N,\Delta}(Q^2)\,,
 \nonumber
\end{eqnarray}
with
\begin{eqnarray}
G^A_{N,\Delta}(Q^2)= (1+aQ^2)exp(-bQ^2)G_A(Q^2) \,,
\end{eqnarray}
where $a = 0.154$ (GeV/c)$^{-2}$, $b=0.166$ (GeV/c)$^2$, and
$G_A(Q^2) = 1/(1+Q^2/M^2_A)^2$ with $M_A = 1.02$ GeV is the
nucleon axial form factor\cite{miss}. 
The corresponding vector form factor $G^V_{\Delta,N}(Q^2)$,
which is associated with $\Gamma^{\mu\nu}_V$ of Eq.(12)
and has been well determined\cite{sl1,sl2} by analyzing the data
of pion electroproduction,
 has the same form as Eq.(15) except that
$G_A(Q^2)$ is replaced by the
usual proton form factor $G_p(Q^2)=1/(1+Q^2/\Lambda_p^2)^2$
with $\Lambda_p^2 = 0.71$ (GeV/c)$^2$.

We further mention 
that the  electroweak
pion production amplitudes (denoted as $J$ from nowon) 
 calculated within the SL model
can be written schematically
in the center of mass (c.m.) frame of the final $\pi N$ system as
\begin{eqnarray}
J_{\pi N, a N}(k,q,W) = J_{\pi N, a N}^{(n.r.)}(k,q,W) +
 \frac{\bar{\Gamma}^\dagger_{\Delta,\pi N}(k,W)
\bar{\Gamma}_{\Delta, a N}(q,W)}{W-m_\Delta -\Sigma(W)} \,,
\end{eqnarray}
where $a = \gamma $, W or $Z$ bosons, $q$ and $k$ are the relevant initial
and final momenta, 
$\Sigma(W)$ is the $\Delta$ self-energy.
The non-resonant amplitudes $J^{(n.r.)}_{\pi N, a N}(k,q,W)$ are calculated 
from
\begin{eqnarray}
J^{(n.r.)}_{\pi N, a N}(k,q,W) = v_{\pi N,aN} + t_{\pi N,\pi N}G_{\pi N}(W)
v_{\pi N,aN} \,,
\end{eqnarray}
where $v_{\pi N,aN}$ are the  non-resonant amplitudes illustrated in Fig.1,
$t_{\pi N,\pi N}$ is the non-resonant
$\pi N$ scattering amplitude, and $G_{\pi N}(W)$ is
the $\pi N$ propagator. Note that the second term of Eq.(17) is the 
consequence of the unitarity condition. This is neglected in the previous
investigations\cite{hwang, hammer,nimi-3,rekalo} of the effects
due to the non-resonant amplitudes on the
parity violating asymmetry of $N(\vec{e},e^\prime)$ reaction.
Furthermore, the non-resonant mechanisms considered in those works also differ
from what are illustrated in Fig.1 and well
tested in extensive investigations of
exclusive $N(\gamma,\pi)$, $N(e,e^\prime\pi)$, and $N(\nu_\mu,\mu^- \pi)$ 
reactions. For example, the vector meson exchanges are not included in
Refs.\cite{hammer,nimi-3} and pseudo-scalar $\pi NN$ coupling,
instead of the pseudo-vector coupling used in the SL model, is used in
Ref.\cite{rekalo}.

The resonant term in Eq.(16)
is defined by the dressed vertices which contain the
influence of the non-resonant interactions as given by the following
equation
\begin{eqnarray}
\bar{\Gamma}_{\Delta, a N}(q,W) 
&=&  \Gamma_{\Delta,aN}(q) + \int d {\bf k}
\Gamma_{\Delta,\pi N}(k)G_{\pi N}(k,W) J_{\pi N, aN}^{(n.r.)}(k,q,W) \,,
\end{eqnarray}
where 
$\Gamma_{\Delta,aN}(q)$ is the $bare$  $N$-$\Delta$ form factor.
The second term in Eq.(18) is commonly called the meson cloud 
contributions to the $N$-$\Delta$ transition. Explicit calculations of
these meson cloud effects also mark an important difference between this work
and all of the previous 
investigations\cite{cg,jones,isha,hwang,nath,musolf,hammer,nimi-3,rekalo} of 
parity violating asymmetry of 
$N(\vec{e},e^\prime)$ reactions. We emphasize that Eqs.(16)-(18) satisfy the
unitarity condition which is essential in interpreting the meson production
data, as explained in Refs.\cite{sl1,sl2,sl3} as well as 
in many works reviewed in Ref.\cite{burkertlee}.

\section{Calculations of Structure Functions}
With the Lagrangian Eq.(1), the formula for calculating the exclusive
cross sections for $(e,e^\prime \pi)$ and
$(\nu, e \pi)$ processes are given in details in Refs.\cite{sl1,sl2,sl3}. The
formula for calculating $(\nu,\nu^\prime \pi)$ cross sections can be easily
obtained from Ref.\cite{sl3} with minor modifications. In this paper, we focus
on inclusive processes.
Their cross sections can be more simply written in terms
of structure functions.

Following Ref.\cite{tw}, the symmetry properties require that
the lepton scattering structure functions
$W_1$, $W_2$ and $W_3$ are in general related to the hadron tensor $W^{\mu\nu}$ by
\begin{eqnarray}
W^{\mu\nu} & = & - W_1 g^{\mu\nu} + \frac{W_2}{m_N^2}p^\mu p^\nu
           + i \frac{W_3}{2m_N^2}
\epsilon^{\mu\nu\alpha\beta}p_\alpha q_\beta \,,
\end{eqnarray}
where the convention $\epsilon^{0123}=-1$ is chosen. The structure functions
are functions of two independent invariant variables. One usually
chooses $Q^2$ and the invariant mass $W=\sqrt{s}=\sqrt{(p+q)^2}$
 in the resonance region,
but chooses Bjorken scaling variable $x=Q^2/(2p\cdot q)$ 
and $Q^2$ in the deeply inelastic region. Here $q$ and $p$ are the
momenta of the exchanged boson ($\gamma$, $W$ or $Z$)
and the initial nucleon, respectively.
The unpolarized inclusive $N(e,e^\prime)$ cross sections can then
be written as
\begin{eqnarray}
\frac{d\sigma}{d\Omega dE^\prime} =\frac{4\alpha^2E^{\prime 2}}{Q^4}
[2 W_1\sin^2\frac{\theta}{2}+W_2\cos^2\frac{\theta}{2}
] \,,
\end{eqnarray}
with
\begin{eqnarray}
Q^2=-q^2 = 4 E E^\prime \sin^2\frac{\theta}{2} \,,
\end{eqnarray}
where $E$ and $E^\prime$ are the incident and outgoing lepton energies,
respectively, $\theta$ is the lepton scattering angle with respect to
the incident lepton	.
For inclusive $(\nu, e^-)$ and $(\bar{\nu}, e^+)$ processes,
we have
\begin{eqnarray}
\frac{d\sigma^{\nu,\bar{\nu}}}{dE^\prime d\Omega^\prime}
=\frac{G^2_F\mid V_{ud}\mid^2}{2\pi^2}E^{\prime 2}
[2W_1 \sin^2\frac{\theta}{2}
+W_2 \cos^2\frac{\theta}{2}
\pm W_3 \frac{E+E^\prime}{m_N}\sin^2\frac{\theta}{2}] \,,
\end{eqnarray}
where the sign in front of $W_3$ is $+$ (-) for $(\nu,e^-)$ ($(\bar{\nu},e^+)$).
For $(\nu,\nu^\prime)$, the cross section formula is the same as Eq.(22)
except that the factor $\mid V_{ud}\mid^2$ is removed.

Within the hadronic models, such as the SL model considered in this paper,
the hadron tensor $W^{\mu\nu}$ 
in the region near the $\Delta$ excitation
can be calculated from summing
all contributions from $a(q)+ N(p) \rightarrow \pi(k)+ N(p)$ with
$a= \gamma, W, Z$.
We can write in general
the hadron tensor $W^{\mu\nu}$ for electroweak pion reactions as
\begin{eqnarray}
W^{\mu\nu} & = & \sum <f|J_{\alpha}^\mu|i>^*<f|J_\alpha ^\nu|i> \,, 
\end{eqnarray}
where $\alpha= em, cc, nc$ denotes the considered
current, and we have introduced 
concise notations
\begin{eqnarray}
\sum = \sum_{\bar{i}}\sum_f (2\pi)^3 \delta^4(p + q - p'-k)\frac{E_N}{m_N} \,,
\end{eqnarray}
and
\begin{eqnarray}
<f\mid J_{\alpha}^\mu \mid i >  & = &
     \frac{1}{(2\pi)^3}\sqrt{\frac{m_N^2}{E_N E_N' 2E_\pi}}
     <k(t_\pi), p^\prime(s',t')|J_{\alpha}^\mu(q)|p(s,t)> \,,
\end{eqnarray}
Here $(s,t)$ denote the z-components of the nucleon spin-isospin,
$t_\pi$ the z-component of the pion isospin, and
$<k(t_\pi), p^\prime (s',t')|J_\alpha^\mu(q)|p(s,t)>$
is the  $a + N \rightarrow \pi+ N$ amplitude with $a= \gamma$,
$ W$, $Z$
for $\alpha=em$, $cc$, and $nc$, respectively. In the dynamical approach of
Refs.\cite{sl1,sl2,sl3}, the current matrix element Eq.(25) has the form
of Eq.(16), consisting of a non-resonant term and a resonant term.

From Eqs.(19) and (23), one can 
derive the expressions for calculating
the structure functions $W_1$, $W_2$, and $W_3$ from the current matrix 
elements defined by Eq.(25).
It is convenient to calculate these structure functions
 in the center of mass (c.m.)
 frame of the initial $a N$ and
final $\pi N$ systems. The direction of the momentum-transfer 
 is chosen to be the quantization
$z$-direction : i.e. $q = (\omega_c,0,0,\mid {\bf q}_c \mid)$ for
the exchanged boson and
$p=(E_N,0,0,-\mid {\bf q}_c \mid)$ for the initial nucleon.
We then have for the contribution from current $j_{\alpha}$ 
with $\alpha= em, cc, nc$, 
\begin{eqnarray}
W_1 &= & \frac{1}{2}\sum[ |<f\mid J_\alpha^x\mid i>|^2 + 
|<f\mid J_\alpha^y\mid i>|^2] \,, \\
W_2 & = & \frac{Q^2}{{\bf q}^2}
 \sum [ \frac{1}{2}(|<f\mid J_\alpha^x\mid i>|^2 + 
|<f\mid J_\alpha^y\mid i>|^2)
 + \frac{Q^2}{{\bf q}_c^2}|<f\mid \bar{J}_\alpha \mid i >|^2] \,, \\
W_3 & = & - \frac{2m_N}{\mid {\bf q}\mid }\sum Im[<f\mid J_\alpha^x\mid i>
<f\mid J_\alpha ^y \mid i >^*] \,,
\end{eqnarray}
where $\bar{J}_\alpha= J_\alpha^0 + \frac{\omega_c}{Q^2}J_\alpha\cdot q$,
and $ {\bf q} $ is three-momentum transfer in the laboratory frame
(i.e. $q = (\omega, {\bf q})$).
In practice, the calculation of any term of the structure functions
defined by Eqs.(26)-(28) can be obtained from appropriate combinations of 
the following integrations
\begin{eqnarray}
I^{\mu\nu}_{\alpha,\beta} & = & \sum 
[<f\mid J_\alpha^\mu\mid i><f\mid J_\beta^\nu\mid i>^*] \nonumber \\
&=& \frac{1}{2}\sum_{s,s',t',t_{\pi}}
  \int d\Omega_{\pi}\frac{|{\bf k}_{c}|m_N}{16\pi^3W}
 <k(t_\pi)p^\prime(s',t')|J_{\alpha}^\mu(q)|p(s,t)>
<k(t_\pi)p^\prime(s',t')|J_{\beta}^\nu(q)|p(s,t)>^* \,.
\nonumber \\
\end{eqnarray}
were $\alpha,\beta = em, cc, nc$, and ${\bf k}_{c}$ is final pion momentum in
the final $\pi N$ c.m. system.

We will also examine the spin dependent structure functions of $(e,e^\prime)$.
They are defined\cite{tw} by writing the hadron tensor for a polarized target
with spin vector $S^\mu$ ($S^2=-1, p\cdot S=0$) as
\begin{eqnarray}
W^{\mu\nu} = W^{\mu\nu}_S + W^{\mu\nu}_A \,,
\end{eqnarray}
where
\begin{eqnarray}
W^{\mu\nu}_S&=& W_1 (-g^{\mu\nu}+\frac{q^\mu q^\nu}{q^2})
+\frac{W_2}{M^2}(p^\mu-\frac{p\cdot q}{q^2}q^\mu)
(p^\nu-\frac{p\cdot q}{q^2}q^\nu) \,,  \\
W^{\mu\nu}_A &=& i\frac{\epsilon^{\mu\nu\alpha\beta}q_\alpha}{p\cdot q}
[g_1 S_\beta + g_2 (S_\beta-\frac{S\cdot q}{p\cdot q} p_\beta)] \,. 
\end{eqnarray}
With some derivations, one can show that
\begin{eqnarray}
g_1&=&\frac{1}{1+\frac{Q^2}{\omega^2}}[\frac{Q^2}{\omega^2}X^{y0} +X^{xy}] 
\,, \\
g_2&=&\frac{1}{1+\frac{Q^2}{\omega^2}}[X^{y0} - X^{xy}] \,,
\end{eqnarray}
with
\begin{eqnarray}
X^{y0} &=&
 - m_N\frac{\omega}
{\mid {\bf q}\mid} Im[I^{y0}_{em,em}\mid_{s=S(x)}] \,, \\
X^{xy}&=& - m_N Im[I^{xy}_{em,em}\mid_{s=S(z)}] \,,
\end{eqnarray} 
where $I^{\mu\nu}_{em,em}\mid_{s=S(i)}$ is the same as that defined in
Eq.(29) except that
the initial nucleon projection $s$ is not summed over but is
fixed in the
chosen direction defined by $S(i)$. $S(x)$ ($S(z)$)
 means that the initial nucleon spin is
polarized in the direction perpendicular (parallel) to the incident electron
direction.

For investigating Quark-Hadron Duality, we would like to compare the
structure functions calculated from using the above formula for
hadronic models with those calculated from the quark currents
 Eqs.(2)-(4)
in the deeply inelastic region.  With the standard definitions 
\begin{eqnarray}
F_1(x,Q^2) &=& m_N W_1(x,Q^2) \,, \nonumber \\
F_2(x,Q^2)&=& \omega W_2(x,Q^2) \,, \nonumber \\
F_3(x,Q^2)&=& \omega W_3(x,Q^2) \,,
\end{eqnarray}
where 
$x= Q^2/(2p\cdot q)=Q^2/(2m_N\omega)$, the Parton Model gives\cite{tw}
(keeping only the contributions from the
$u$ and $d$ quarks in the considered $\Delta$
region)
\subsection{$p(e,e')$}
\begin{eqnarray}
F_{2}(x,Q^2)
& =& x[\frac{4}{9}(u(x,Q^2)+\bar{u}(x,Q^2))
+\frac{1}{9}(d(x,Q^2)+\bar{d}(x,Q^2))] \,,  \\
F_{1}(x,Q^2) &=& F_{2}(x,Q^2)/2x \,, \\
g_1(x,Q^2) &=& \frac{4}{18}[u^\uparrow(x,Q^2)-u^\downarrow(x,Q^2)
+\bar{u}^\uparrow(x,Q^2)-\bar{u}^\downarrow(x,Q^2)] \nonumber \\
& &+\frac{1}{18}
[d^\uparrow(x,Q^2)-d^\downarrow(x,Q^2)
+\bar{d}^\uparrow(x,Q^2)-\bar{d}^\downarrow(x,Q^2)] \,.
\end{eqnarray}
\subsection{$p(\nu,e^-)$}
\begin{eqnarray}
F_2(x,Q^2)&=&2xF_1(x,Q^2)= 2x(d(x,Q^2)+\bar{u}(x,Q^2)) \,, \\
F_3(x,Q^2)&=&2(d(x,Q^2)-\bar{u}(x,Q^2)) \,.
\end{eqnarray}
\subsection{ $p(\nu,\nu^\prime)$}
\begin{eqnarray}
F_2(x,Q^2)&=&2xF_1(x,Q^2)= x[\frac{1}{4}+(\frac{1}{2}-\frac{3}{4}
\sin^2\theta_W)^2]
(u(x,Q^2)+\bar{u}(x,Q^2))
\nonumber \\
& &+x[\frac{1}{4}+(-\frac{1}{2}
+\frac{2}{3}\sin^2\theta_W)^2](d(x,Q^2)+\bar{d}(x,Q^2)) \,, \\
F_3(x,Q^2)&=& [\frac{1}{2}-\frac{4}{3}\sin^2\theta_W](u(x,Q^2)-\bar{u}(x,Q^2))
\nonumber \\
& & + [\frac{1}{2}-\frac{2}{3}\sin^2\theta_W](d(x,Q^2)-\bar{d}(x,Q^2)) \,.
\end{eqnarray}
Here $q(x,Q^2)$, $\bar{q}(x,Q^2)$ with $q=u,d$
 are the parton distribution functions (PDF)
determined from fitting the data in deeply inelastic region.
In Eq.(40) $q^{\uparrow \downarrow}(x,Q^2)$  
are the spin dependent parton distribution functions.
The structure functions for the neutron target
can be obtained from Eqs.(38)-(44) by interchanging the
$u$ and $d$ parton distribution functions.
In actual calculations, the strange quark and sea quark contributions are
included, but are found to be very small in the considered $\Delta$ region.
Thus our results presented below
 are from Eqs.(38)-(44). This is consistent with the
considered hadronic SL model
 which also neglects any
possible reaction mechanisms involving intermediate
strange hadrons.

\section{Quark-Hadron Duality}

We now turn to exploring the
Quark-Hadron Duality in the $\Delta$ excitation region. 
This is done by comparing
the structure functions calculated from Eqs.(38)-(44) using the
CTEQ6 parton distribution functions \cite{cteq6}
with those from Eqs.(26)-(28) and (33)-(36) using the 
hadronic SL model described in section II.  
It is well known\cite{rgp,ji-1,barbieri}
that more quantitative tests of Quark-Hadron Duality need to
include target mass corrections. Furthermore, the
role of the higher-twist effects must be better understood.
For simplicity, we  will not take such a more involved procedure and
will only compare all results from the SL model with the
Parton Model predictions at $Q^2=10$ (GeV/c)$^2$.
Thus our goal here is more qualitative. We will focus on exploring
whether the neutrino-induced reactions show the similar Quark-Hadron Duality 
observed in $(e,e^\prime)$. Furthermore we will also consider the
spin dependent structure functions and parity violating asymmetry of
$p(\vec{e},e)$ to which the procedures for including the
target mass corrections have not been developed.
We follow the usual criterion\cite{bg,rgp,meln}
 that the local Quark-Hadron Duality is 
seen if the
the predictions from our hadronic model are "oscillating"
around the predictions from
Parton Model such that their averaged values could be very close after
target mass corrections are included.

Following the previous works, as reviewed in Ref.\cite{meln}, we present
the calculated structure functions as functions of the
Nachtmann scaling variable defined by
\begin{eqnarray}
\xi =\frac{2x}{1+(1+4x^2m_N^2/Q^2)^{1/2}}
\end{eqnarray}
where $x=\frac{Q^2}{2p\cdot q}=\frac{Q^2}{2m_N\omega}$ is the Bjorken
scaling variable.
One can show\cite{rgp} that $\xi = p_q^+/p^+$ is the fraction of the
plus light-front momentum of the nucleon carried by the struck quark
in the infinite momentum frame. The use of this variable includes
some of the target mass corrections, as discussed in Ref.\cite{rgp}.
We note here that in the $\xi < $ about 0.7  region the
Parton Model results at $Q^2=10$ (GeV/c)$^2$
will correspond to $ W >$ about 2 GeV which
is much larger than the considered $\Delta$ region 
(1.1 GeV $\leq $ W $\leq $ 1.4 GeV). 
For example, for $Q^2=1.5$ (GeV)$^2$ the $\Delta$ peak (
$W=1.232$ GeV) occurs at $\xi = 0.57$ which corresponds
to $W=2.8$ GeV at $Q^2=10$ (GeV/c)$^2$. 

First we consider the $p(e,e')$ processes. 
As reported in Refs.\cite{nicu,liang}, the recent data of structure
functions $F_1$ and $F_2$ from JLab have further
established the Quark-Hadron Duality in the entire resonant region.
This is illustrated in Fig.3 along with the results calculated from
the SL model (solid curves near $\xi \sim 0.6$ ) in the $\Delta$ region.
From now on, we will only consider the data in the $\Delta$ region.
In Fig.4 we compare our hadronic model calculations at
$Q^2=0.7, 1.5, 2.5, 3.5$ (GeV/c)$^2$ (solid curves,
 from left to right) with these data. Each solid curve covers the
same $\Delta$ region with $1.1 $ GeV $\leq$  W $\leq 1.4$ GeV.
We see that the predictions from the employed hadronic model
(solid curves)
agree well with the data and oscillate around the Parton Model
 predictions (dashed curve).

To pursue further, it is natural to ask, from the point of view of
Standard Model, whether the 
Quark-Hadron Duality observed in $p(e,e')$
should also be seen in neutrino-induced  processes.
Experimental data for such investigations are still absent, but could
be obtained at new neutrino facilities in the near future.
To facilitate these developments, we here  present predictions
for $p(\nu,e^-)$ and $p(\nu,\nu^\prime)$ processes.
Since our model allows us to also predict the structure functions 
for the neutron target and hence we will also provide
predictions of $F_{i,I=0} = (F_{i,p} + F_{i,n})/2$
for the isospin $I=0$ deuteron-like target.
Calculations for general nuclear targets, such as those considered 
in Ref.\cite{arrington}, are beyond the scope of this work.

Our predictions for  structure functions  are shown in Fig.5
for $(e,e^\prime)$,
Fig.6 for $(\nu,e^-)$, and Fig.7 for $(\nu,\nu^\prime)$. 
Clearly all cases exhibit the similar feature of
Quark-Hadron Duality.
Experimental confirmations of our predictions (solid curves)
shown in Figs.5-7 will
be useful for making further progress.
If they are confirmed, our next step is to
examine the local Quark-Hadron Duality $quantitatively$ by including the
target mass corrections\cite{gp,ji-1,barbieri} and  considering the
 available information about the role of the higher twist effects. 

As reviewed in Ref.\cite{meln}, an another criterion of 
the Quark-Hadron Duality is that the
structure functions in the resonance region should slide along the
Parton Model predictions as $Q^2$ increases. This is clearly
the case in Figs.5-7 for $Q^2 < 4$ (GeV/c)$^2$. 
In Fig.8 we further show within the employed SL hadronic
model that this should also be the case up to rather high
$Q^2=20 $(GeV/c)$^2$ where the high precision data for the $\Delta$ region
could be obtained from the experiments 
with 12-GeV upgrade of JLab. The existing SLAC data\cite{brasse,stoler}
do not have enough high accuracy for investigating the
$\Delta$ at $ Q^2 \geq$ about 6 (GeV/c)$^2$.
 
We now turn to discussing the spin dependent structure functions $g_1$ and
$g_2$.
We first note from Eqs.(33)-(36) that
these functions depend on the interference terms 
$I^{y0}_{em,em} \propto J^y_{em} J^0_{em}$
and $I^{xy}_{em,em} \propto J^x_{em} J^y_{em}$. Such terms also determine
various polarization observables of the exclusive $p(\vec{e},e^\prime \pi)$
and $\vec{p}(\vec{e},e^\prime\pi)$ precesses, as discussed in, for example,
Ref.\cite{nl}. It is therefore important to test 
the SL model predictions against the recent data from such polarization
measurements\cite{biselli,joo}. It is clear from Figs.9-12 that the SL
model can describe the JLab data very well and certainly can be
used here to investigate the spin dependent structure functions.
The details of these comparisons can be found in Refs.\cite{biselli,joo}
and also in the captions of Figs.9-12.

Our results for $g_1$ and
$g_2$ of $p(e,e^\prime)$ are shown in Fig.13. In the left side of Fig.13 
we again see that our predictions for $g_1$ (solid curves)
agree reasonably well 
with the data\cite{fatemi}.
(Note that the determination
 of  these $g_1$ data from the polarization data
 of Ref.\cite{fatemi}
involved some model dependent input\cite{fatemi-a}.)
To get the Parton Model predictions using Eq.(40), we need
to known the spin dependent parton distribution functions. However
the determination of such information
in the considered large $x$ region is still in developing stage and
therefore no attempt is made here to do calculation using Eq.(40).
Instead, we assume that the Parton model predictions can be identified with
the dashed curve which is from fitting the data of $g_1$
measured in  deeply inelastic scattering \cite{g1data}. 
The left side of Fig.13 then indicates that $g_1$ clearly does not
exhibit the local Quark-Hadron Duality.
Such a disagreement was in fact expected\cite{anse} by considering the 
constraints imposed by the Ellis-Jaffe integral\cite{ellis} and
Drell-Hearn-Gerasimov sum rule\cite{dhg}.

It is known\cite{tw} that the 
naive Parton Model, considered here, can not predict
$g_2$. There are however some $g_2$
data\cite{g2data} from deeply inelastic scattering
which can be described well by using the Wandzura-Wilczek formula\cite{wwil}
\begin{eqnarray}
g^{WW}_2(x,Q^2) = [\int_{x}^{1}\frac{dy}{y}g_1(y,Q^2)]-g_1(x,Q^2) \,.
\end{eqnarray}
The dashed curve in the right side of Fig.13 is obtained from the above
formula using the dashed curve for $g_1$ in the left as the input.
Clearly it disagrees with the predictions (solid curves)
from the considered hadronic SL model.
Unlike the result for $g_1$, there is no simple
explanation of such a similar breakdown of local Quark-Hadron Duality.

The Quark-Hadron Duality for the
spin dependent structure functions of $p(\nu,e)$ and
$p(\nu,\nu^\prime)$ are not investigated here since there are no
corresponding PQCD calculations or data from deeply inelastic scattering
data to compare with.

\section{Parity Violating Asymmetry}
The formula for calculating parity violating asymmetry $A$ of 
$N(\vec{e},e^\prime)$ reactions have been given in literatures\cite{cg}.
For our purposes, we write it in terms of structure functions defined in
section III.
With some straightforward derivations, we have
\begin{eqnarray}
A & =& \frac{d\sigma(h_e=+1)-d\sigma(h_e=-1)}
{d\sigma(h_e=+1)+d\sigma(h_e=-1)} \nonumber  \\
&=& - \frac{Q^2 G_F}{\sqrt{2} (4\pi \alpha)}\frac{N}{D} \,,
\end{eqnarray}
with
\begin{eqnarray}
N & = &  \cos^2\frac{\theta}{2} W_2^{em-nc} +
         2 \sin^2\frac{\theta}{2} W_1^{em-nc}
           +  \sin^2\frac{\theta}{2}
(1 - 4 \sin^2 \theta_W)\frac{E + E^\prime }{M}W_3^{em-nc} \,,\\
D & = & \cos^2\frac{\theta}{2} W_2^{em-em} +
        2 \sin^2\frac{\theta}{2} W_1^{em-em} \,.
\end{eqnarray}
Here the electromagnetic
structure functions   $W_i^{em-em}$ are
calculated from Eqs.(26)-(27) with $\alpha=em$. The
 interference term $W_i^{em-nc}$ between
the electromagnetic and neutral currents can also be calculated from 
Eqs.(26)-(28) with the following replacements
\begin{eqnarray}
 \mid <f\mid J^i_{\alpha} \mid i>\mid ^2
&\rightarrow&  <f\mid J^i_{em} \mid i><f\mid J^i_{nc} \mid i>^* 
\nonumber \\
& &+ <f\mid J^i_{nc} \mid i><f\mid J^i_{em} \mid i>^* \,, \nonumber \\
& & \nonumber \\
<f\mid J_\alpha^x\mid >
<f\mid J_\alpha ^y \mid i >^* & \rightarrow& <f\mid J_{em}^x\mid i>
<f\mid J_{nc}^y \mid i >^*  \nonumber \\
& &+<f\mid J_{nc}^x\mid i> <f\mid J_{em}^y \mid i >^*  \,.\nonumber
\end{eqnarray}
Obviously these quantities can also be calculated
from integration $I^{\mu\nu}_{\alpha,\beta}$ defined in
Eq.(29).

We next focus on the first two terms of $N$ of Eq.(48) which depend
on $W_{1,2}^{em-nc}$. 
Here we note that $W_{1,2}^{em-nc}$ is the symmetric part of the hadron 
tensor defined in Eq.(19).  Hence they
can only have the contributions from
the vector parts  of neutral currents because of the vector structure
Eq.(5) of the electromagnetic current.
To see this more clearly, we  use  Eq.(8) to write the neutral current 
as\begin{eqnarray}
J^\mu _{nc} = V^\mu_{nc} - A^\mu_3 \,,
\end{eqnarray}
with \begin{eqnarray}
V_{nc}^\mu = ( 1 - 2 \sin^2 \theta_W)J_{em}^\mu - V_{isoscalar}^\mu \,.
\end{eqnarray}
Obviously the $[-A^\mu_3]$ part of Eq.(50) will not
contribute to $W_{1,2}^{em-nc}$ and hence 
the first two terms of $N$ of Eq.(48) can then be written as
\begin{eqnarray}
\cos^2\frac{\theta}{2} W_2^{em-nc} +
         2 \sin^2\frac{\theta}{2} W_1^{em-nc} &=&
(1-2\sin^2\theta_W)(\cos^2\frac{\theta}{2} W_2^{em-em} +
         2 \sin^2\frac{\theta}{2} W_1^{em-em}) \nonumber \\
& &
+(\cos^2\frac{\theta}{2} W_2^{em-is} +
         2 \sin^2\frac{\theta}{2} W_1^{em-is}) \nonumber \\
&=& (1-2\sin^2\theta_W) D +(\cos^2\frac{\theta}{2} W_2^{em-is} +
         2 \sin^2\frac{\theta}{2} W_1^{em-is}) \,,\nonumber 
\end{eqnarray}
where
$D$ is defined in Eq.(49) and
$W_i^{em-is}$ is the same
as $W_i^{em-nc}$ with $J^\mu_{nc}$  replaced by the isoscalar
vector current [$-V^\mu_{isoscalar}$] defined in Eq.(51). Thus
only the non-resonant 
isospin $I=1/2 $ $\pi N$ state contribute to $W_i^{em-is}$.
With the above relation, Eq.(47) for the  asymmetry can then be written as
\begin{eqnarray}
A & = & - \frac{Q^2 G_F}{\sqrt{2} 4\pi \alpha}
{}[2 - 4\sin^2 \theta_W + \Delta_V + \Delta_A] \nonumber \\
&=& - Q^2 [8.99 \times 10^{-5}] (1.075 + \Delta_V +\Delta_A) \,,
\end{eqnarray}
where $Q^2$ is in unit of (GeV/c)$^2$ and  
\begin{eqnarray}
\Delta_V & = &  \frac{\cos^2\frac{\theta}{2} W_2^{em-is} +
         2 \sin^2\frac{\theta}{2} W_1^{em-is}}{D} \,, \\
\Delta_A & = & \frac{\sin^2 \frac{\theta}{2}
(1 - 4 \sin^2 \theta_W)\frac{E + E'}{M}W_3^{em-nc}}{D} \,.
\end{eqnarray}
In Eq.(52) we have
evaluated ($2-4\sin^2\theta_W) \sim 1.075$ which is a model-independent
constant and is 
the main contribution to the asymmetry.
This term was identified by Cahn and Gilman\cite{cg} and later
investigators\cite{jones,isha,hwang,nath,musolf,hammer,nimi-3,rekalo}.

We now note that the term $\Delta_A$ 
defined by (54) depends only on  $W_3^{em-nc}$ which is
antisymmetric as defined in Eq.(19). Because of the vector structure 
of electromagnetic current Eq.(5), $W_3^{em-nc}$
can only have the contributions from
the axial vector parts of neutral currents. Clearly, the information about
the axial vector $N$-$\Delta$ form factor, which is the quantity we
hope to learn about as discussed in section I, is isolated in
$\Delta_A$ through its dependence on $W_3^{em-nc}$. Therefore
it is important to
identify the region where $\Delta_A$ is much larger than $\Delta_V$ such that
the extracted axial $N$-$\Delta$ form factor has less model dependence. 

As mentioned above, only the non-resonant
isospin $I=1/2$ $\pi N$ final states contributes to
$W_i^{em-is}$. Thus we expect that
$\Delta_V$ is weaker than $\Delta_A$ near the $\Delta$ resonance.
This is illustrated in  
Fig.14 for incident electron energy $E=1$ GeV. 
We see that $\Delta_A$ (solid curve) is indeed 
much larger than $\Delta_V$ at energies near $W=1.232$ GeV of the
$\Delta$ position. 
This is also the case for other typical electron kinematics 
which can be conducted at JLab, as illustrated in
Fig.15. The contribution from $\Delta_A$ (solid curves) are clearly
much larger than $\Delta_V$ (dashed curves).
Therefore the asymmetry data at resonance position
$W=1.232$ GeV can be used to extract
the contributions from the axial vector currents.

We now focus on examining how $\Delta_A$ depends on the
axial $N$-$\Delta$ form factor.
Recalling Eqs.(16)-(18) for the current matrix elements, 
$\Delta_A$, like other observables, contains contributions from
 the resonant  and  non-resonant 
amplitudes. As seen in Fig.16, 
the non-resonant contribution (dotted curve near the bottom)
is very small compared with the full model prediction (solid curve).
This agrees with previous investigations\cite{hwang,hammer,nimi-3,rekalo}, 
although those earlier
works neglected the unitary condition as discussed in section I.
The large difference between the solid and the dotted curves in Fig.16
indicates that $\Delta_A$ is a useful quantity for extracting
the dressed axial $N$-$\Delta$ form factor defined by Eq.(18).
When the pion cloud effects
(second term of Eq.(18)) on the $N$-$\Delta$ transition
is turned off, we obtain the dash curves. This is consistent with our 
previous findings in Refs.\cite{sl1,sl2,sl3} that pion cloud effects 
on $N$-$\Delta$ transition are
significant. 

We present in Fig.17
our predictions of asymmetry $A$ for several incident electron energies
$E= 0.8, 1.5, 2, 4$ GeV and at the
$\Delta$ peak. Here we note that our result for $E=0.8$ GeV is about
10 $\%$ larger than that  shown in Fig.6 of Ref.\cite{hammer}.
Furthermore, their $A$ decreases with $Q^2$ in striking
difference with our predictions. This is perhaps
mainly due to the use of different
$N$-$\Delta$ form factor, but the differences in
treating the non-resonant amplitudes, which interfere with
the dominant
$N$-$\Delta$ transition amplitude, also play some roles.
Experimental tests of our predictions in Fig.17 will be useful in 
examining the extent to which the axial $N$-$\Delta$ form factor Eq.(15),
which was determined in our study\cite{sl3} of $(\nu_\mu,\mu^-\pi)$ reaction, is
valid. 

As pointed out in Ref.\cite{sl3}, the form factor Eq.(15)
is rather different from the form 
$G^A_{N,\Delta}(Q^2) = (1 - \alpha Q^2/(\beta + Q^2))G_A(Q^2)$ with
$\alpha= 1.21$ and $\beta=2$ (GeV/c)$^2$
 which was determined in earlier 
works\cite{kitagaki}. 
It is therefore
interesting to see how these two form factors can be distinguished by
the parity violation asymmetry of $p(\vec{e},e^\prime)$. This is
illustrated in Fig.18.
Experiment test of our prediction shown in Fig. 18
will help distinguish these two axial
$N$-$\Delta$ form factors.
                                                                                
To end this section, 
We return briefly
to the investigation of Quark-Hadron Duality presented in 
section IV. It is also interesting to
explore\cite{bosted}  whether the
Quark-Hadron Duality also exists in the parity violating asymmetry $A$.
This can be done here by comparing the predictions of our hadronic model
and that of Parton model.
The formula for calculating $A$ within the Parton Model
was derived in Ref.\cite{cg}.
 Keeping only the $u$ and $d$ quarks in the considered
large $x$ region, one finds for a proton target
\begin{eqnarray}
A =\frac{3G_FQ^2}{\pi\alpha 2\sqrt{2}}
\frac{2C_{1u}[u(x)+\bar{u}(x)]-C_{1d}[d(x)+\bar{d}(x)]
+Y[2C_{2u}(u(x)-\bar{u}(x))-C_{2d}(d(x)+\bar{d}(x))]}
{4(u(x)+\bar{u}(x))+(d(x)+\bar{d}(x))} \nonumber \\
\end{eqnarray}
where
\begin{eqnarray}
Y&=&\frac{1-(1-y)^2}{1+(1-y)^2-y^2R/(1+R)} \,, \\
y&=& \frac{\omega}{E} \,, \\
R&=& \frac{\sigma_L}{\sigma_T} \,.
\end{eqnarray}
Note that $R$ is the ratio between the longitudinal and transverse
total cross sections. We use the values of $R$ calculated from the SL model.
The variable $y$ depends on incident electron
energy $E$ and the energy transfer $\omega$.
The other coefficients in Eq.(55) depend on the
electroweak coupling constants
for leptons and quarks
\begin{eqnarray}
C_{1u}&=& g^e_Ag^u_V =
-\frac{1}{2}+\frac{4}{3} \sin^2\theta_W,  \nonumber \\
C_{1d}&=&g^e_Ag^d_V=
\frac{1}{2}-\frac{2}{3} \sin^2\theta_W,  \nonumber \\
C_{2u}&=& g^e_Vg^u_A =-\frac{1}{2}+2 \sin^2\theta_W, \nonumber \\
C_{2d}&=&g^e_Vg^d_A=\frac{1}{2}-2 \sin^2\theta_W.
\end{eqnarray}
When the radiative corrections within the standard model is included
\cite{el,bosted}
$C_{1u} \sim -0.1886, C_{1d} \sim 0.3414, C_{2u} \sim -0.0359,
C_{2d} \sim 0.0265$. These values are only slightly different from those calculated from
 Eq.(59).

The asymmetry of $p(\vec{e},e^\prime)$
predicted from our hadronic model (solid curves) and 
Parton Model (dashed curves) for several typical electron kinematics
are compared in Fig.19. 
If our hadronic model results (solid curves) are confirmed experimentally,
the parity violating
asymmetry in the $\Delta$ region
obviously does not show local Quark-Hadron Duality.
The situation is similar to the results for the
spin dependent structure functions shown in Fig.14.

\section{Summary}

The dynamical model
developed in Refs.\cite{sl1,sl2,sl3} (the SL model) has been extended
 to include the weak 
neutral current contributions for investigating all possible
electroweak pion production reactions in the region near the $\Delta$ excitation. 
The main purpose is to examine
the Quark-Hadron Duality in the neutrino-induced reactions and
to explore how the axial $N$-$\Delta$ transition form factor 
can be determined by using
the parity violating asymmetry of $p(\vec{e},e^\prime)$.
The experimental data for testing our predictions can be
 obtained at JLab and new neutrino facilities.

We have found that the $(e,e^\prime)$ structure functions $F_1$ and
$F_2$ predicted by
the SL model are in good agreement with the recent
data\cite{nicu,liang} which had verified more quantitatively
the  Quark-Hadron Duality first observed by Bloom and Gilman\cite{bg}. 
The predicted structure functions for $(\nu,e)$ and $(\nu,\nu^\prime)$ processes
also show Quark-Hadron Duality to the extent similar to what has been
observed in $(e,e^\prime)$. Furthermore, we also predict that
Quark-Hadron Duality should also be seen in all electroweak
reactions on the neutron or equivalently the isospin $I=0$ deuteron-like target.
These results suggest that
the SL model can be  a  candidate hadronic model for developing
a theoretical understanding of Quark-Hadron Duality within the Standard Model.
For example, it will be interesting to explore which parts of the
predictions from the SL model can be related to the predictions from
the Parton Model.  
This is clearly a difficult question to answer and
is beyond the scope of this paper.

We have also investigated the spin dependent structure 
functions $g_1$ and $g_2$ of $(e,e^\prime)$.
It is found that the Quark-Hadron Duality is not seen in
the calculated  $g_1$ and $g_2$, while our results for $g_1$ and
some polarization observables associated with the exclusive
$p(\vec{e},e^\prime \pi)$
and $\vec{p}(\vec{e},e^\prime\pi)$ reactions
are in  good agreement with the recent data.
Experimental data for $g_2$ in the $\Delta$ region
are clearly needed.  If our predictions are confirmed, we perhaps have
more complete information for exploring
why the Quark-Hadron Duality breaks down in the spin dependent structure
functions. 
 
We should also emphasize that our investigations of Quark-Hadron Duality
are rather qualitative since they are based on the 
naive Parton Model. To test local Quark-Hadron Duality $quantitatively$,
we need to include target mass corrections\cite{gp,barbieri}
and consider the roles of higher
twist effects. In particular, the procedures for including these effects in
the calculations of spin dependent structure functions $g_1$ and $g_2$
 and parity violating asymmetry $A$ must
also be developed.

In the investigation of parity violating asymmetry $A$ of the inclusive
$p(\vec{e},e^\prime)$, we have shown
 that the non-resonant contribution is small
at the $W=1.232$ GeV $\Delta$ peak and hence a precise measurement
of $A$ can be used to improve the determination of the axial
$N$-$\Delta$ transition form factor. We have also predicted that
the parity violating asymmetry $A$, like the spin dependent structure
functions $g_1$ and $g_2$, does not exhibit Quark-Hadron Duality. 
Predictions for the experiments which can be conducted at JLab have been
given.

\section*{Acknowledgments}

We would like to thank A. Biselli, R. Fatemi, Y. Liang and C. Smith for
useful discussions on the analyses of their experimental data.
This work was supported by the U.S. Department of Energy, Office of
Nuclear Physics Division, under contract no. W-31-109-ENG-38, 
and by Japan Society for the Promotion
of Science, Grant-in-Aid for Scientific Research (C) 15540275.

\newpage

\begin{figure}
\centering
\includegraphics[width=8cm]{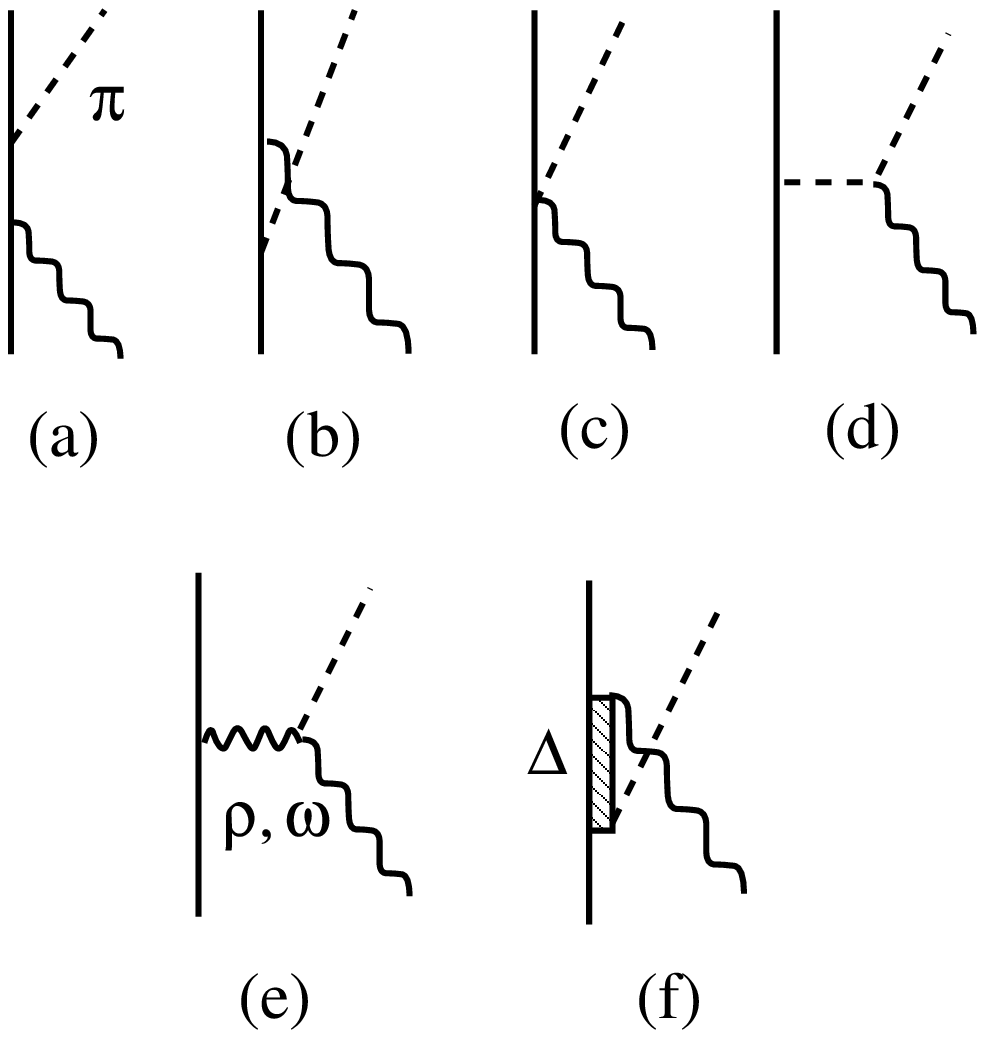}
\caption{ Non-resonant pion production mechanisms of the SL model.
See text for explaining
how these mechanisms are related to the electrmagnetic current,
charged current, and neutral current.}
\end{figure}

\begin{figure}
\centering
\includegraphics[width=2cm]{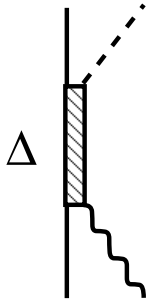}
\caption{The $\Delta$ excitation mechanism.}
\end{figure}

\begin{figure}
\centering
\includegraphics[width=6cm]{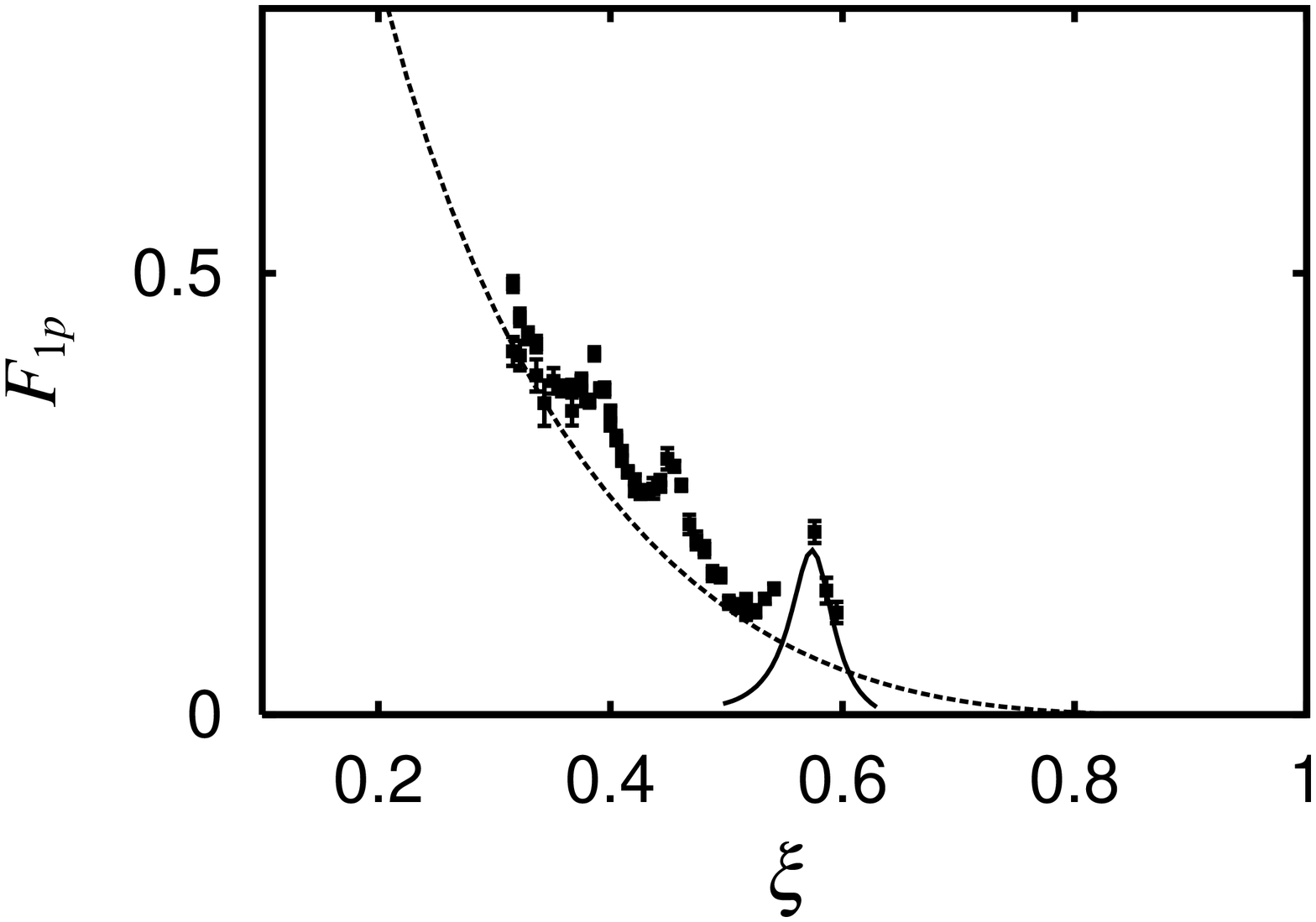}
\includegraphics[width=6cm]{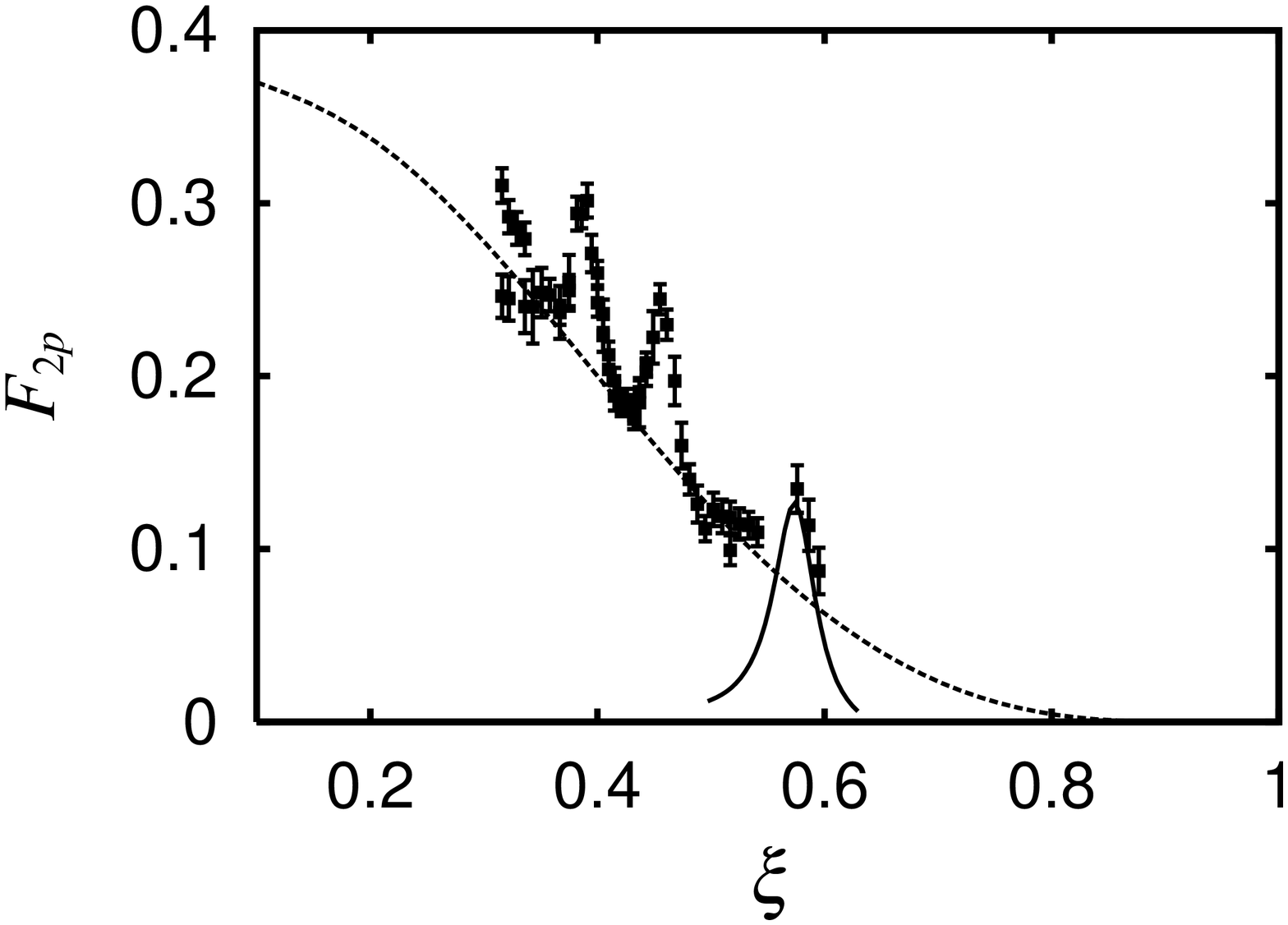}
\caption{Structure functions $F_1$ (left) and
$F_2$ (right) of $(e,e^\prime)$ for the proton
target at $Q^2=1.5$ (GeV/c)$^2$. The dashed curves are
calculated from using the CTEQ6 parton distribution functions at $Q^2=10$
(GeV/c)$^2$. The solid curves in the region near $\xi \sim 0.6$  are
calculated from the SL model.
  The data are from
Liang et al.\cite{liang}.}
\end{figure}

\begin{figure}
\centering
\includegraphics[width=6cm]{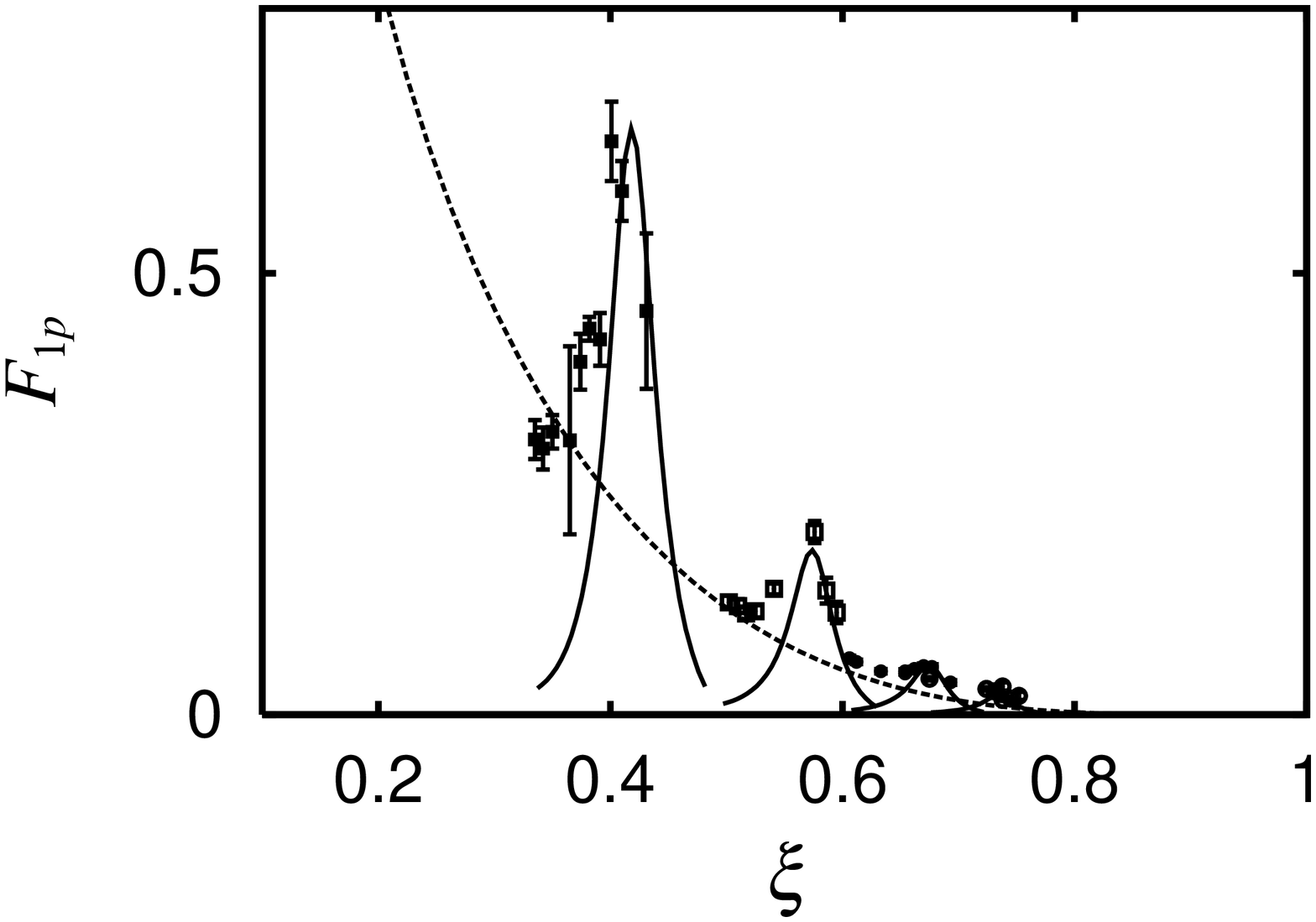}
\includegraphics[width=6cm]{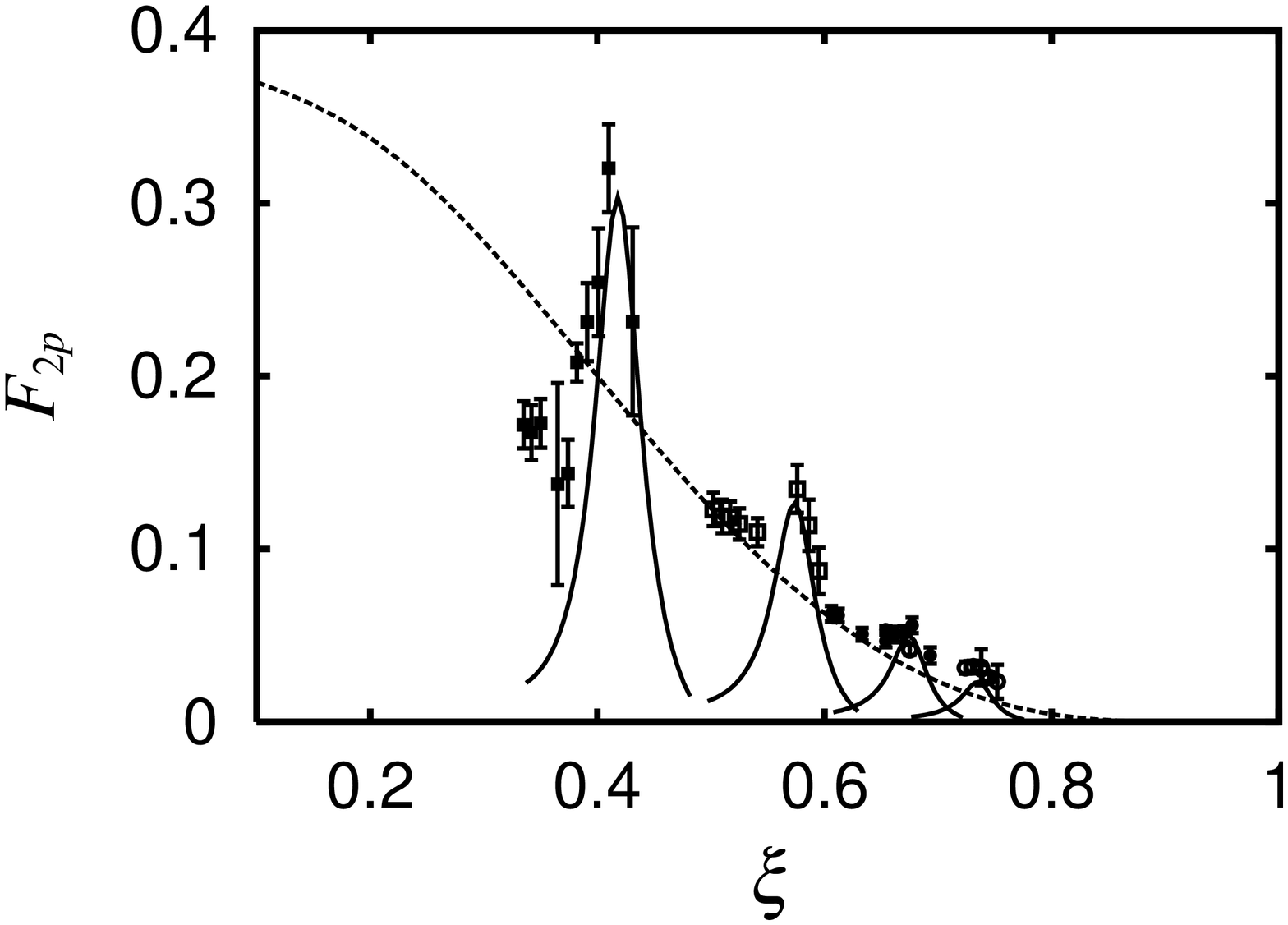}
\caption{Structure functions $F_1$ (left) and
$F_2$ (right) of $(e,e^\prime)$ for the proton
target. The dashed curves are
calculated from using the CTEQ6 parton distribution functions at $Q^2=10$
(GeV/c)$^2$. The solid curves are the results at
$Q^2= 0.7,1.5,2.5,3.5$ (GeV/c)$^2$
(from left to right) calculated from the SL model.
  The data are from
Liang et al.\cite{liang}.}
\end{figure}

\begin{figure}
\centering
\includegraphics[width=6cm]{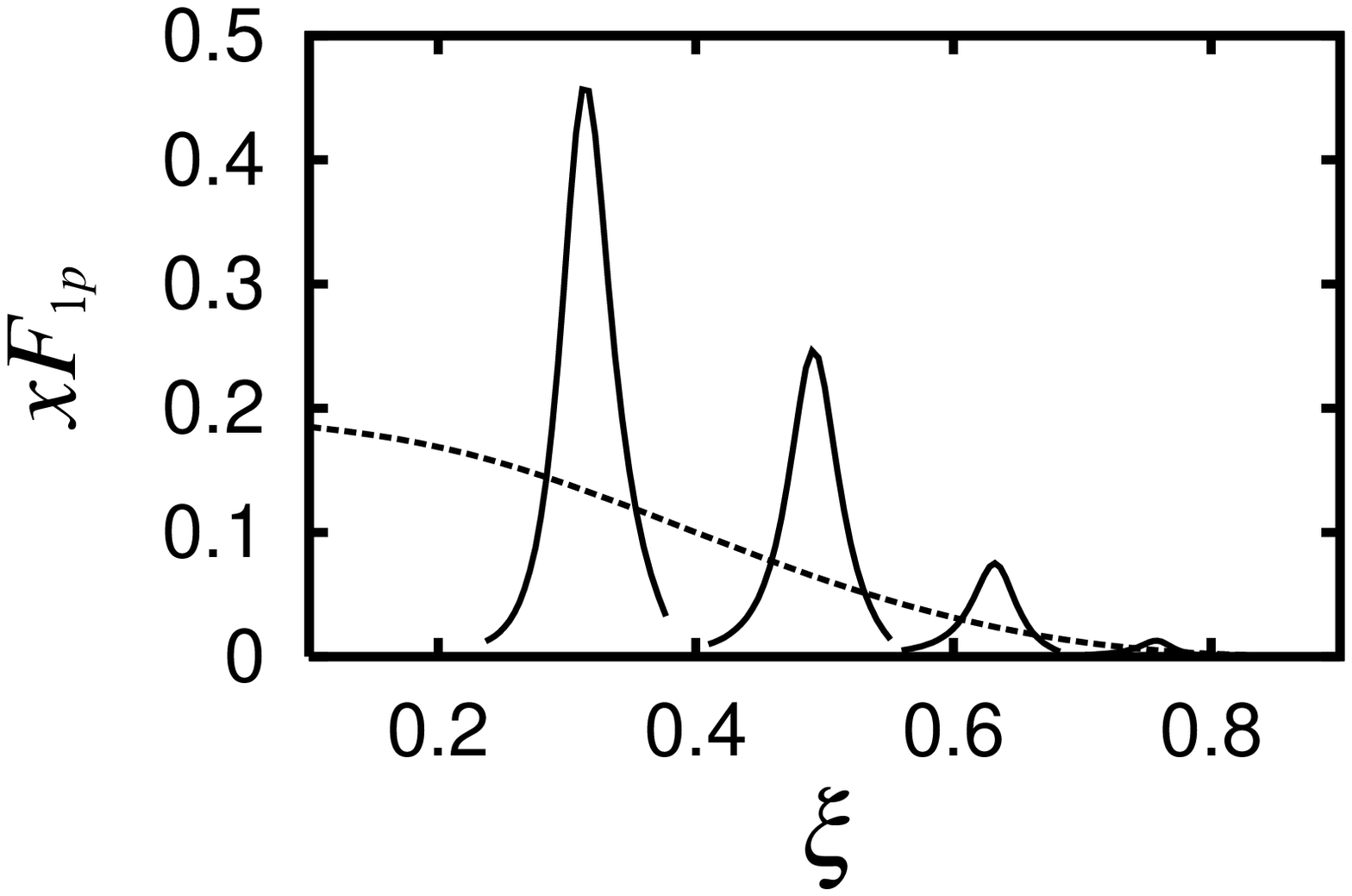}
\includegraphics[width=6cm]{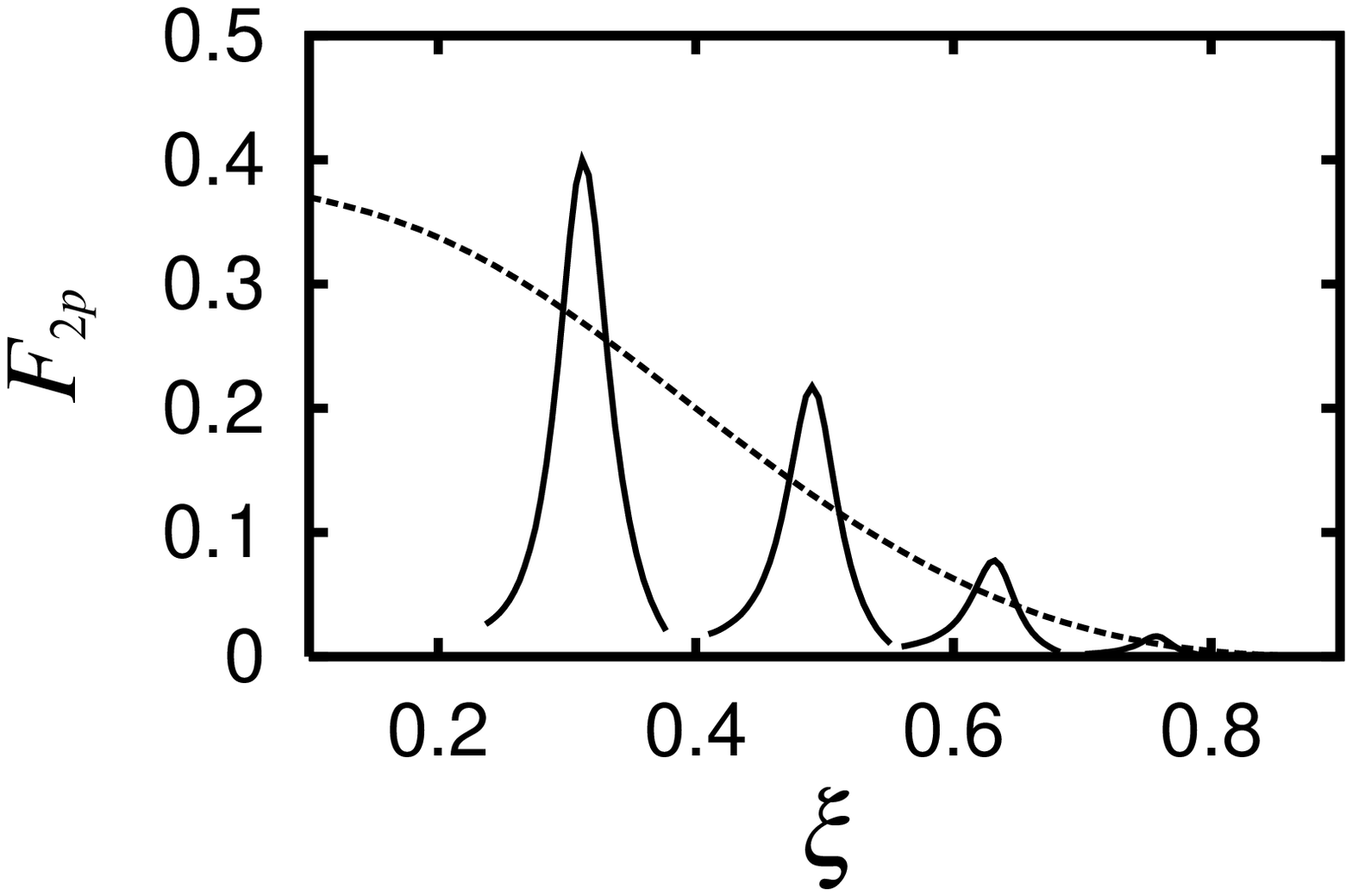}

\includegraphics[width=6cm]{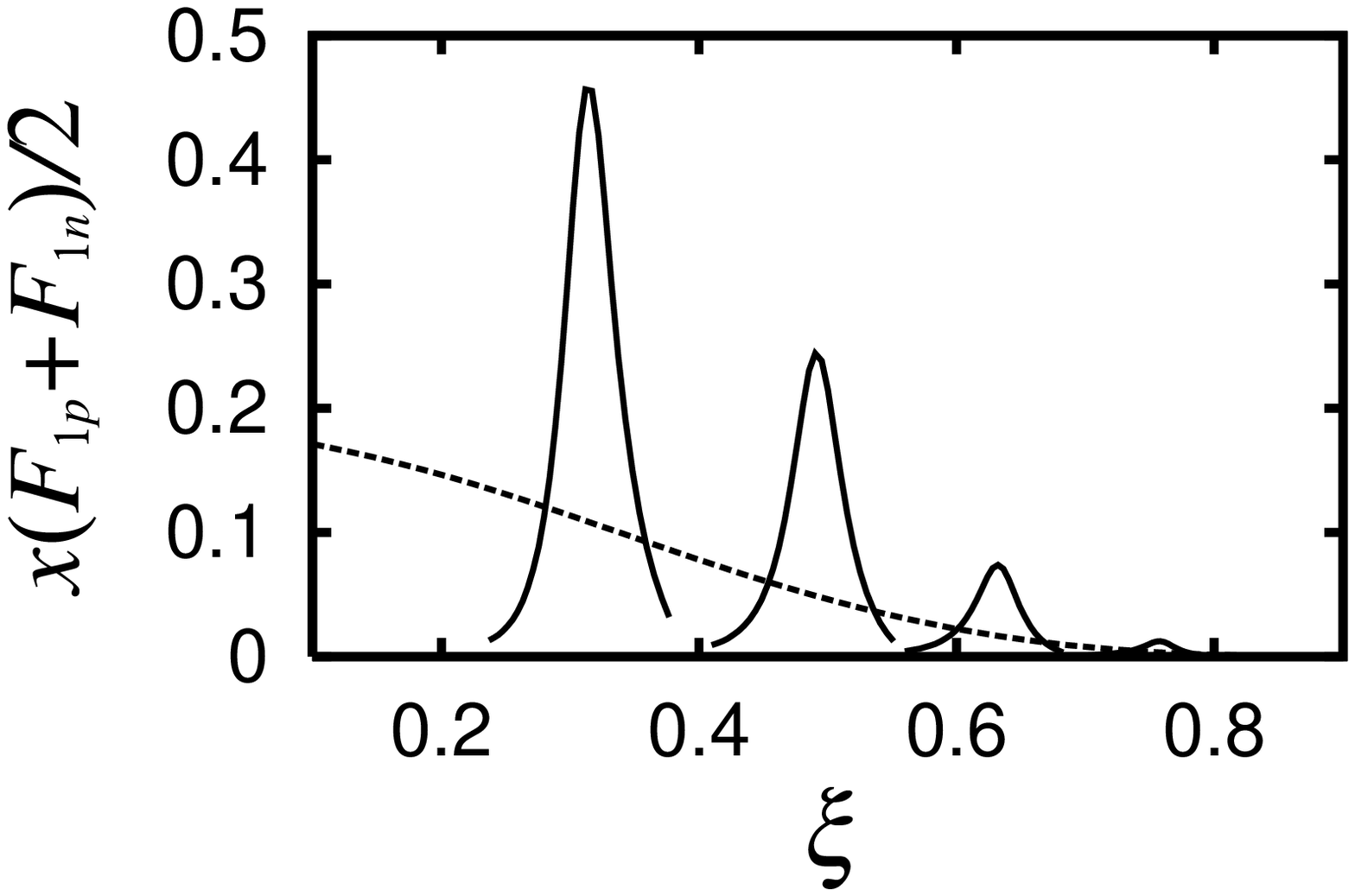}
\includegraphics[width=6cm]{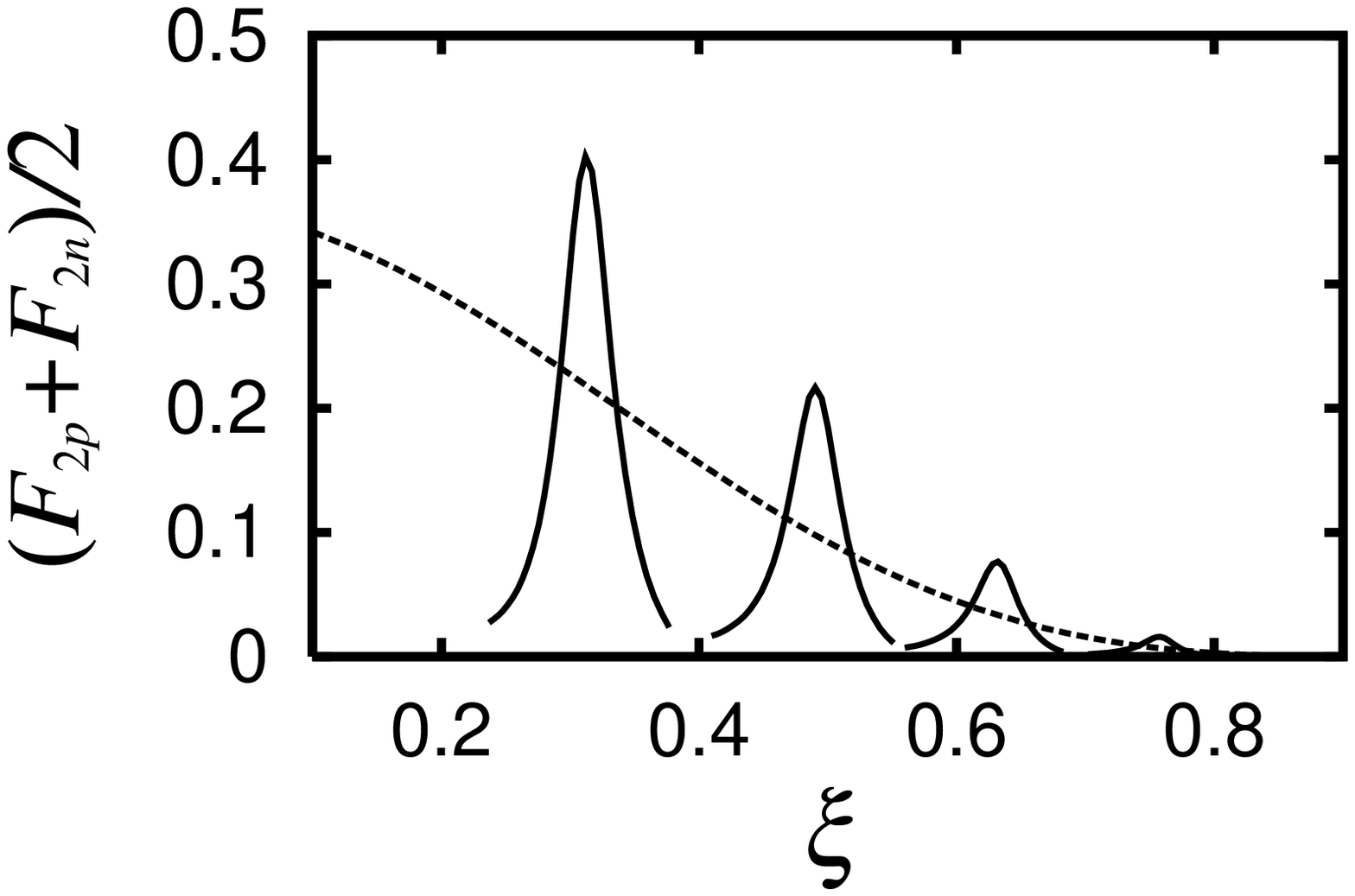}

\caption{Structure functions $xF_1$ (left) and
$F_2$ (right) of $(e,e^\prime)$ for the proton (top)
and an I=0 deuteron-like target (bottom).
 The dashed curves are
calculated from 
 using the CTEQ6 parton distribution functions at $Q^2=10$
(GeV/c)$^2$. The solid curves are the results at $Q^2= 0.4,1,2,4$ (GeV/c)$^2$
(from left to right) calculated from the SL model.}
\end{figure}

\begin{figure}
\centering
\includegraphics[width=4.5cm]{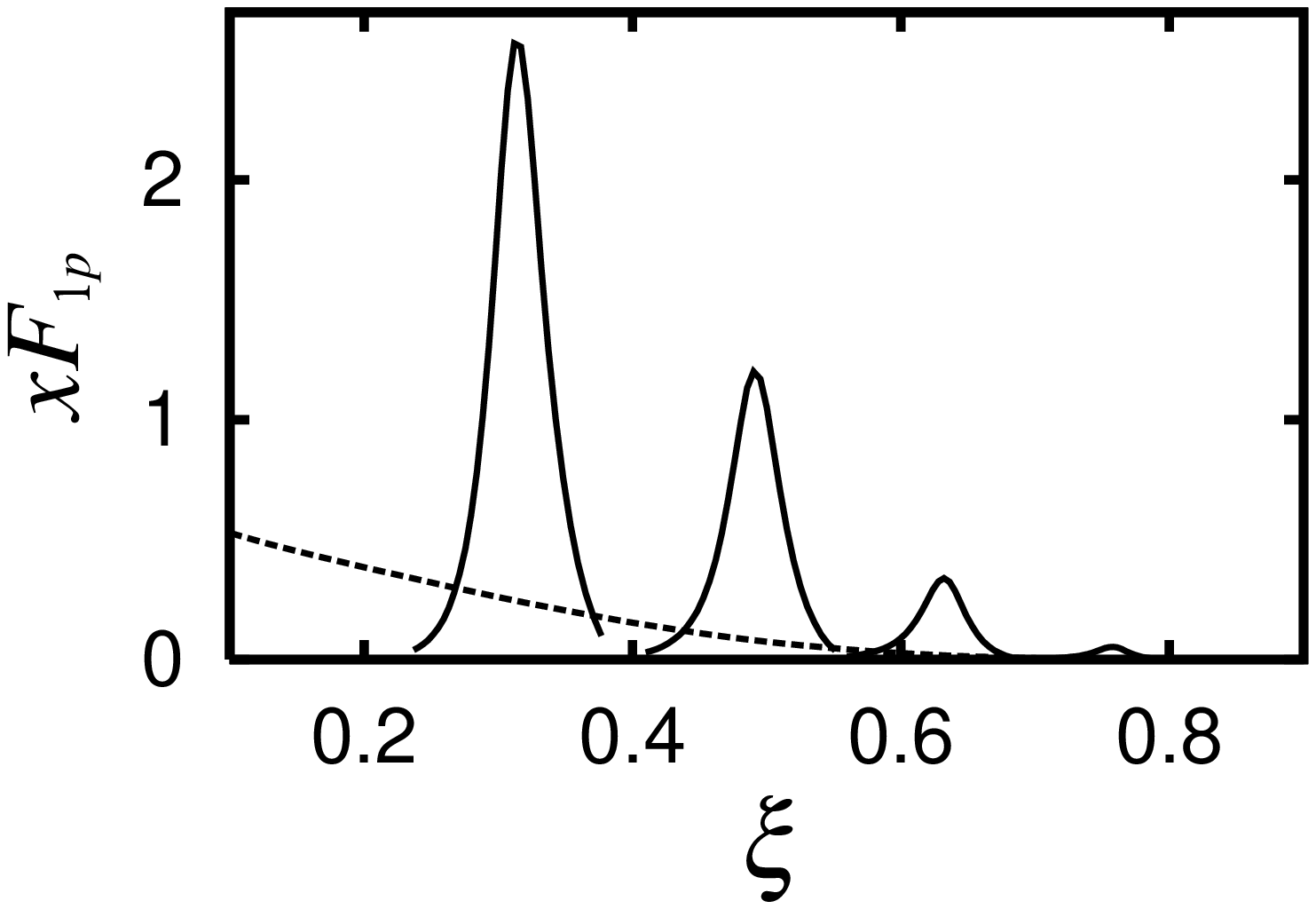}
\includegraphics[width=4.5cm]{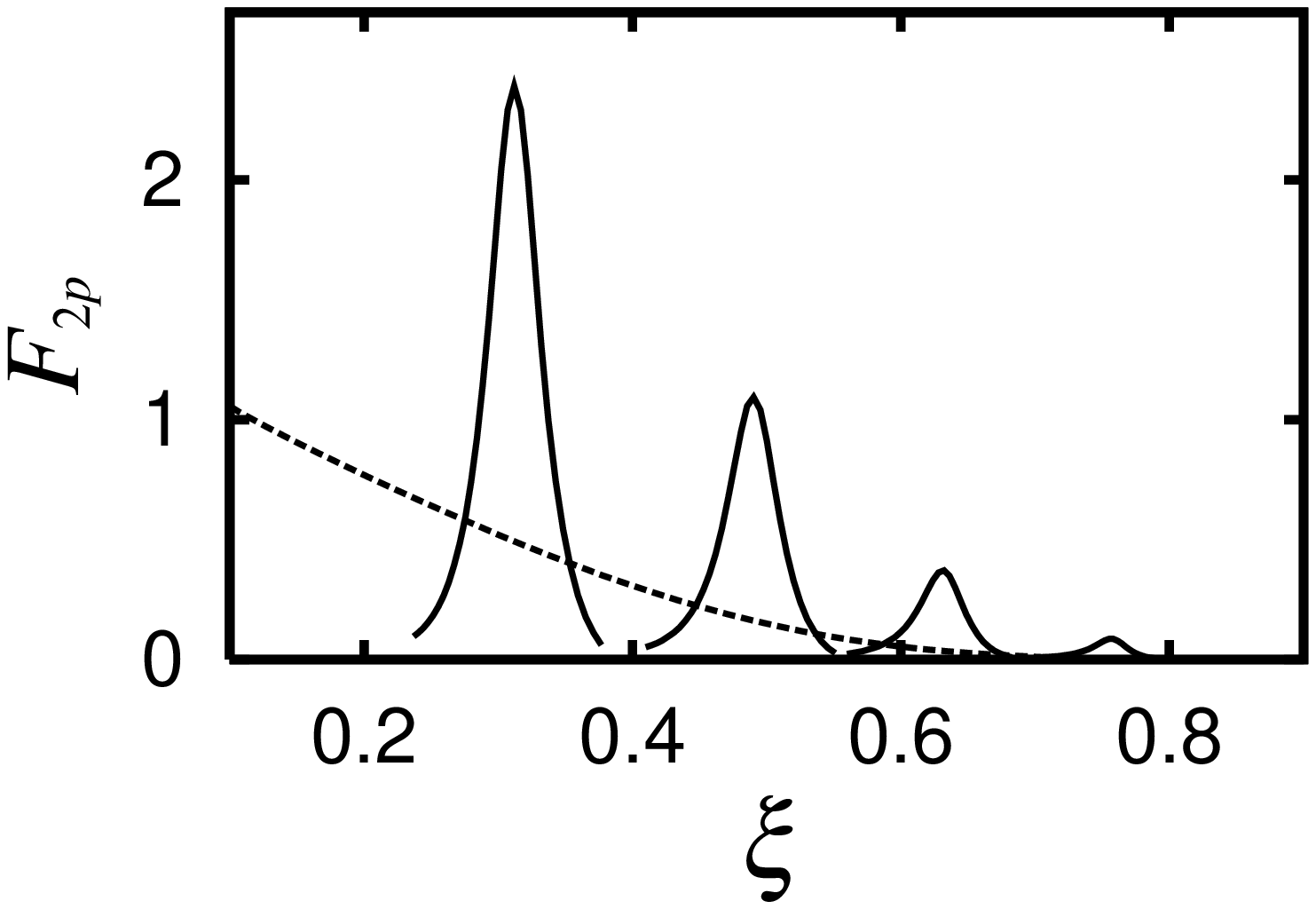}
\includegraphics[width=4.5cm]{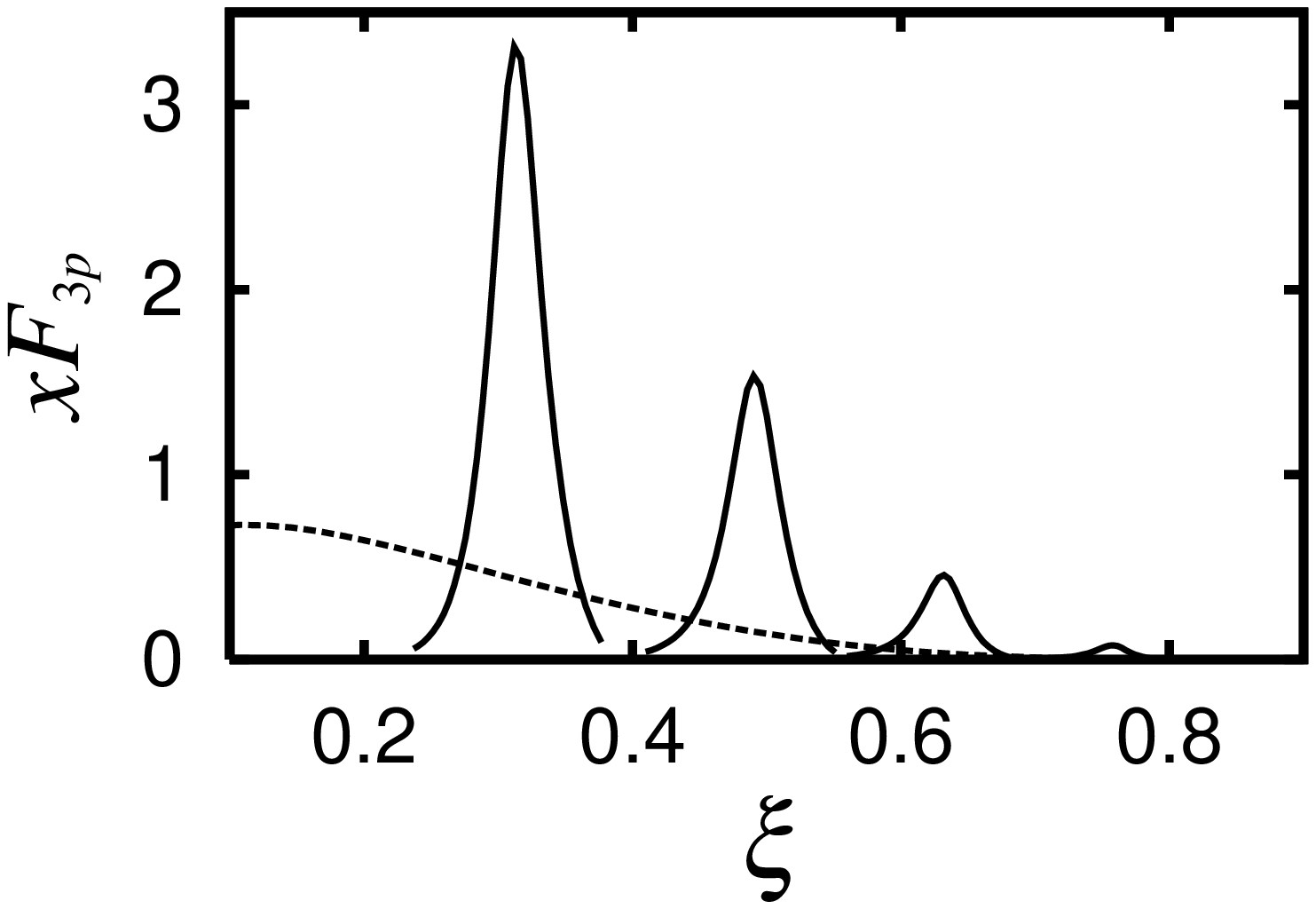}

\includegraphics[width=4.5cm]{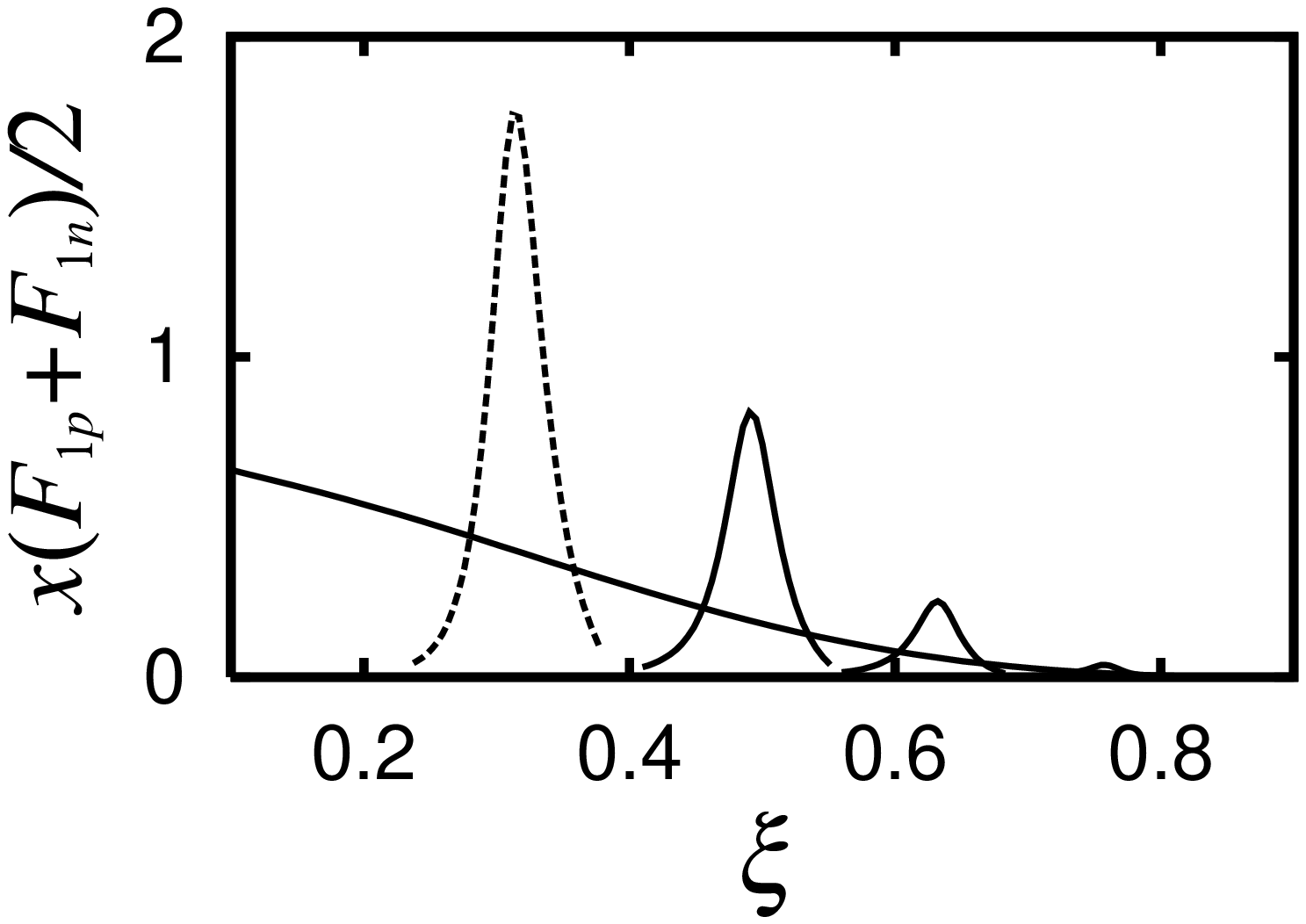}
\includegraphics[width=4.5cm]{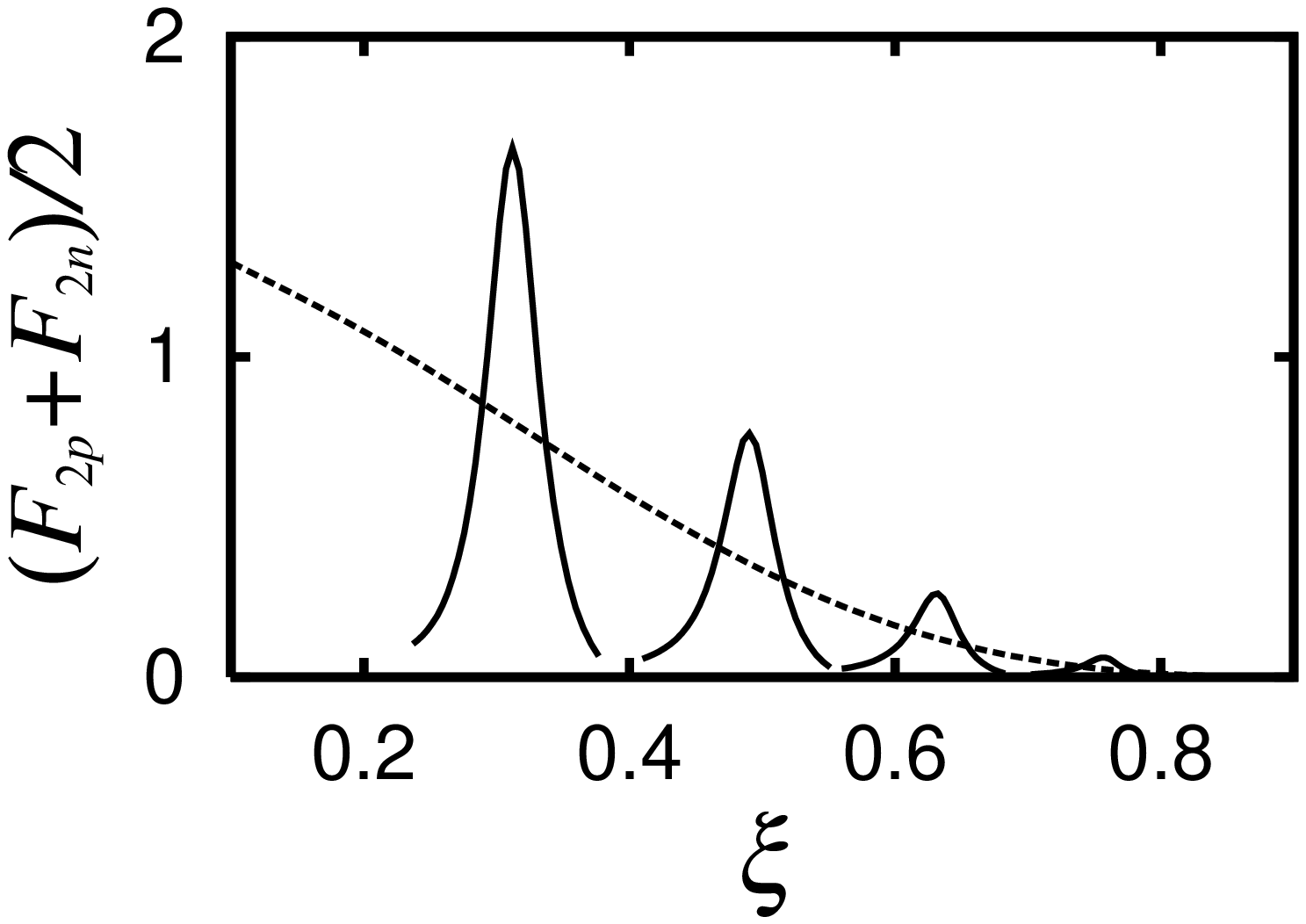}
\includegraphics[width=4.5cm]{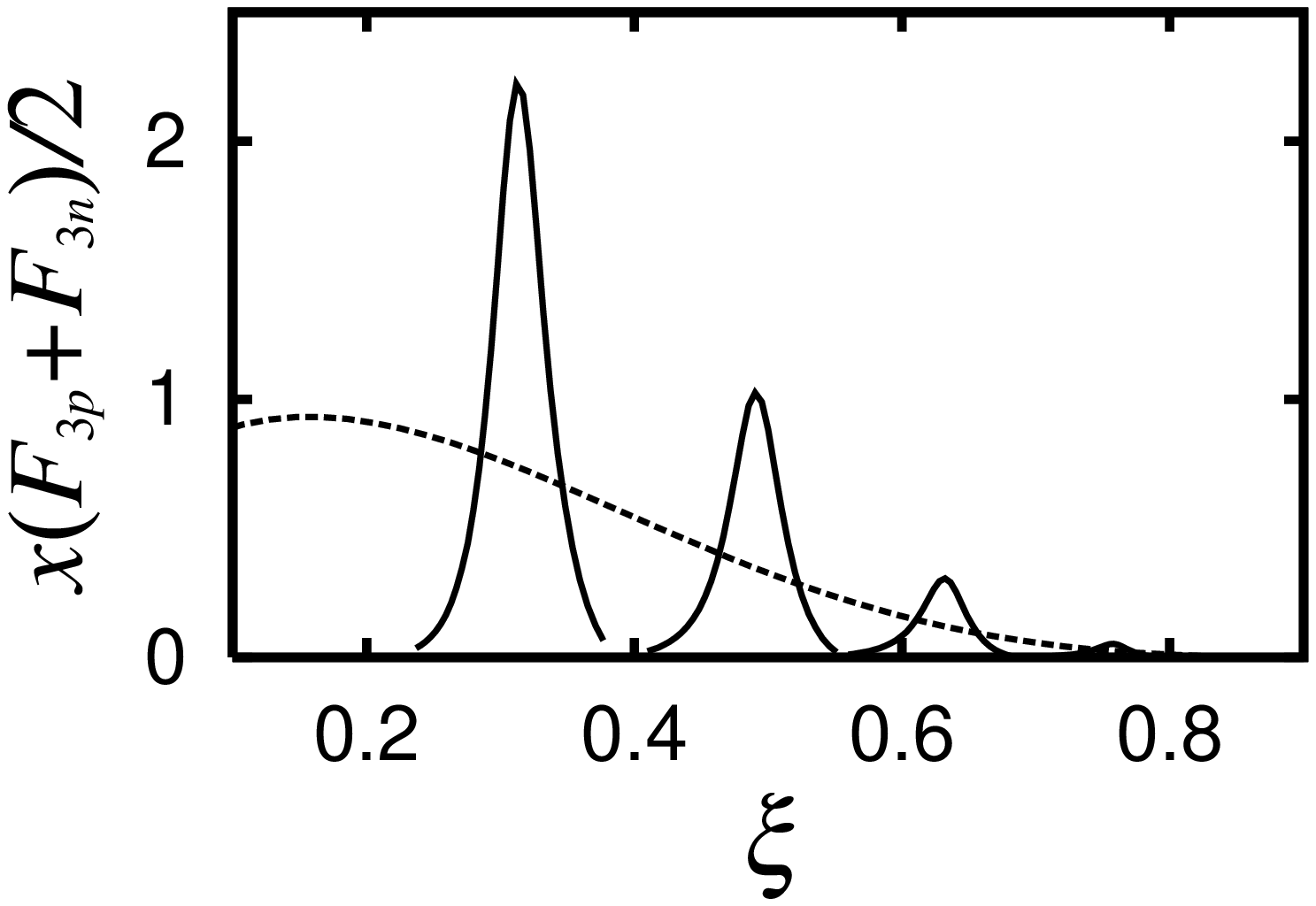}

\caption{Structure functions $xF_1$ (left),
$F_2$ (center), and $xF_3$(right) of $(\nu,e)$ for the proton (top)
and an I=0 deuteron-like target (bottom).
The dashed curves are
calculated from  
using the CTEQ6 parton distribution functions at $Q^2=10$
(GeV/c)$^2$. The solid curves are the results at $Q^2= 0.4,1,2,4$ (GeV/c)$^2$
(from left to right) calculated from the SL model.}
\end{figure}

\begin{figure}
\centering
\includegraphics[width=4.5cm]{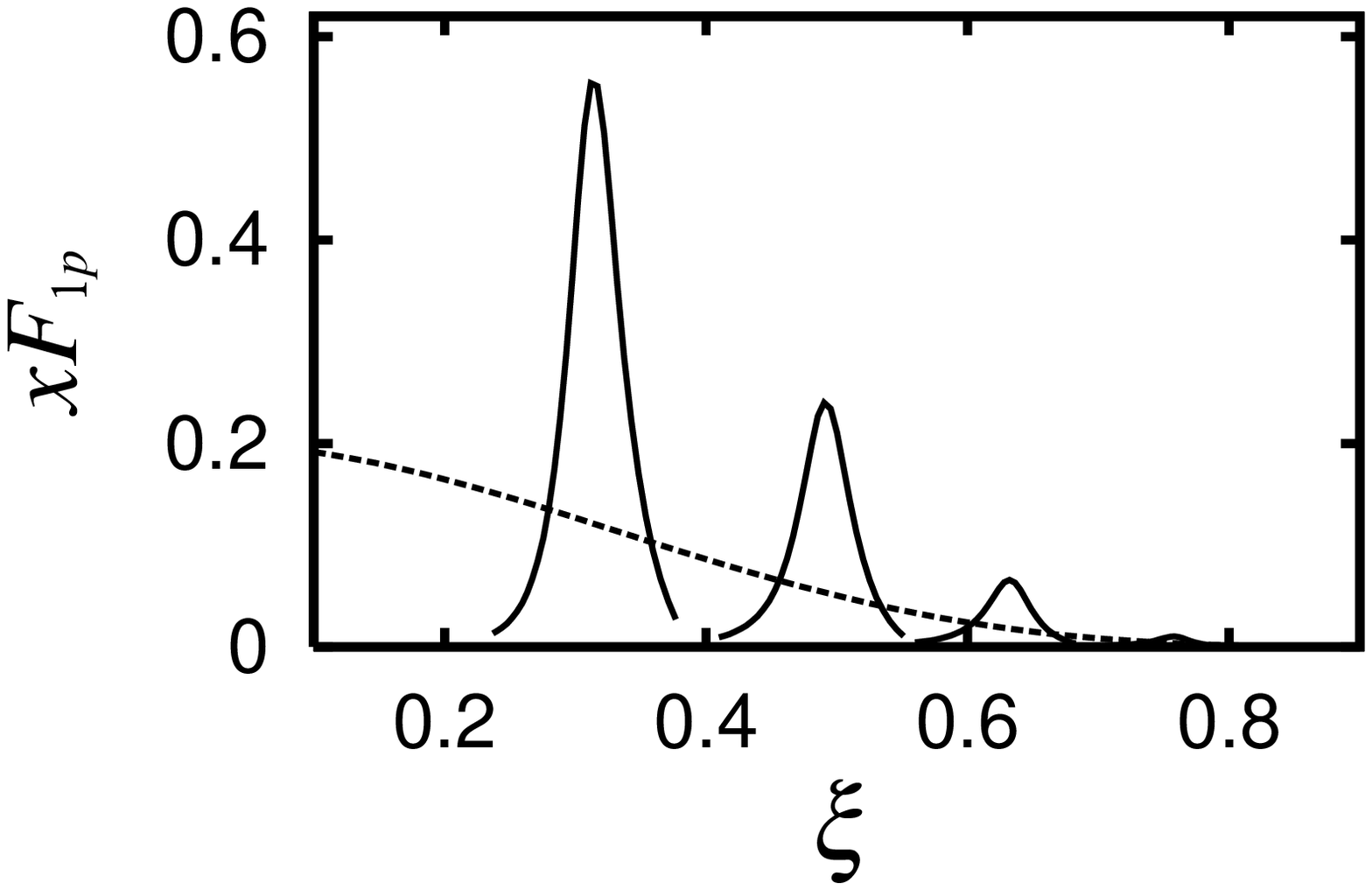}
\includegraphics[width=4.5cm]{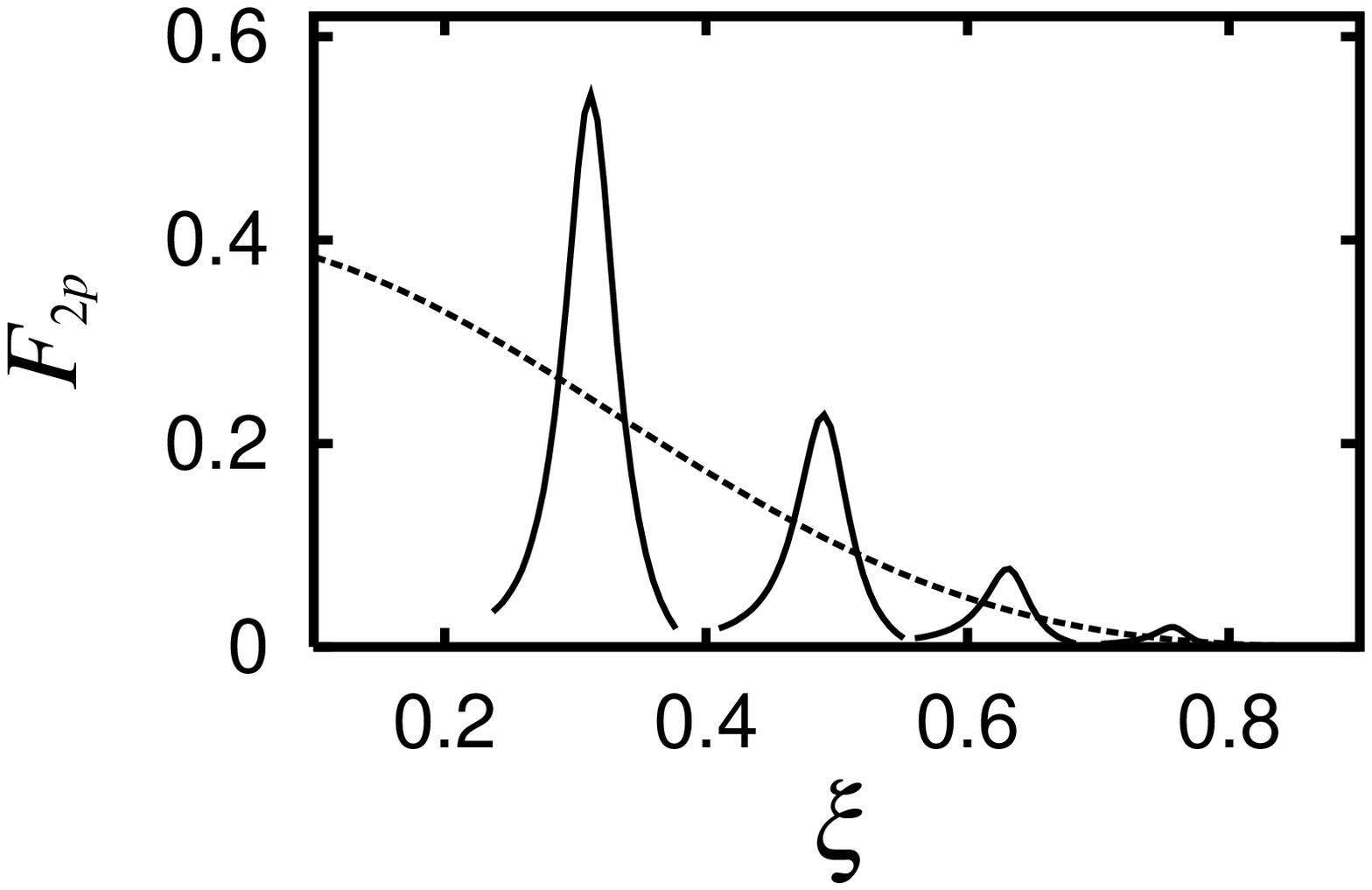}
\includegraphics[width=4.5cm]{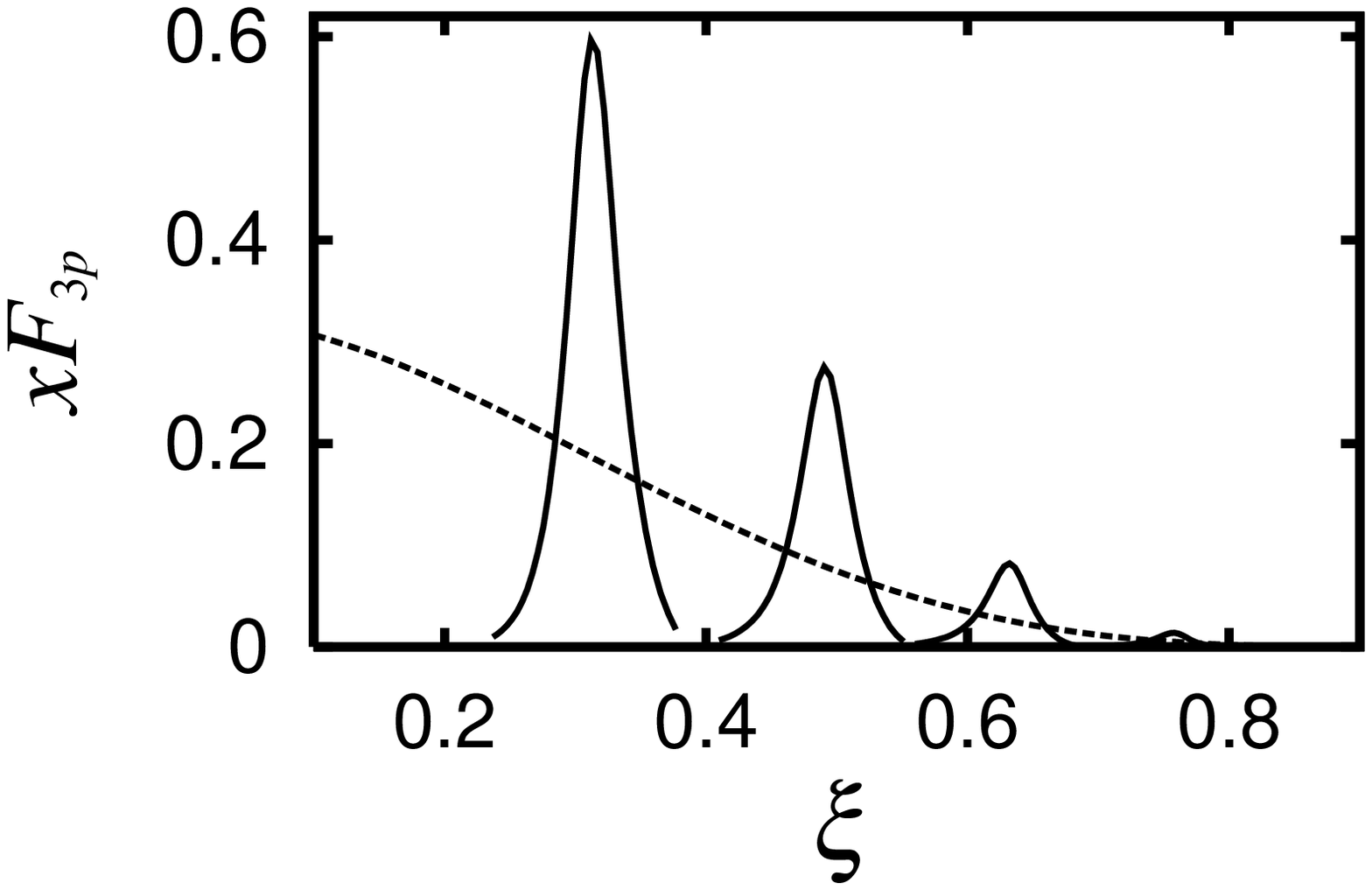}

\includegraphics[width=4.5cm]{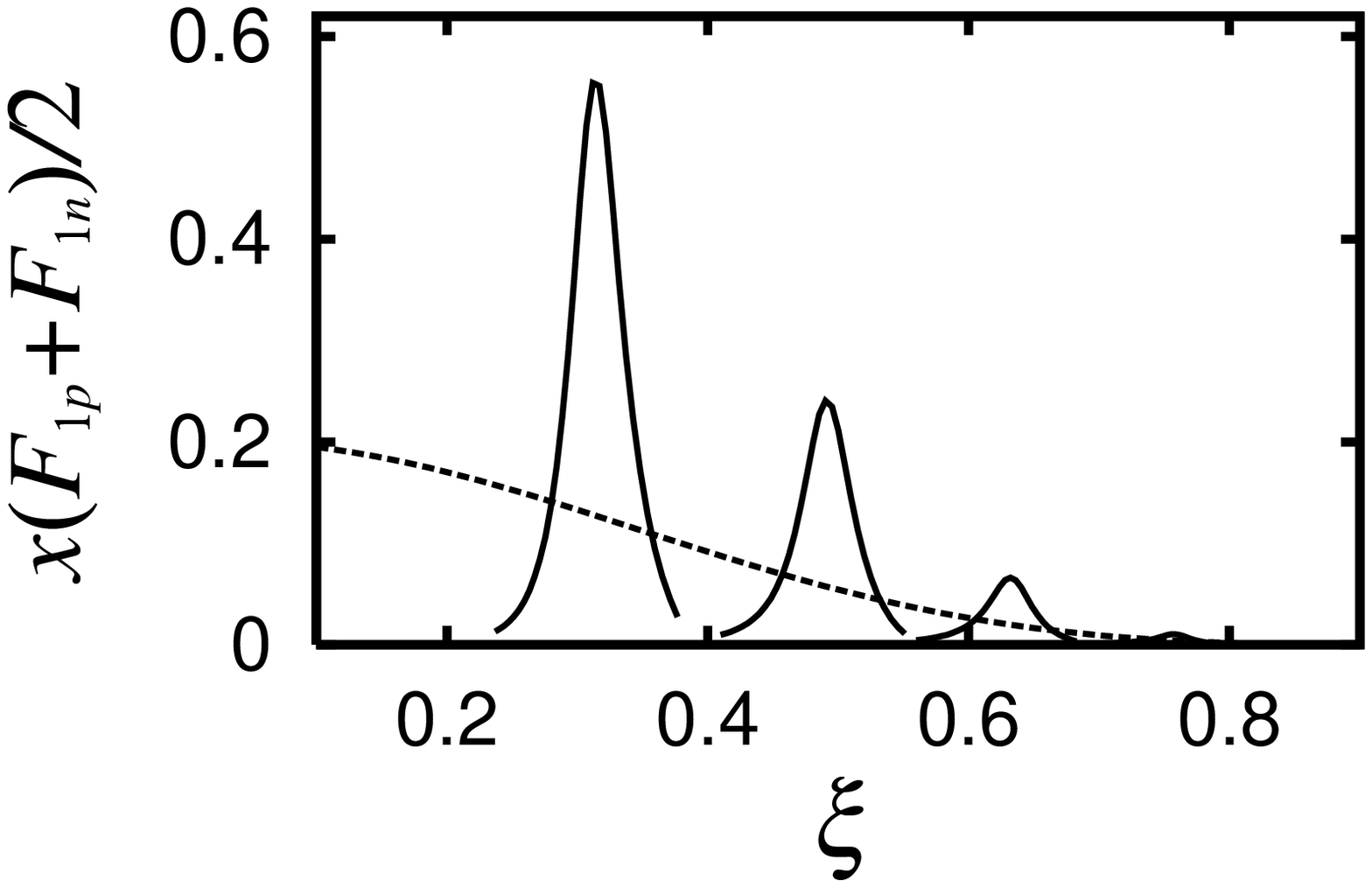}
\includegraphics[width=4.5cm]{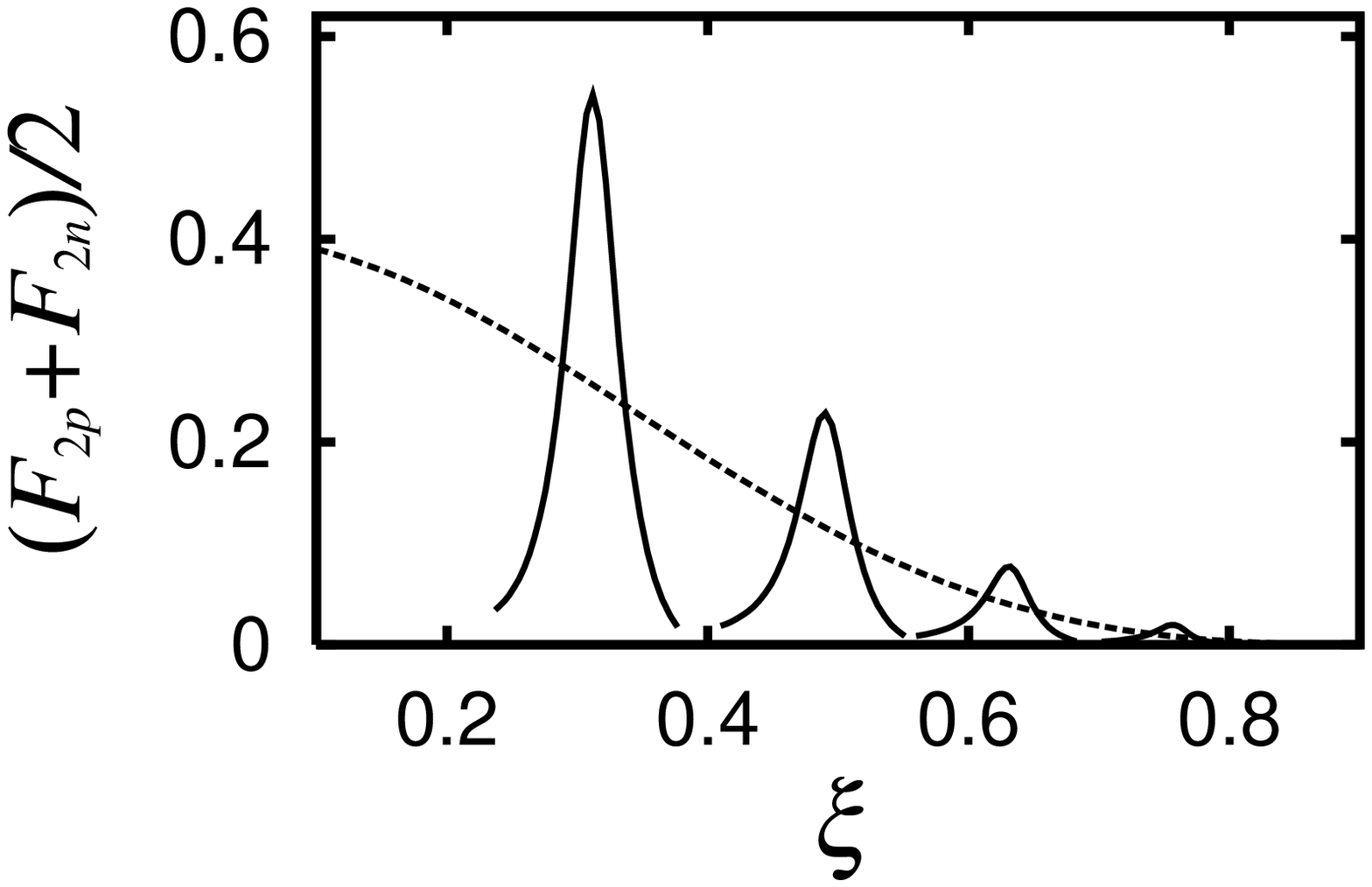}
\includegraphics[width=4.5cm]{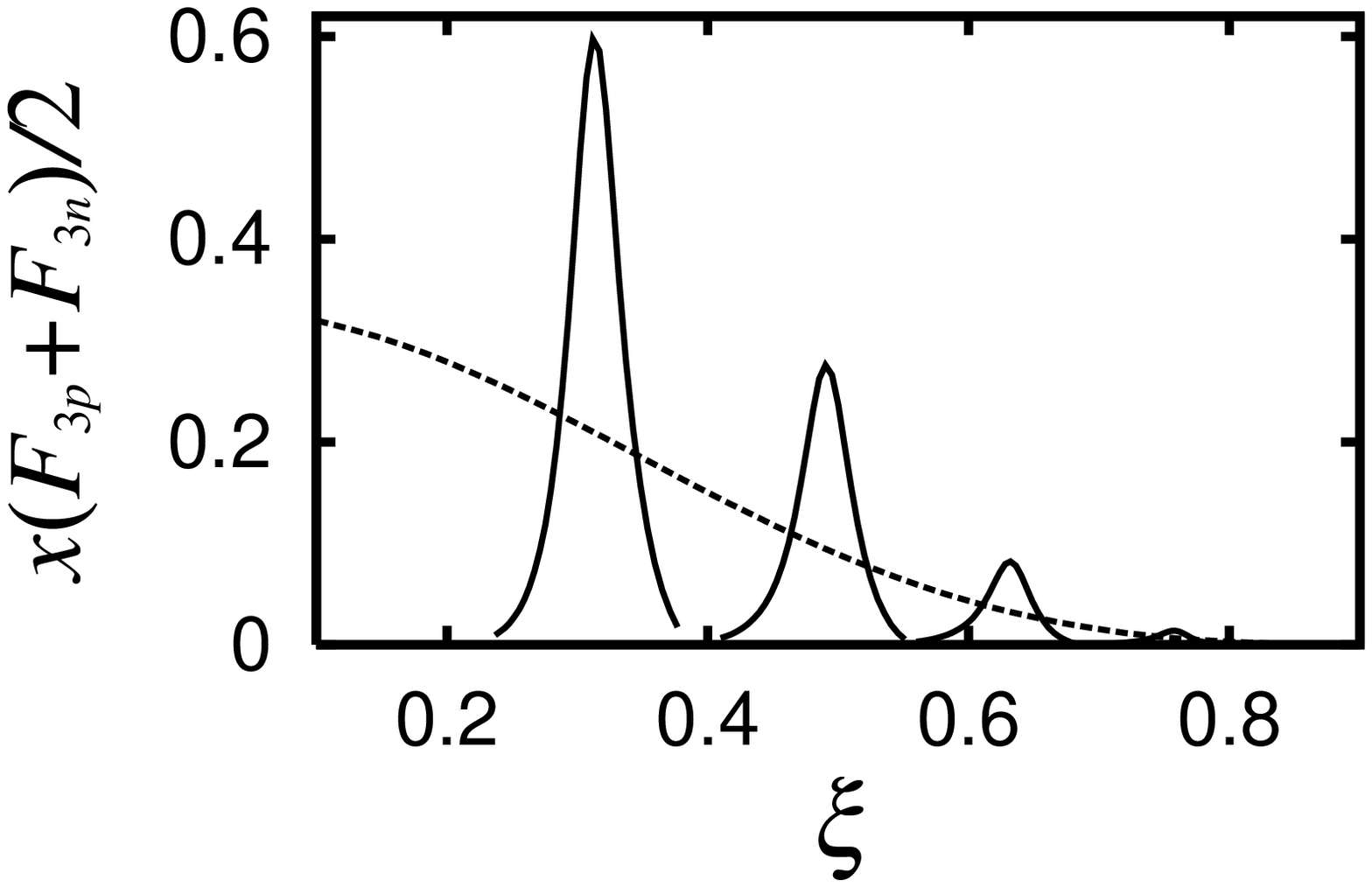}
\caption{Structure functions $xF_1$ (left),
$F_2$ (center), and $xF_3$(right) of $(\nu,\nu^\prime)$ 
for the proton (top)
and an I=0 deuteron-like target (bottom).
The dashed curves are
calculated from  using the
CTEQ6 parton distribution functions at $Q^2=10$
(GeV/c)$^2$. The solid curves are the results at $Q^2= 0.4,1,2,4$ (GeV/c)$^2$
(from left to right) calculated from the extended SL model described in this
work.}

\end{figure}

\begin{figure}
\centering
\includegraphics[width=6cm]{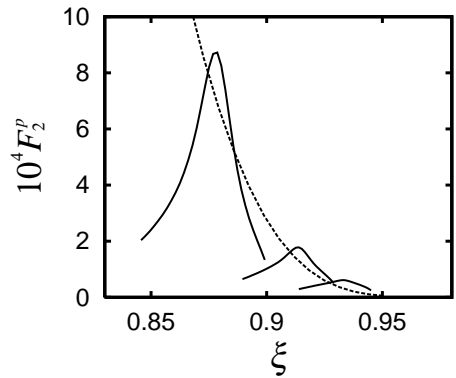}
\caption{The solid curves are the structure function $F_2$ of
$p(e,e^\prime)$ as a function of the scaling variable $\xi$
calculated at $Q^2=10,15,20$ (GeV/c)$^2$ (from left to right) from the
SL model. The dashed curve is calculated from
using
the CTEQ6 parton distributions at $Q^2=20$
(GeV/c)$^2$.}
\end{figure}

\begin{figure}
\centering
\includegraphics[width=10cm]{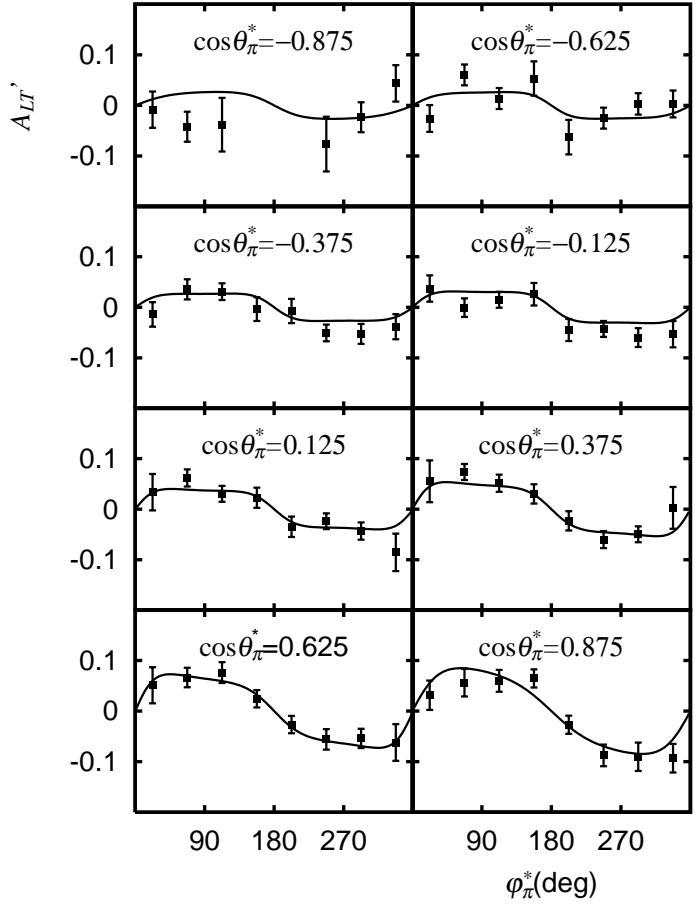}
\caption{The beam asymmetry $A_{LT'}$ (defined in Ref.\cite{joo})
 vs  $\phi^*_\pi$ 
of $p(e,e^\prime \pi^+)n$ reaction 
at $Q^2=0.4$ (GeV/c)$^2$ and $W=1.22$ GeV 
predicted by
the SL model are compared with the data of Joo et al. \cite{joo}.
$(\theta^*_\pi, \phi^*_\pi$) are the pion angles in the center of mass frame of
the final $\pi N$ system.}.
\end{figure}

\begin{figure}
\centering
\includegraphics[width=10cm]{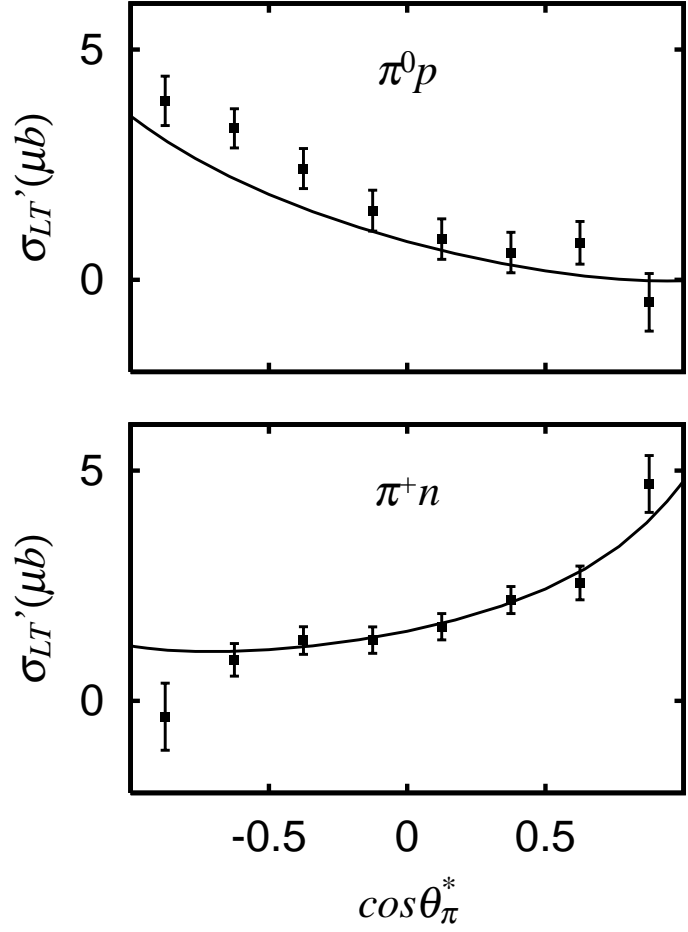}
\caption{The transverse-longitudinal interference term
 $\sigma_{LT'}$ (defined in Ref.\cite{joo}) vs $cos\theta^*$  for
$p(e,e^\prime \pi^0)p$ (top) and $p(e,e^\prime \pi^+)n$ (bottom) 
at $W=1.22GeV$
predicted by
the SL model are compared with the data of Joo et al. \cite{joo}.
$\theta^*_\pi$ is the pion scattering angle in the center of mass frame of
the final $\pi N$ system.}
\end{figure}

\clearpage

\begin{figure}
\centering
\includegraphics[width=6cm]{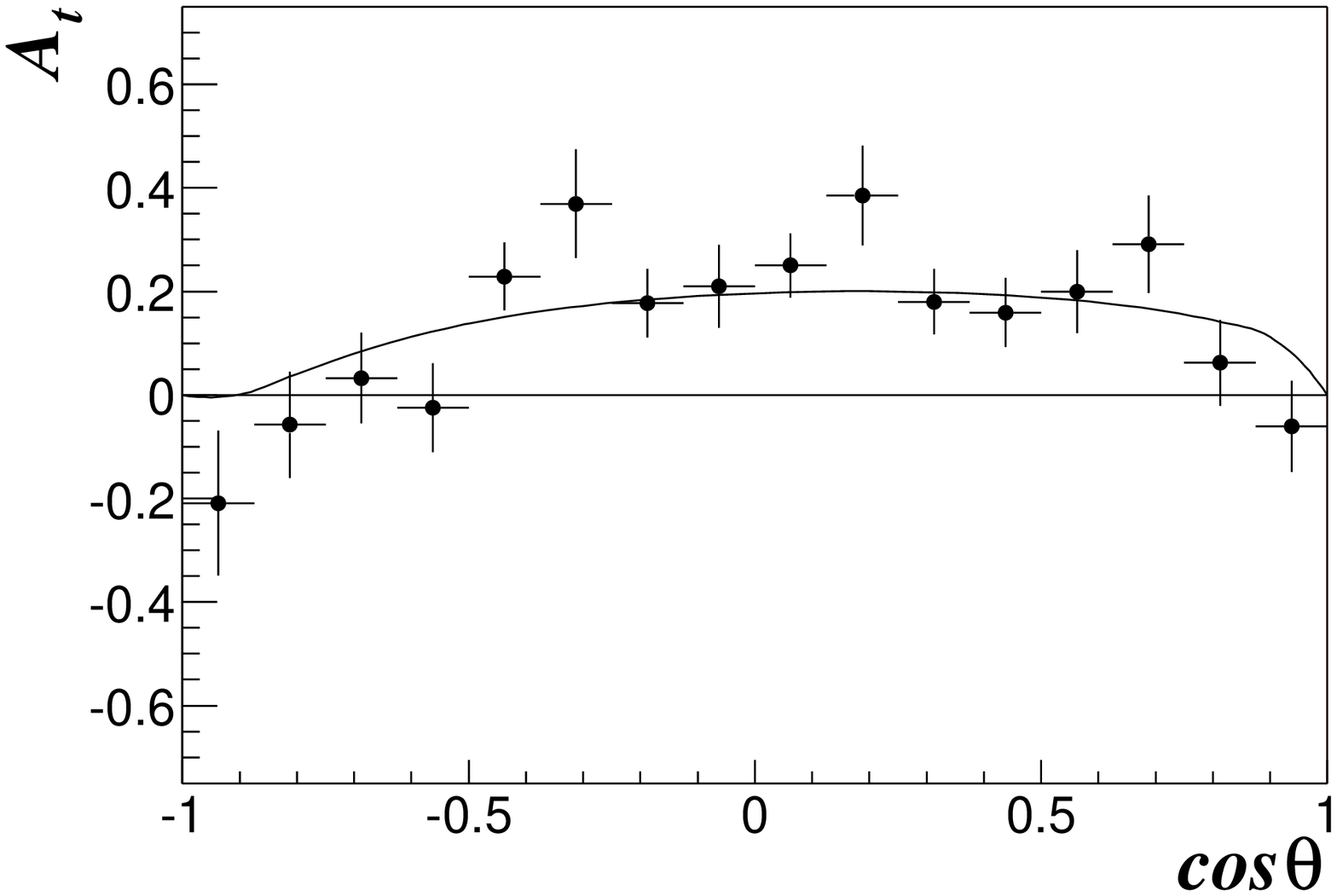}
\includegraphics[width=6cm]{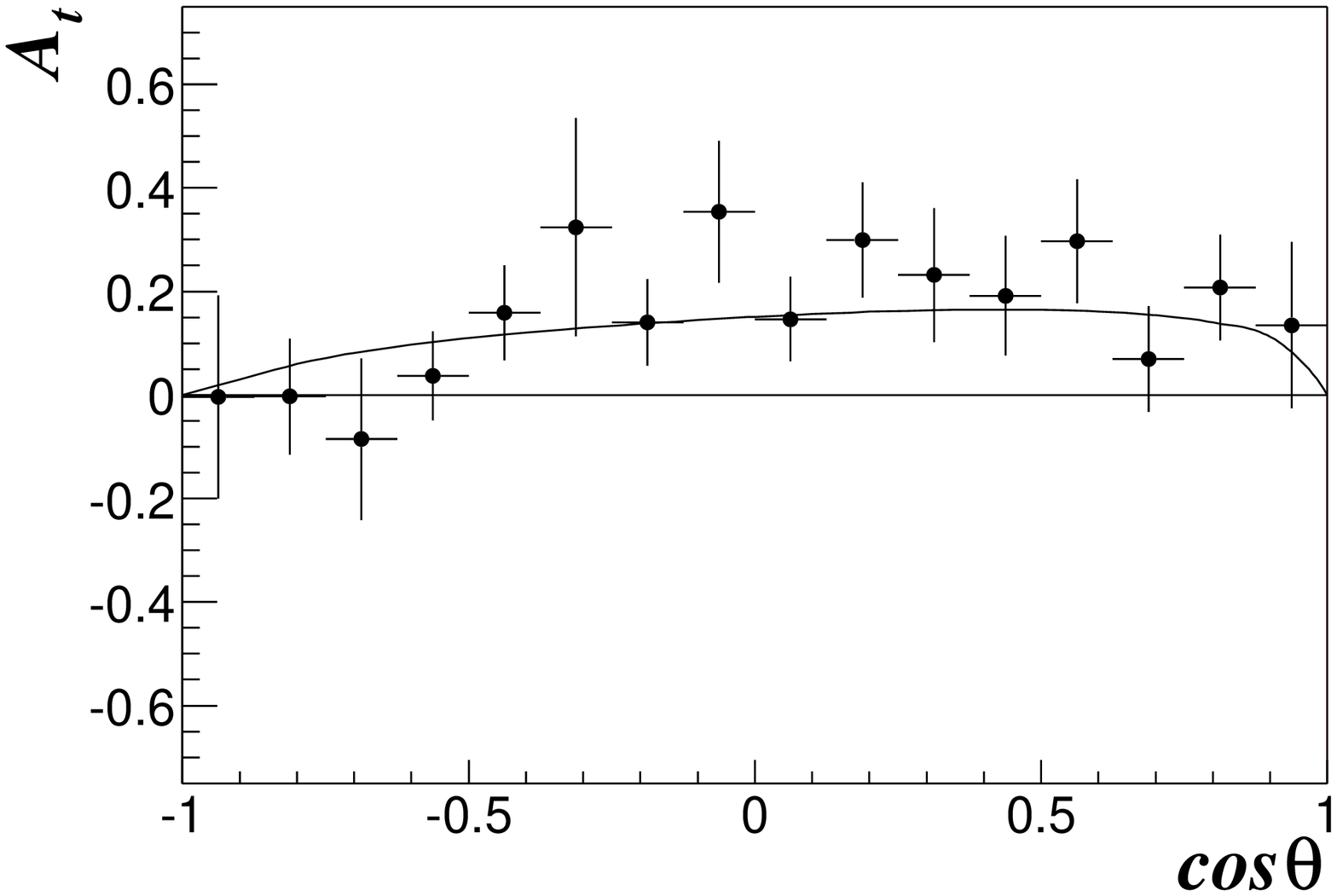}

\includegraphics[width=6cm]{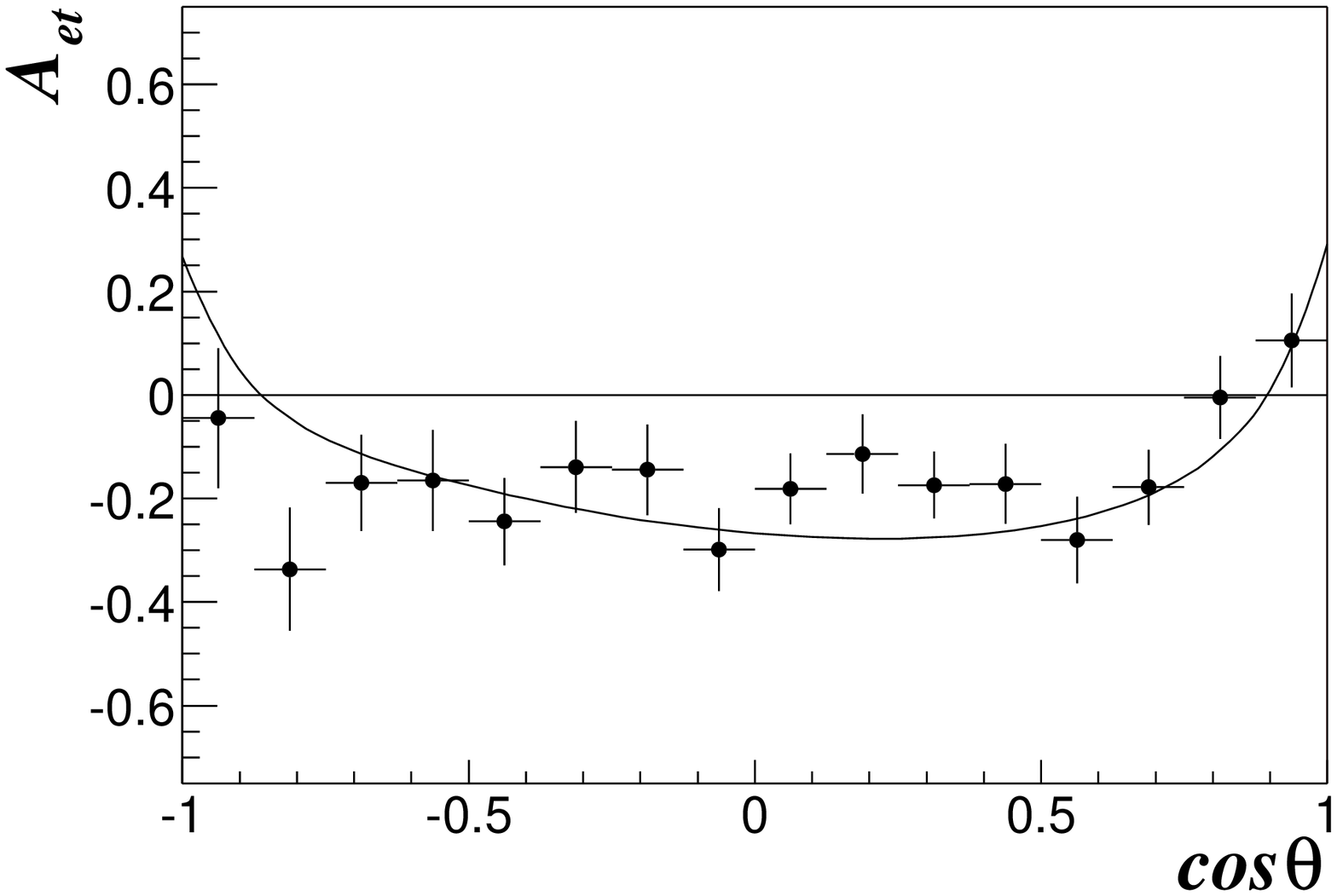}
\includegraphics[width=6cm]{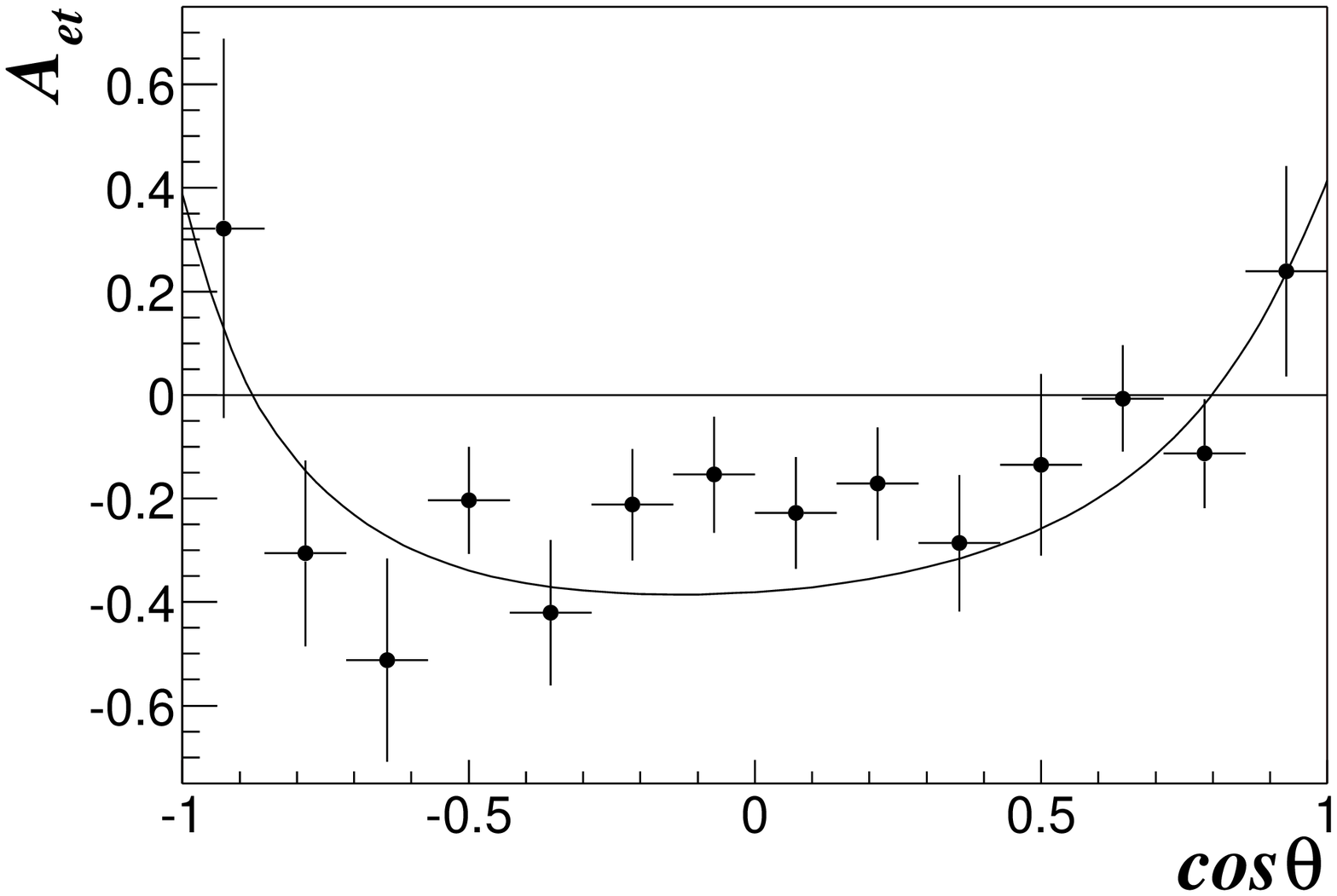}
\caption{The beam-target asymmetries $A_{t}$(upper half))
 and $A_{et}$(lower half)  (as defined in Ref.\cite{biselli})
of $p(e,e^\prime \pi^0)p$ reaction  as a 
function of $\cos \theta^*$ integrated over $\phi^*$ at $W=1.22$ GeV and \\
$ 0.5 $ (GeV/c)$^2$ $<$ Q$^2$ $< 0.9$ (GeV/c)$^2$ (left),
$ 0.9 $ (GeV/c)$^2$ $<$ Q$^2$ $< 1.5$ (GeV/c)$^2$ (right).
$(\theta^*, \phi*$) are the pion angles in the center of mass frame of
the final $\pi N$ system. The solid curves are the predictions of the
SL model. The data are from
Biselli et al. \cite{biselli}}
\end{figure}

\begin{figure}
\centering
\includegraphics[width=6cm]{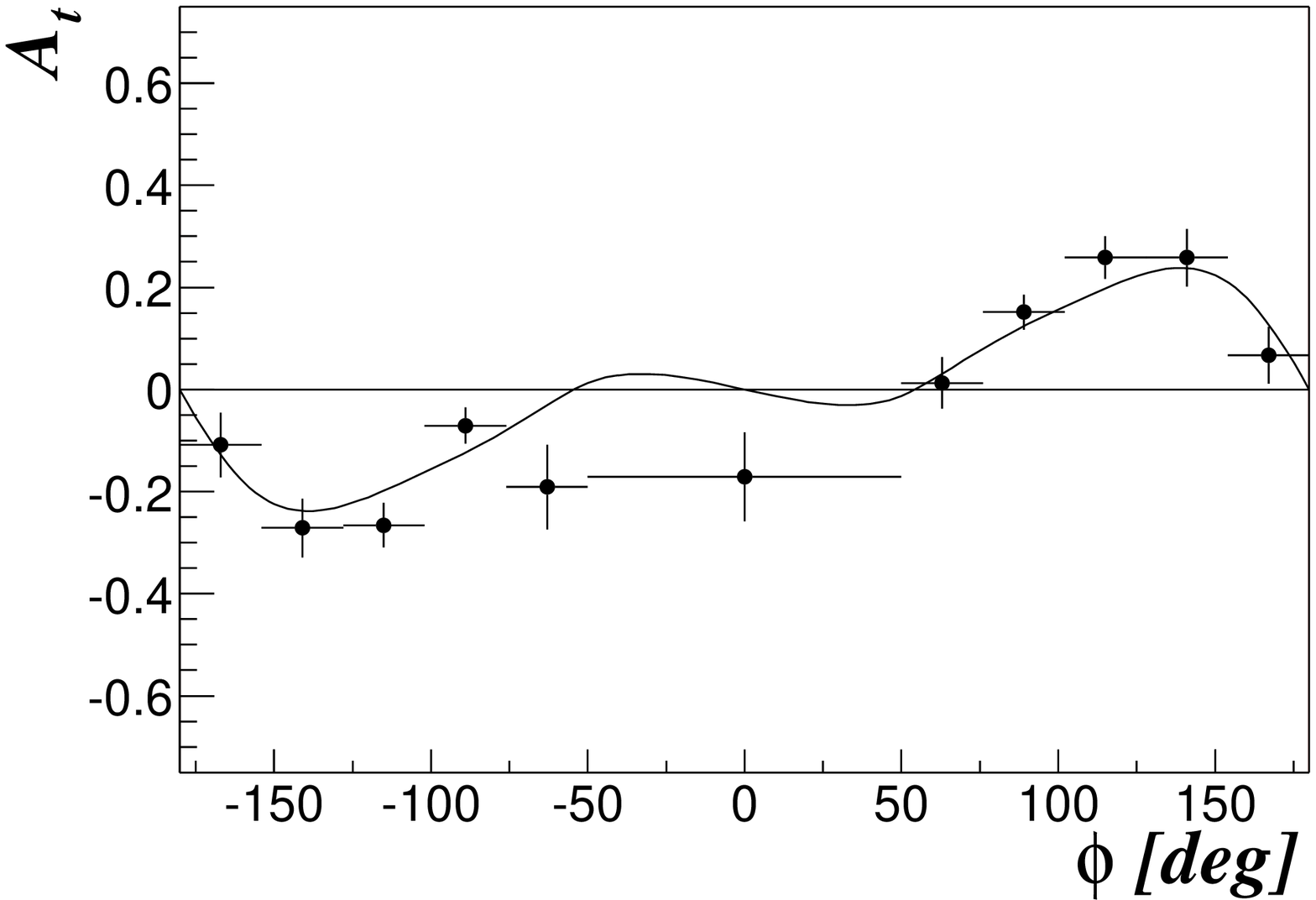}
\includegraphics[width=6cm]{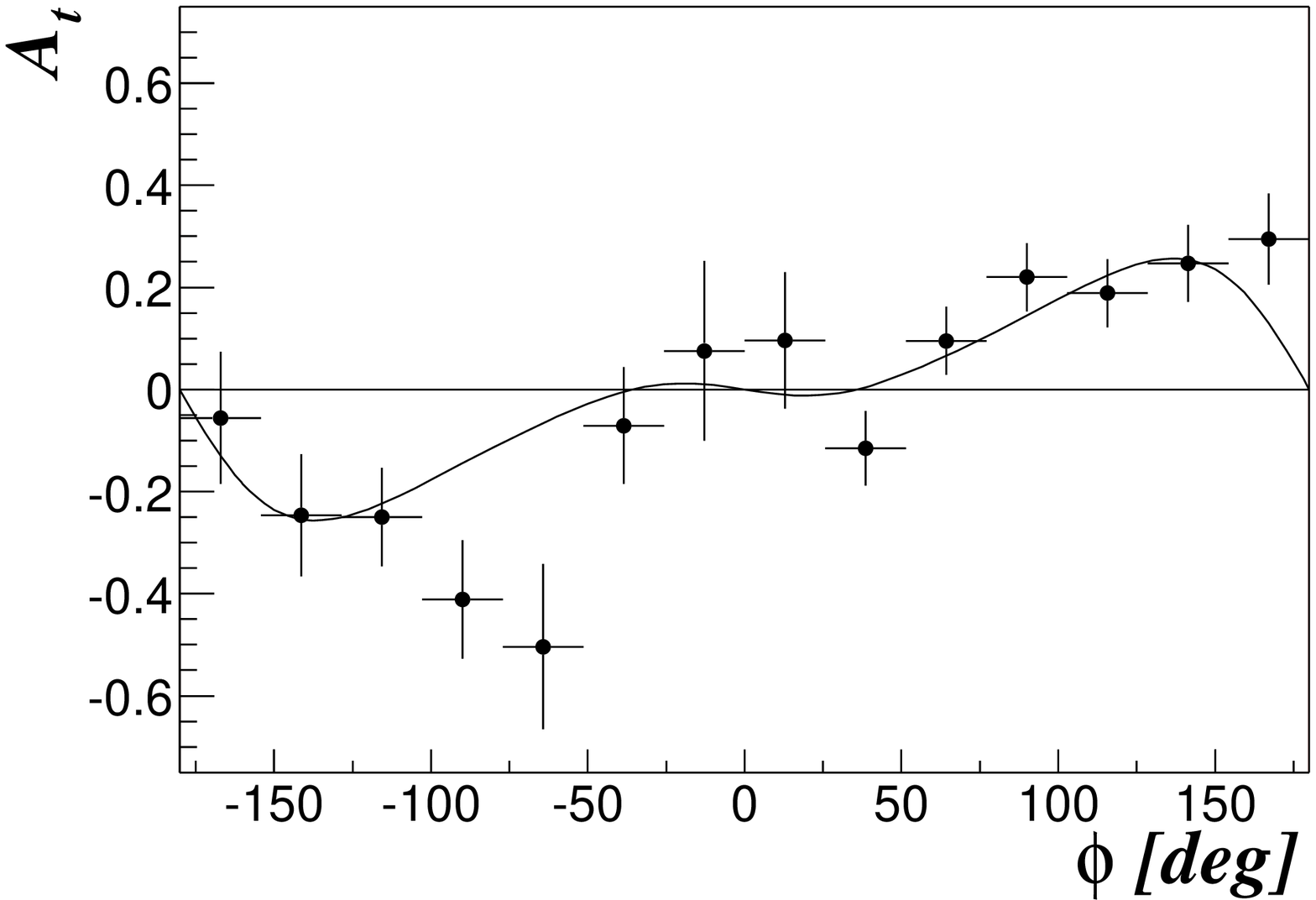}

\includegraphics[width=6cm]{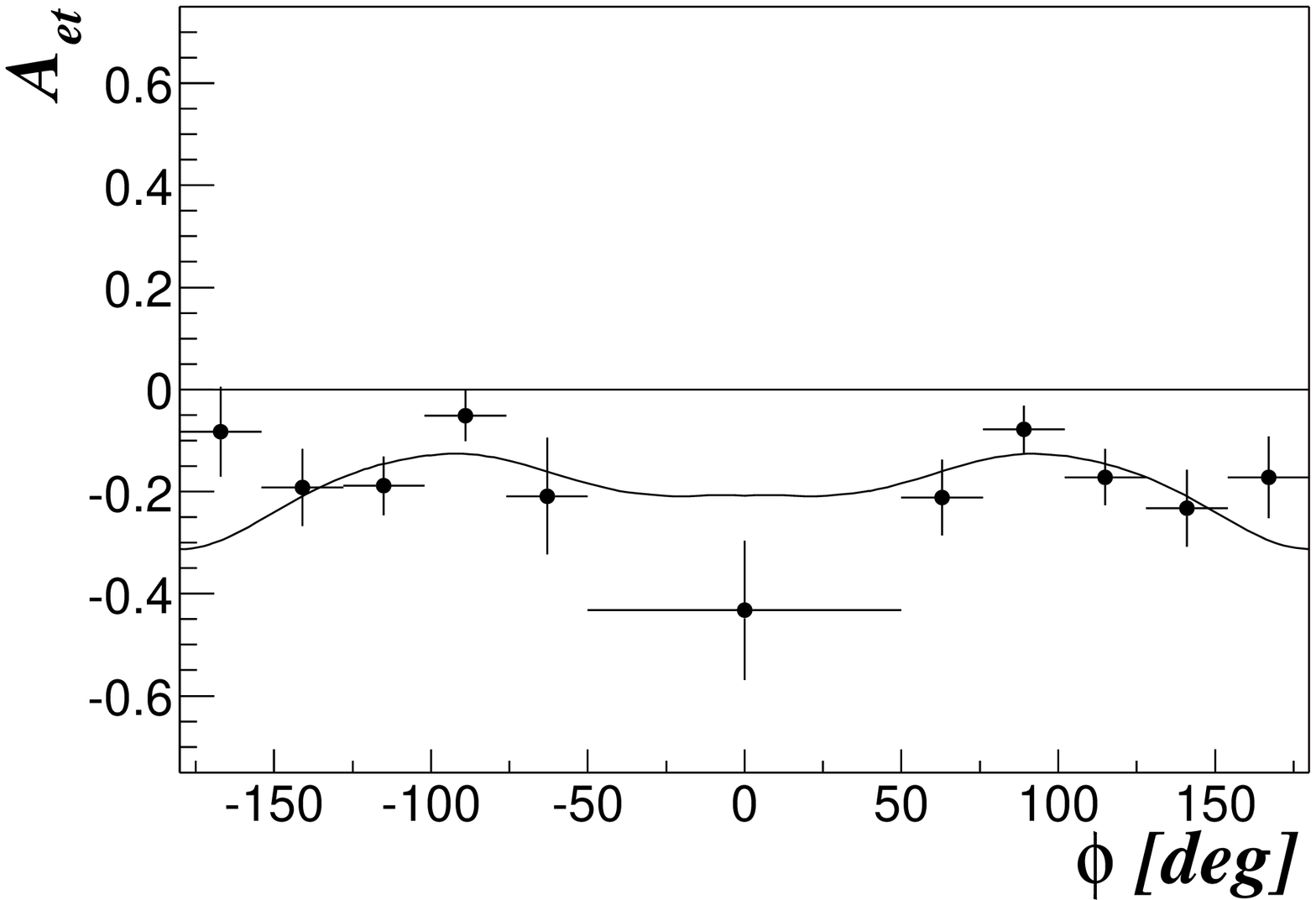}
\includegraphics[width=6cm]{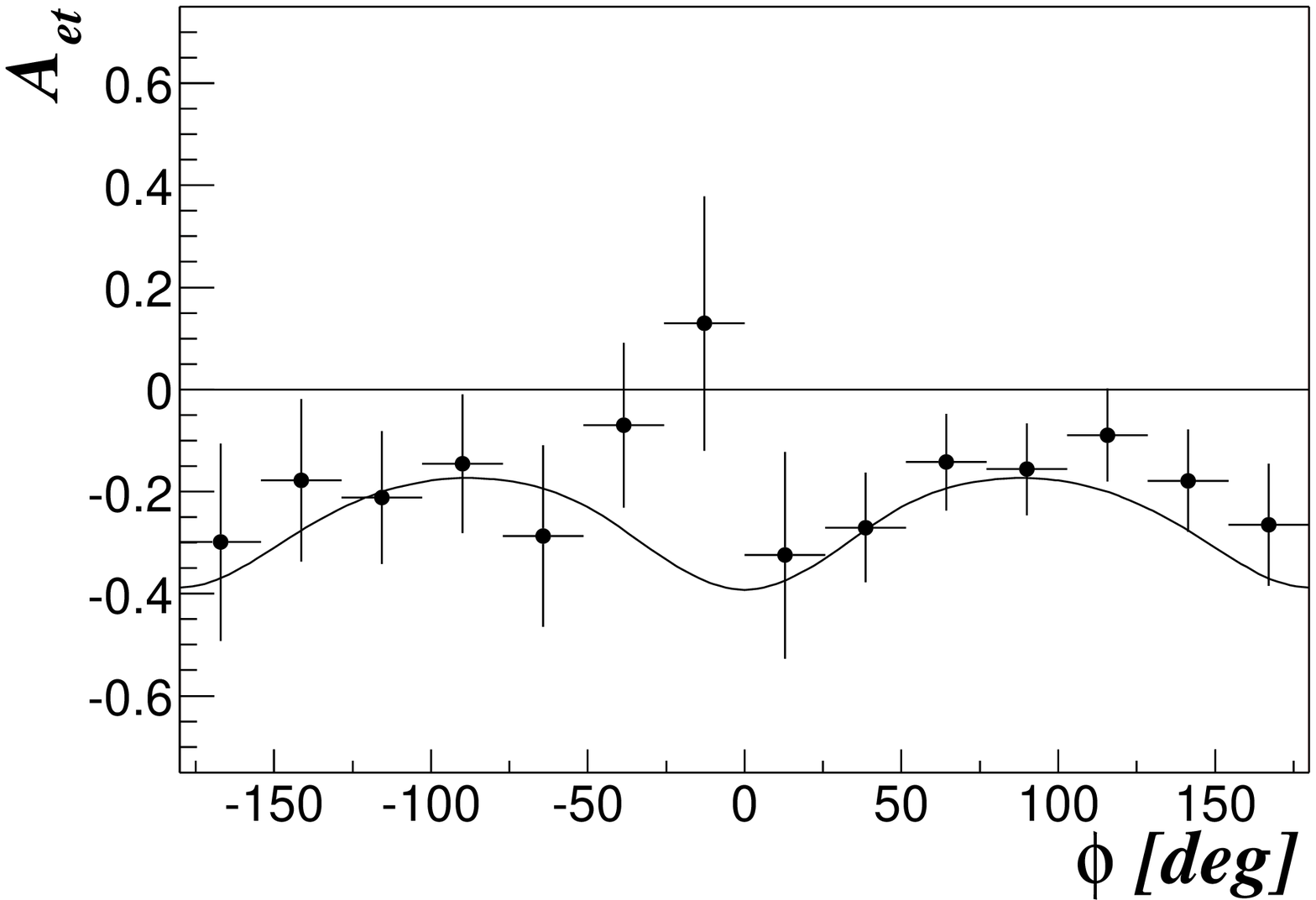}
\caption{The beam-target asymmetries $A_{t}$(upper half)) 
 and $A_{et}$(lower half) (as defined in Ref.\cite{biselli}) 
of $p(e,e^\prime \pi^0)p$ reaction as a function
of $\phi^*$ integrated over $\cos\theta^*$ at $W=1.22$ GeV and \\
$ 0.5 $ (GeV/c)$^2$ $<$ Q$^2$ $< 0.9$ (GeV/c)$^2$ (left),
$ 0.9 $ (GeV/c)$^2$ $<$ Q$^2$ $< 1.5$ (GeV/c)$^2$ (right).
$(\theta^*, \phi*$) are the pion angles in the center of mass frame of 
the final $\pi N$ system. The solid curves are the predictions of the
SL model.  The data are from
Biselli et al. \cite{biselli}.}
\end{figure}

\begin{figure}
\centering
\includegraphics[width=6cm]{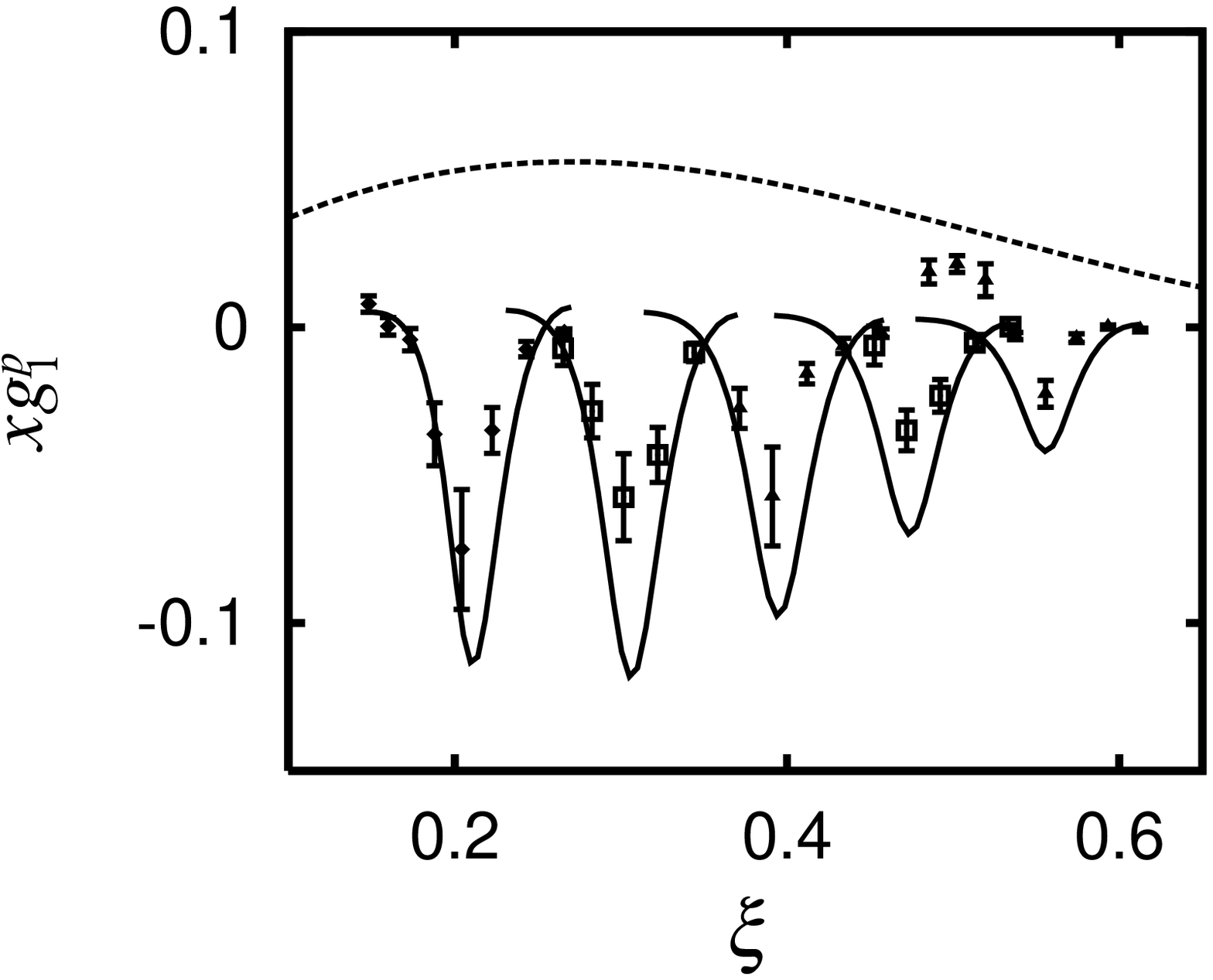}\hspace*{0.5cm}
\includegraphics[width=6cm]{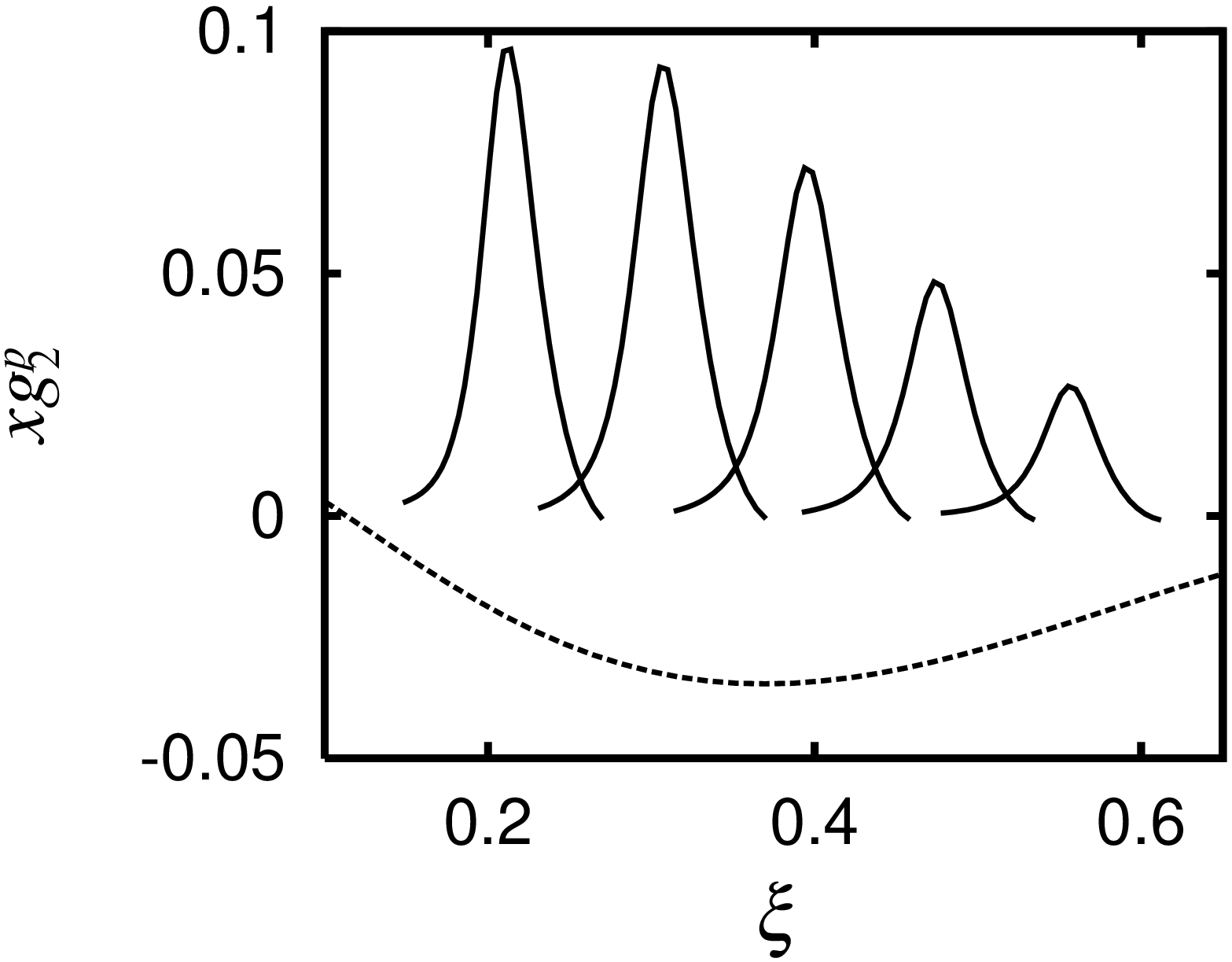}
\caption{Spin dependent structure functions $g_1$ (left) and $g_2$ (right)
for the proton target.
The dashed curves are from the fits to the deeply inelastic scattering data as
explained in the text. The solid curves are the results
at $Q^2 =0.21, 0.35, 0.62, 0.92, 1.37$ (GeV/c)$^2$ (from left to right)
calculated from the  SL model. 
 The data for $g_1$ are from Fatemi et al. \cite{fatemi}.}
\end{figure}

\begin{figure}
\centering
\includegraphics[width=6cm]{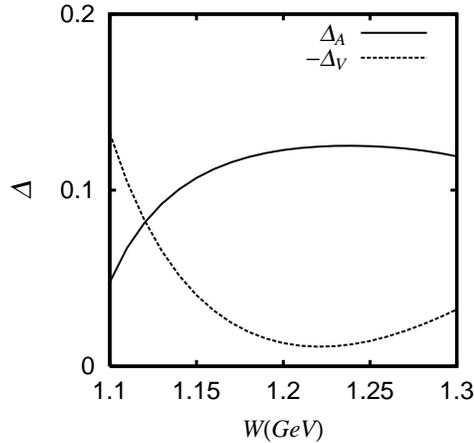}
\caption{
The parity violating aymmetry 
terms $\Delta_A$(solid curve) and $-\Delta_V$(dotted curve)
(defined in Eqs.(53)-(54)) of $p(\vec{e},e ^\prime)$ as a
function of invariant mass $W$ calculated from the extended SL model
described in this work. The results are for
incident electron energy $E=1$ GeV
and scattering angle $\theta=110^o$.}
\end{figure}

\begin{figure}
\centering
\includegraphics[width=6cm]{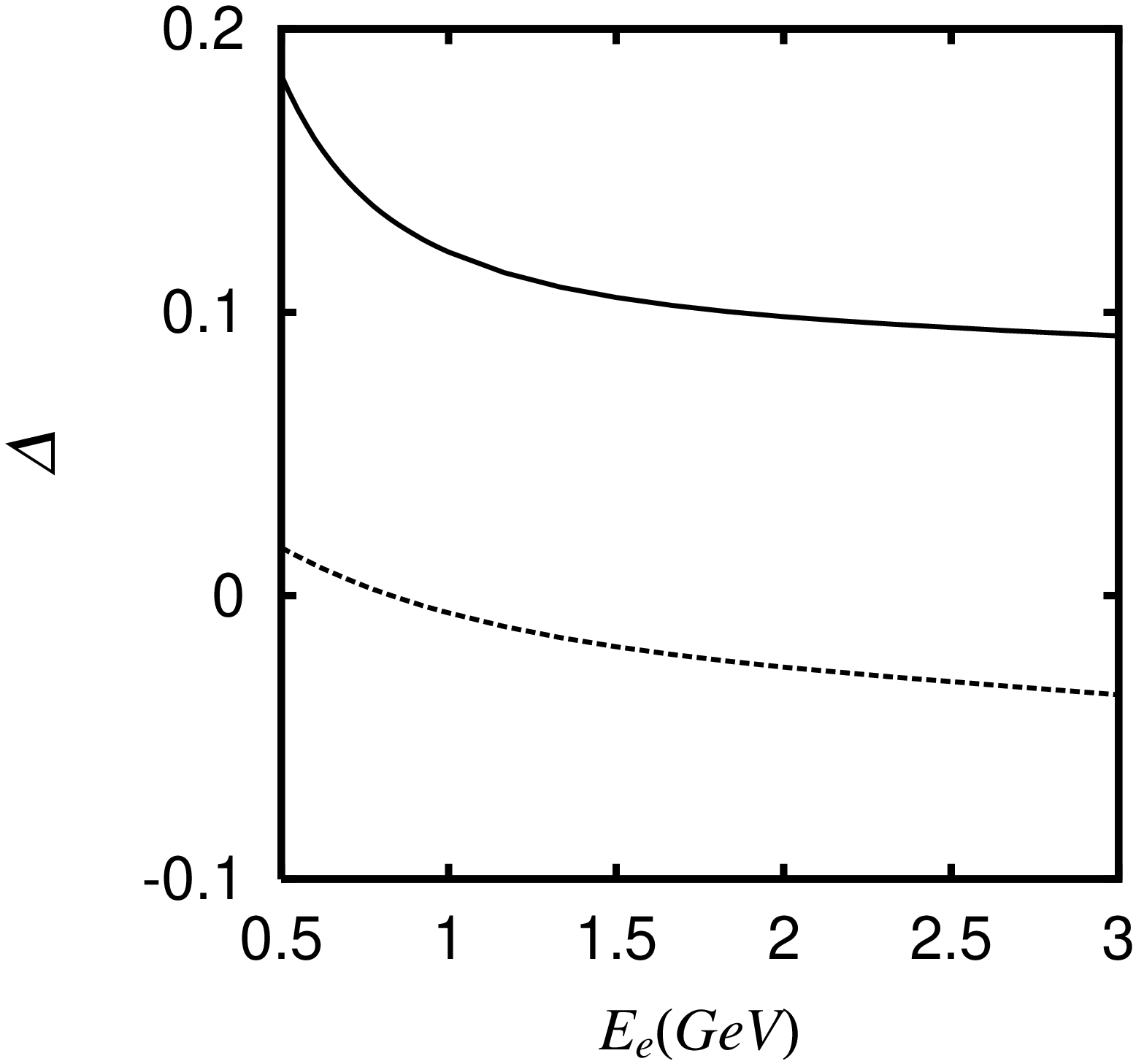}\hspace*{0.5cm}
\includegraphics[width=6cm]{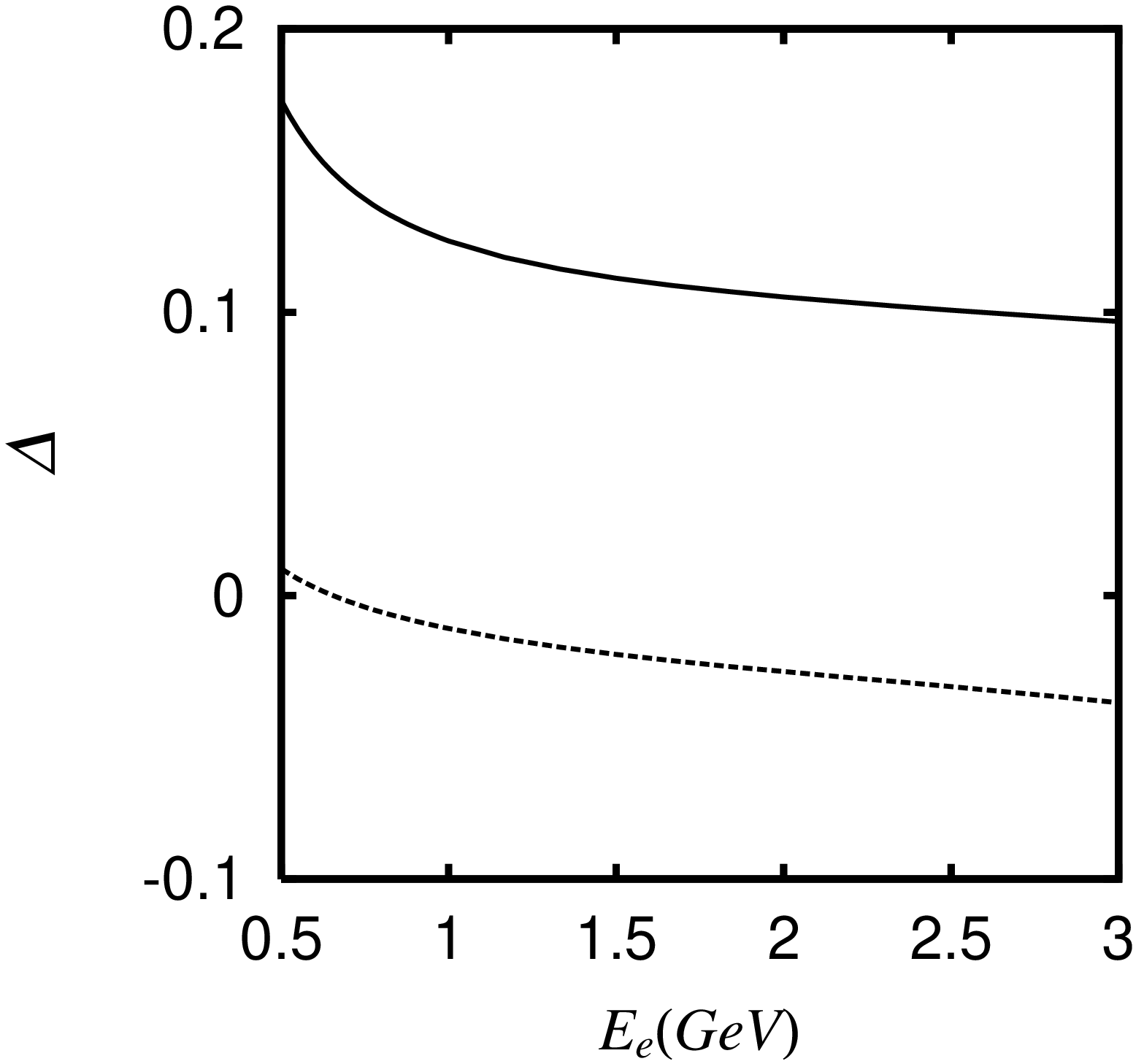}
\caption{The parity violating asymmetry terms
$\Delta_A$(solid curves) and $\Delta_V$(dotted curves) 
(defined in Eqs.(53)-(54)) of $p(\vec{e},e ^\prime)$ as
a function of the incident electron energy $E_e$ 
calculated from the extended SL model described in this work.
The results are for
invariant mass $W=1.232$ GeV and 
electron scattering angles
 $\theta=60^o$(left) and  $\theta=110^o$(right). }
\end{figure}

\begin{figure}
\centering
\includegraphics[width=8cm]{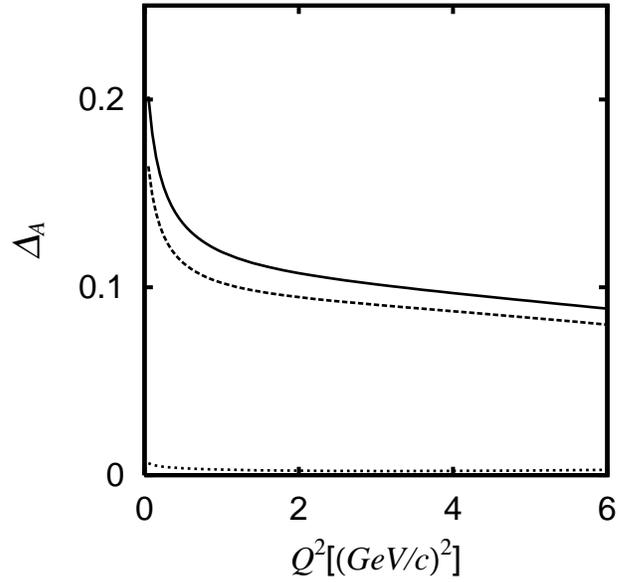}
\caption{
The parity violating asymmetry term
$\Delta_A$ (defined in Eq. (54)) of $p(\vec{e},e ^\prime)$ as
a function of $Q^2$ evaluated with
an invariant mass $W=1.232GeV$ and
electron scattering angle
$\theta=110^o$. The solid curve is from the extended SL model described in
this work. The dashed curve is from turning off the
 pion cloud effect on the axial $N$-$\Delta$ transition.
The dotted curve near the bottom is from keeping only the
non-resonant amplitude in the calculation.} 
\end{figure}

\begin{figure}
\centering
\includegraphics[width=8cm]{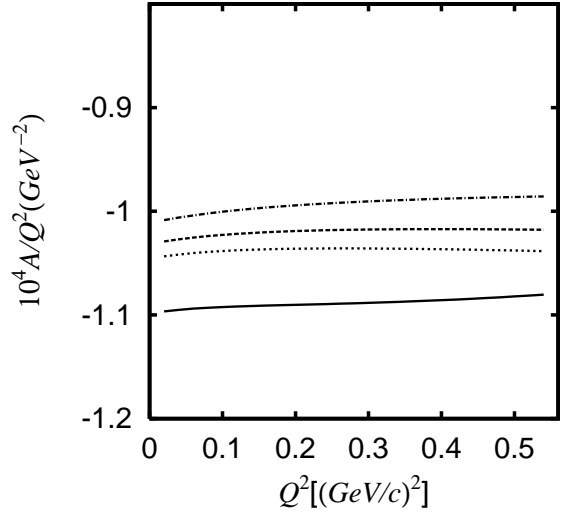}
\caption{
The $Q^2$-dependence of the scaled
parity violating asymmetry $A/Q^2$ of $p(\vec{e},e^\prime)$
calculated from the extended SL model described in this work.
The results are for $W=1.232$ GeV and incident electron energy E=
       0.8 (solid curve),
       1.5 (dotted),
       2.0 (dashed),
       4.0 (dash-dotted) }
\end{figure}
%

%
\begin{figure}
\centering
\includegraphics[width=8cm]{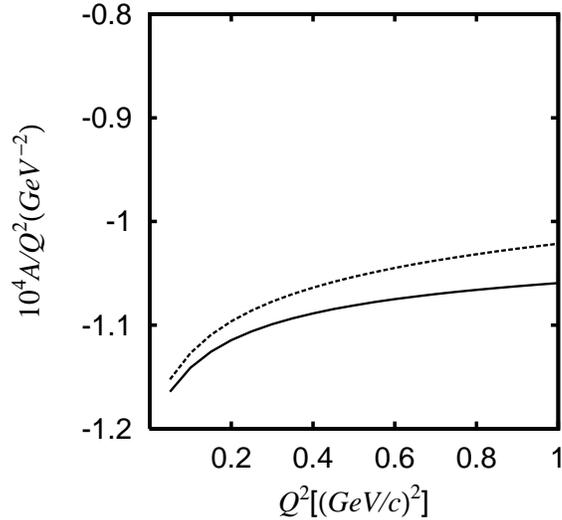}
\caption{
The scaled
parity violating asymmetry $A/Q^2$ of $p(\vec{e},e^\prime)$
calculated from the extended SL model described in this work.
The solid (dashed) curve is from using
the axial $N$-$\Delta$ form factor
$G^A_{N,\Delta}$ of Ref.\cite{sl3} ( Refs.\cite{kitagaki}).
The results are for invariant $W$=1.232 GeV and electron angle $\theta=110^o$.}
\end{figure}
%
\begin{figure}
\centering
\includegraphics[width=6cm]{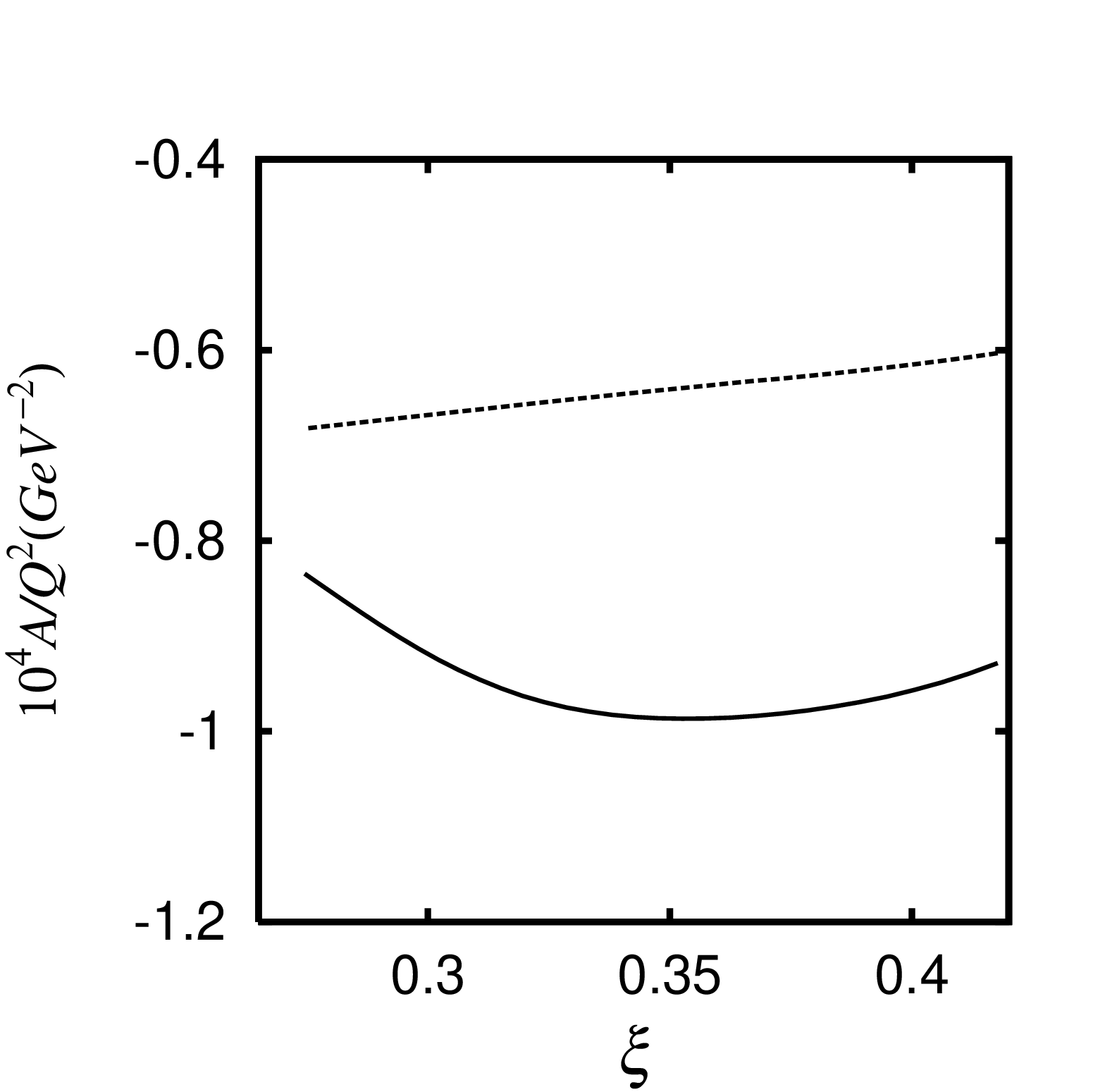}\hspace*{0.5cm}
\includegraphics[width=6cm]{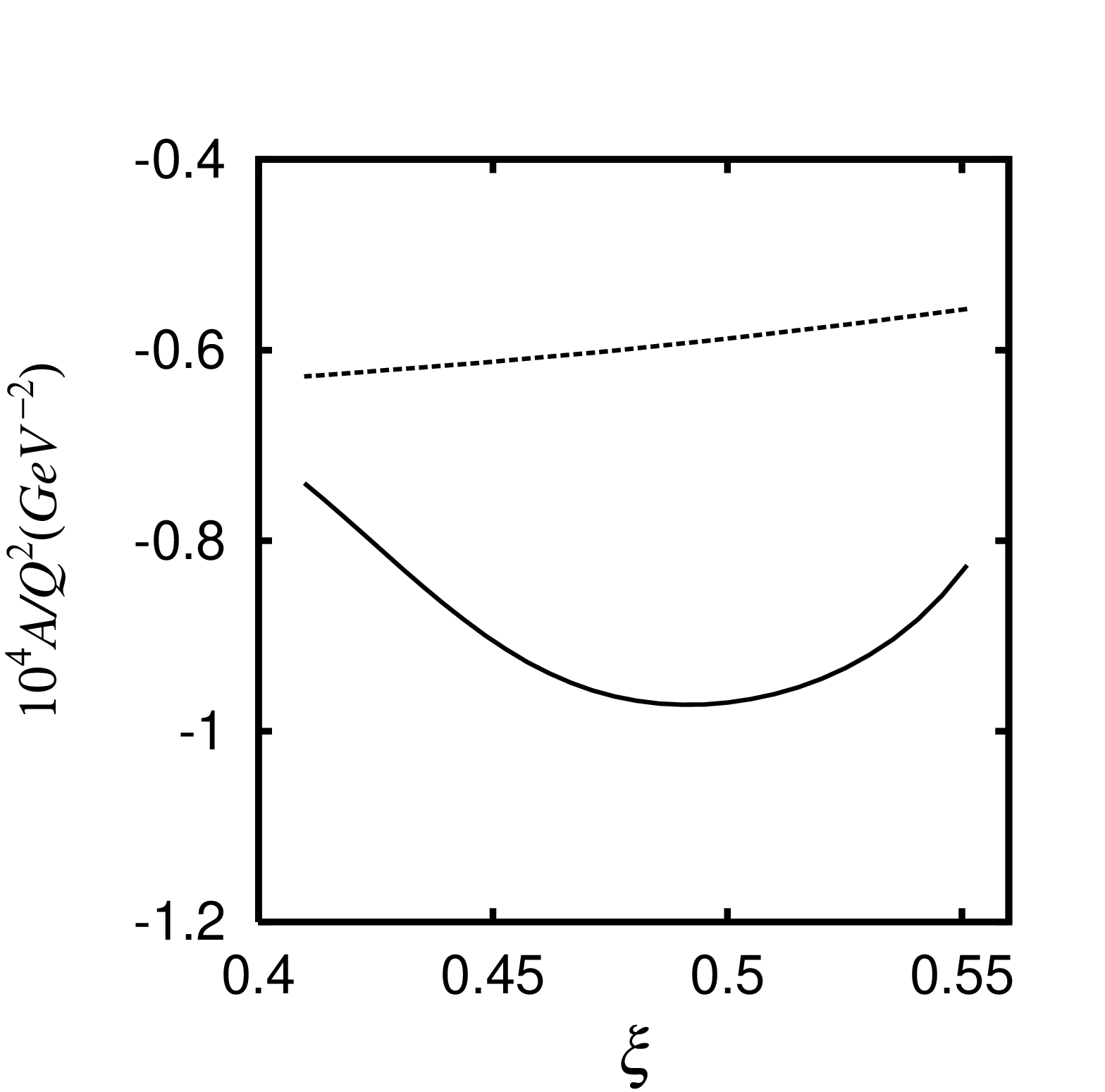}

\includegraphics[width=6cm]{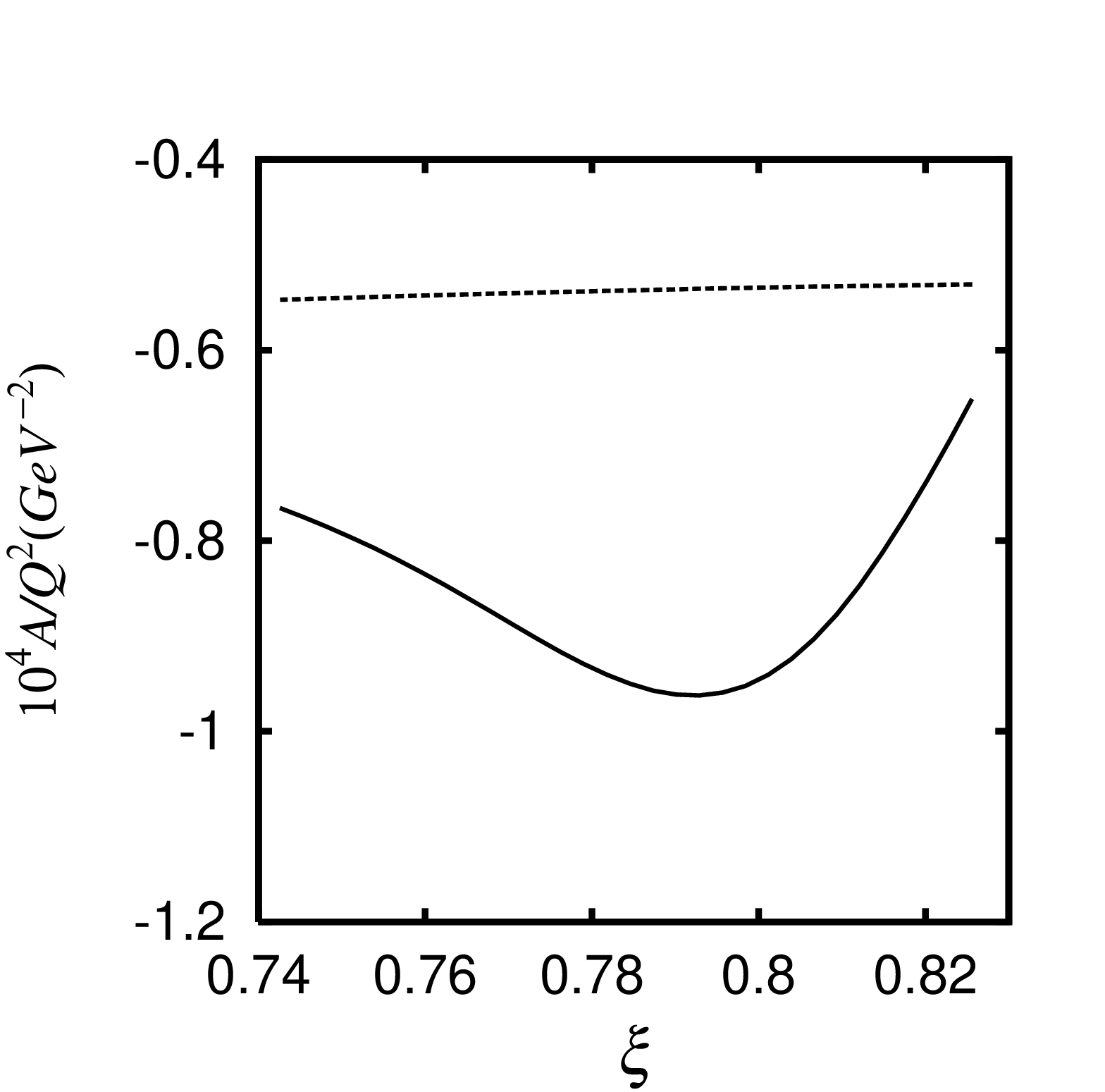}\hspace*{0.5cm}
\includegraphics[width=6cm]{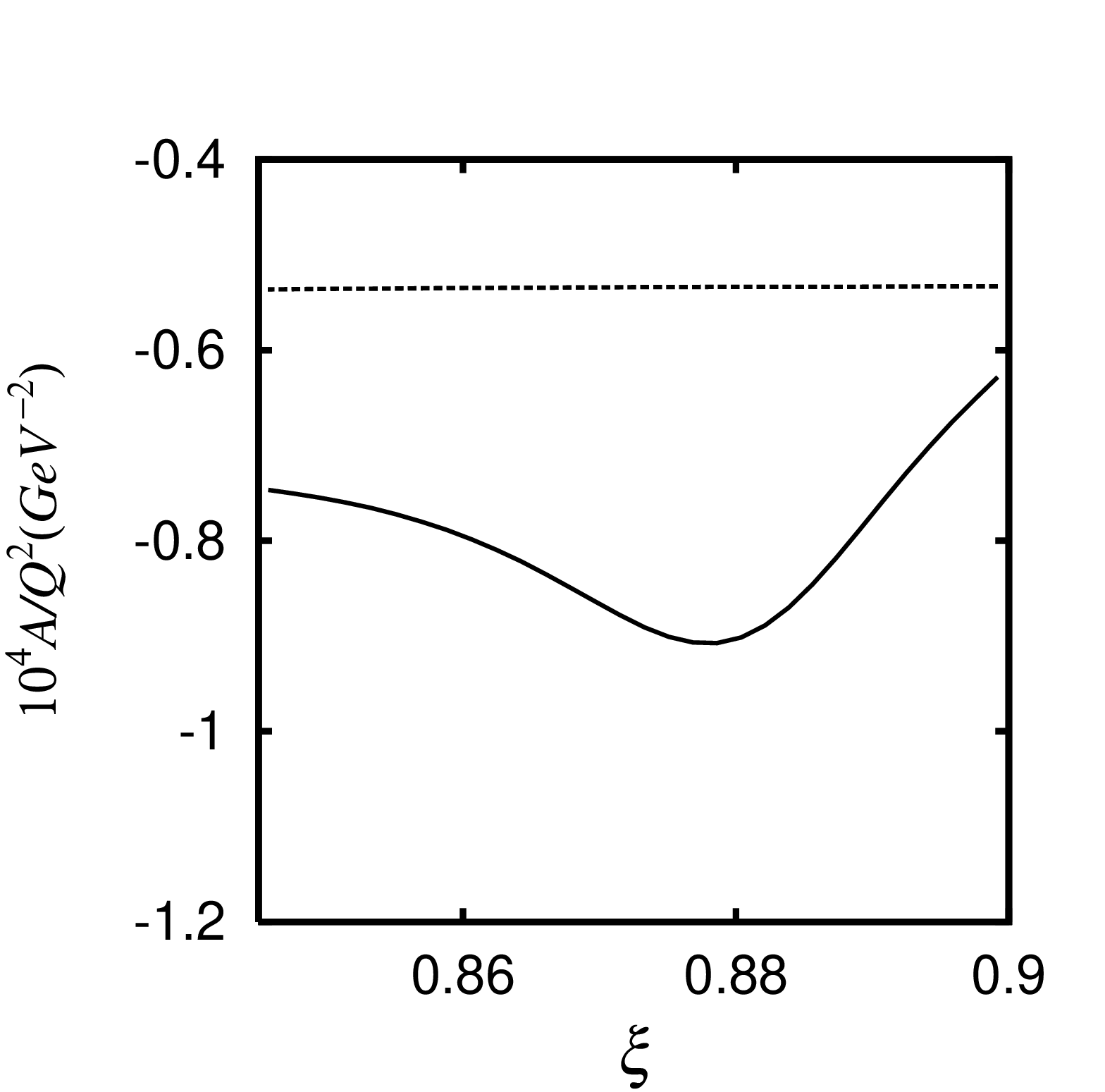}
\caption{
The  scaled
parity violating asymmetry $A/Q^2$ of $p(\vec{e},e^\prime)$ as
a function of the scaling variable $\xi$ 
predicted by the extended SL model described in this work (solid curves)
is compared with  that calculated from Eqs.(55)-(59) 
using the CTEQ6 parton distribution
functions (dashed curves). The results are for
 $E=4$ GeV ,$Q^2=0.5$ (GeV/c)$^2$ (top left), $E=6$ GeV
,$Q^2=1$ (GeV/c)$^2$ (top right),
$E=8$ GeV, $Q^2=5$ (GeV/c)$^2$ (bottom left), 
and $E=12$ GeV, $Q^2=10$ (GeV/c)$^2$ (bottom right).}
\end{figure}


\begin{thebibliography}{99}
\bibitem{sl1}
T.~Sato and T.-S.~H. Lee, Phys. Rev. C {\bf 54}, 2660 (1996).

\bibitem{sl2}
T.~Sato and T.-S.~H. Lee, Phys. Rev. C {\bf 63}, 055201 (2001).

\bibitem{sl3}
T. Sato, D. Uno and T.-S. H. Lee, Phys. Rev. C{\bf 67}, 065201 (2003).

\bibitem{bg}
E. D. Bloom and F. J. Gilman, Phys. Rev. Lett. {\bf 25}, 1140 (1970).

\bibitem{fermi}
J.G. Morfin, Nucl. Phys. {\bf B112}, 251 (2002).

\bibitem{well}
S.P. Well, N. Simicevic, K. Johnson and The G0 Collaboration, proposal
PR-04-101, Jefferson Laboratory (2004).

\bibitem{bosted}
P. Bosted et al, proposal PR-05-005, Jefferson Laboratory (2005).

\bibitem{meln}
 W. Melnitchouk, R. Ent, and C. E. Keppel,
Phys. Rep. {\bf 406}, 127 (2005).

\bibitem{nicu}
I. Niculescu et al., Phys. Rev. Lett. {\bf 85}, 1186 (2000).

\bibitem{liang}
Y. Liang et al. , arXiv:nucl-ex/0410027 v1 (2004).

\bibitem{nach}
O. Nachtmann, Nucl. Phys. {\bf B63}, 237 (1973).

\bibitem{gp}
H. Georgi and H. D. Politzer, Phys. Rev. D {\bf 14}, 1829 (1976).

\bibitem{rgp}
A. De Rujula, H. Georgi, and H. D. Politzer, Phys. Lett. {\bf B 64}, 428 (1976);
Annuals Phys. {\bf 103}, 315 (1977).

\bibitem{xi1}
X. Ji and P. Unrau, Phys. Rev. D {\bf 52}, 72 (1995).

\bibitem{xi2}
X. Ji and W. Melnitchouk, Phys. Rev. D {\bf 56}, R1 (1997).

\bibitem{mueller}
A. Mueller, Phys. Lett. {\bf B 308}, 355 (1993).

\bibitem{carlson}
C.E. Carlson and N.C. Mukhopadhyay, Phys. Rev. Lett. {\bf 74}, 1288 (1995);
Phys. Rev. D {\bf 58}, 094029 (1998).

\bibitem{edel}
J. Edelmann, G. Piller, N. Kaiser, and W. Weise,
Nucl. Phys. {\bf A665}, 125 (2000).

\bibitem{meln-1}
W. Melnitchouk, K. Tsushima, A.W. Thomas,
     Eur. Phys. J. {\bf A14}, 105 (2002).

\bibitem{holstein}
For Standard model, see, for example, {\bf Dynamics of the Standard Model},
by J. F. Donoghue, E. Golowich, and B. R. Holstein,
1992 (Cambridge University Press)

\bibitem{biselli}
A. Biselli et al., Phys. Rev. C {\bf 68}, 035202 (2003).

\bibitem{joo}
K. Joo et al., Phys. Rev. C {\bf 70}, 042201 (2004).

\bibitem{fatemi}
R. Fatemi et al., Phys. Rev. Lett. {\bf 91}, 222002 (2003).

\bibitem{abe}
K. Abe et al., Phys. Rev. D {\bf 58}, 112003 (1998).

\bibitem{wwil}
S. Wandzura and F. Wilczek, Phys. Lett. {\bf B 72}, 195 (1977).

\bibitem{burkertlee}
V. Burkert and T.-S. H. Lee, Int.J.Mod.Phys {\bf E13}, 1035-1112 (2004).

\bibitem{kitagaki}
T. Kitagaki et al., Phys. Rev. D {\bf 42}, 1331 (1990);
an other references there in.

\bibitem{nimi-1}
T.R. Hemmert, B.R. Holstein, and N.C. Mukhopadhyay, Phys. Rev. D {\bf 51}, 
158 (1995).

\bibitem{nimi-2}
J. Liu, N.C. Mukhopadhyay, and L. Zhang, Phys. Rev. C {\bf 52} 1630 (1995).

\bibitem{golli}
B. Golli, S. Sirca, L. Amoreira and M. Fiolhais,
Phys. Lett. B{\bf 553}, 51 (2003).

\bibitem{bruno}
B. Julia-Diaz, D.O. Riska, and F. Coester, Phys. Rev. C{\bf 70}, 045204 (2004).

\bibitem{cg}
 R.N. Cahn and F.J. Gilman, Phys. Rev. {\bf D17}, 1313 (1978).

\bibitem{jones}
D.R.T. Jones and S.T. Petcov, Phys. Lett {\bf B91}, 137 (1980).

\bibitem{isha}
D. Ishankuliev and M. Ya. Safin, Sov. J. Nucl. Phys. {\bf 31}, 512 (1980).

\bibitem{hwang}
S.P. Li, E.M. Henley, and W.-Y. P. Hwang, Ann. Phys. (NY) {\bf 143}, 371 (1982).

\bibitem{nath}
L.M. Nath, K. Schilcher, M. Kretzschmar, Phys. Rev. {\bf D25}, 2300 (1982).

\bibitem{musolf}
M.J. Musolf et al., Phys. Rep. {\bf 239}, 1 (1994).

\bibitem{hammer}
H-W. Hammer and D. Drechsel, Z. Phys. {\bf A353}, 321 (1995).

\bibitem{nimi-3}
N.C.Mukhopadhyay, M.J. Ramsey-Musolf, S.J. Pollock, J. Liu, and H.-W. Hammer,
Nucl. Phys. {\bf A633}, 481 (1998).

\bibitem{rekalo}
M.P. Rekalo, J. Arvieux, and E.Tomasi-Gustafsson, Phys. Rev. {\bf C65},
035501 (2002).

\bibitem{tw}
A.W. Thomas and W. Weise, {\bf The Structure of the Nucleon}, 2001 (Wiley-VCH)

\bibitem{gl}
J. Gasser and H. Leutwyler, Ann. Phys. {\bf 158}, 142 (1984).

\bibitem{park}
T.-S. Park et al., Phys. Rep. {\bf 233}, 341 (1993).

\bibitem{miss}
V. Bernard et al., J. Phys. G {\bf 28}, R1 (2002).


\bibitem{cteq6}
J. Pumplin, D.R. Stump, J. Huston, H.L. Lai, P. Nadolsky, and W.K. Tung,
Journal of High Energy Physics, {\bf 0207}, 012 (2002) (hep-ph/0201195 ).

\bibitem{ji-1}
X. Ji, Phys. Lett. {\bf B309}, 187 (1993).

\bibitem{barbieri}
R. Barbiberi, J. R. Ellis, M. K. Gaillard, and
G.G. Ross, Nucl. Phys. {\bf B 117}, 50 (1976);
S. Dasu et al., Phys. Rev. D {\bf 49}, 5641 (1994).

\bibitem{arrington}
J. Arrington et al., Phys. Rev. {\bf C64}, 014602 (2001).

\bibitem{brasse}
F. W. Brase et al. Nucl. Phys. {\bf B110}, 410 (1976).

\bibitem{stoler}
P. Stoler, Phys. Rev. D {\bf 44}, 73 (1991).

\bibitem{nl}
S. Nozawa and T.-S. H. Lee, Nucl. Phys. {\bf A 513}, 543 (1990).

\bibitem{fatemi-a}
R. Fatemi, priviate communications.


\bibitem{g1data}
Y. Goto et al, Phys. Rev. {\bf D62}, 034017 (2000);
M. Hirai, S. Kumano and N. Saito,
Phys. Rev. {\bf D69}, 054021 (2004).

\bibitem{anse}
M. Anselmino, B.L. Ioffe, and E. Leader, Yad. Fiz. {\bf 49}, 214 
(1989)[Sov. J. Nucl. Phys. {\bf 49}, 136 (1989)]; F. Close, in
{\it Excited Baryons 1988}, Proceedings of the Topical Workshop on
Excited Baryons, edited by G. Adams, N.C. Mukhopadhyay, 
and P. Stoler (World Scientific, Singapore, 1989); 
V. D. Burkert and B. L. Ioffe, J. Exp. Theore. Phys. {\bf 78}, 619 (1994).

\bibitem{ellis} J. Ellis and R. Jaffe, Phys. Rev. D {\bf 9}, 1444 (1974).

\bibitem{dhg}
S. D. Drell and A.C. Hearn, Phys. Rev. Lett. {\bf 16}, 908 (1966);
S. B. Gerasimov, Sov. J. Nucl. Phys. {\bf 2}, 598 (1966).

\bibitem{g2data}
P.L. Anthony et al. ,
Phys. Lett. {\bf B553}, 18 (2003).

\bibitem{el}
 J. Erler and P. Langacker, Particle Data Group, Phys. Rev. D {\bf 66},
 010001 (2002).


\end{thebibliography}
\end{document}